%% file: mis.tex
\documentclass[floatfix,nofootinbib,showpacs,showkeys,preprintnumbers]{revtex4}
\usepackage{dcolumn}
\usepackage{bm}
\usepackage[centertags]{amsmath}
\allowdisplaybreaks[1]
\usepackage{amsbsy}
\usepackage{amsfonts}
\usepackage{amssymb}
\usepackage[dvips,monochrome]{color}
\usepackage[dvips]{graphicx}
\newcommand{\bin}{\ensuremath{b}}
\begin{document}
\preprint{KIAS-P01051}
\preprint{hep-ph/0111254}
\title{
Model Independent Information On Solar Neutrino Oscillations
}
\author{M.V. Garzelli}
\email{garzelli@to.infn.it}
\affiliation{
Dipartimento di Fisica Teorica, Universit\`a di Torino,
and
INFN, Sezione di Torino,
\\
Via P. Giuria 1, I--10125 Torino, Italy
}
\author{C. Giunti}
\email{giunti@to.infn.it}
\homepage{http://www.to.infn.it/~giunti}
\affiliation{
INFN, Sezione di Torino,
and
Dipartimento di Fisica Teorica, Universit\`a di Torino,\\
Via P. Giuria 1, I--10125 Torino, Italy
\\
and
\\
School of Physics, Korea Institute for Advanced Study,
Seoul 130-012, Korea
}
\date{13 December 2001}
\begin{abstract}
We present the results of a
Bayesian analysis
of solar neutrino data
in terms of $\nu_e\to\nu_{\mu,\tau}$ oscillations,
independent from the Standard Solar Model predictions
for the solar neutrino fluxes.
We show that such a model independent analysis
allows to constraint the values of the neutrino mixing parameters
in limited regions around the usual SMA, LMA, LOW and VO regions.
Furthermore,
there is a strong indication in favor of
large neutrino mixing and large values of $\Delta{m}^2$
(LMA region).
We calculate also the allowed ranges of the neutrino fluxes
and we show that they are in good agreement with the
Standard Solar Model prediction.
In particular, the ratio of
the $^8\mathrm{B}$ flux with its Standard Solar Model prediction
is constrained in the interval
$[
\protect\input{fig/misga_f5_1.tab}
,
\protect\input{fig/misga_f5_2.tab}
]$
with 99.73\% probability.
Finally,
we show that the hypothesis of no neutrino oscillations
is strongly disfavored in a model independent way
with respect to the hypothesis of neutrino oscillations.
\end{abstract}
\maketitle

\section{Introduction}
\label{Introduction}

An important
model independent
evidence in favor of transitions of solar electron neutrinos
in other active states
has been recently obtained from the comparison
of the results of the Super-Kamiokande \cite{SK-sun-01}
and SNO \cite{SNO-01}
solar neutrino experiments
\cite{SNO-01,Fogli:2001vr,Giunti-aoe-01,Fiorentini:2001jt}.
This evidence is supported
\cite{Bahcall:2001pe}
also by the results of the
Homestake \cite{Homestake-98},
GALLEX \cite{GALLEX-99},
SAGE \cite{SAGE-nu00}
and
GNO \cite{GNO-00}
experiments,
that measured different suppressions of the solar electron neutrino flux
with respect to the Standard Solar Model (SSM) \cite{BP2000} prediction.

The simplest and most natural mechanism that explains all solar neutrino data
is $\nu_e\to\nu_{\mu,\tau}$ oscillations,
due to neutrino masses and mixing
(see \cite{Bilenky-Pontecorvo-PR-78,%
Bilenky-Petcov-RMP-87,%
CWKim-book-93,%
BGG-review-98}).
Several groups have recently analyzed
the data of solar neutrino experiments
in terms of neutrino oscillations
\cite{Barger:2001zs,Fogli:2001vr,Bahcall:2001zu,%
Bandyopadhyay:2001aa,Creminelli:2001ij-hepph,Barbieri:2000sv-hepph,%
Kachelriess:2001sg,%
Berezinsky-Lissia-01,Berezinsky:2001se,%
Krastev:2001tv,Garzelli:2001zu,Smy:2001yn,Bilenky:2001tf,%
Strumia:2001gi,Bahcall:2001cb}.
The authors of Refs.~\cite{Fogli:2001vr,Bahcall:2001zu,%
Bandyopadhyay:2001aa,Creminelli:2001ij-hepph,Barbieri:2000sv-hepph,%
Kachelriess:2001sg,%
Krastev:2001tv,Garzelli:2001zu,Bahcall:2001cb}
found the allowed regions in the space of the neutrino mixing parameters
assuming that the neutrino fluxes\footnote{
The flux of solar neutrinos
is composed by eight fluxes,
$pp$,
$pep$,
$hep$,
${^7\mathrm{Be}}$,
${^8\mathrm{B}}$,
${^{13}\mathrm{N}}$,
${^{15}\mathrm{O}}$,
${^{17}\mathrm{F}}$,
produced by the corresponding thermonuclear reactions in the core of the Sun
(see \cite{Bahcall-book-89}).
}
produced in the core of the sun
are given by the BP2000 Standard Solar Model \cite{BP2000}.
The authors of
Refs.~\cite{Bahcall:2001zu,Barbieri:2000sv-hepph,Krastev:2001tv,%
Smy:2001yn,%
Bahcall:2001cb}
performed analyses with the
${^8\mathrm{B}}$ neutrino flux
considered as a free parameter to be determined from the data
(in Ref.~\cite{Barbieri:2000sv-hepph}
also the ${^7\mathrm{Be}}$ neutrino flux
has been considered as a free parameter,
whereas the authors of Ref.~\cite{Krastev:2001tv}
considered also the $hep$ neutrino flux
as a free parameter).

In this paper we calculate model independent
allowed regions in the space of the neutrino mixing parameters.
In the following, by ``model independent''
we mean independent from the solar model.
In other words,
we do not assume that the neutrino fluxes produced in the core of the sun
are given by a Standard Solar Model.

We consider the simplest case of mixing of two neutrinos
$\nu_e$ and $\nu_a$,
where $\nu_a$ is a linear combination of $\nu_\mu$ and $\nu_\tau$:
\begin{align}
\null & \null
\nu_e = \cos\vartheta \, \nu_1 + \sin\vartheta \, \nu_2
\,,
\nonumber
\\
\null & \null
\nu_a = -\sin\vartheta \, \nu_1 + \cos\vartheta \, \nu_2
\,.
\label{mixing}
\end{align}
Here,
$\vartheta$ is the mixing angle
and
$\nu_1$, $\nu_2$
are massive neutrinos
with masses
$m_1$, $m_2$,
respectively.
Neutrino oscillations depend on the mixing angle $\vartheta$
and the squared-mass difference
$\Delta{m}^2 \equiv m_2^2 - m_1^2$
(see \cite{Bilenky-Pontecorvo-PR-78,%
Bilenky-Petcov-RMP-87,%
CWKim-book-93,%
BGG-review-98}).

Let us notice,
that
the treatment of solar $\nu_e\to\nu_a$ transitions
in the framework of the simple two-neutrino mixing scheme in Eq.~(\ref{mixing})
is valid with good approximation also in the case of mixing
of three neutrinos
($\nu_e$, $\nu_\mu$, $\nu_\tau$)
with small $U_{e3}$
\cite{Bilenky-Giunti-CHOOZ-98,Fogli:1999zg},
as indicated by the results of the CHOOZ
long-baseline $\bar\nu_e$ disappearance experiment
\cite{Apollonio:1999ae}.
Furthermore,
the treatment is also valid in 3+1 four-neutrino mixing schemes
in which
the sterile neutrino
is practically decoupled from the active ones
($1-|U_{s4}|^2 \ll 1$)
\cite{Barger-Fate-2000}
and
$U_{e3}$
is small, as indicated
\cite{Giunti-Laveder-3+1-00}
by CHOOZ data.

Our statistical analysis of the data is Bayesian.
We have already explained in Ref.~\cite{Garzelli:2001zu}
the advantages of using
Bayesian Probability Theory in the analysis of solar neutrino data\footnote{
Bayesian Probability Theory
has been applied to the analysis of solar neutrino data
also in Refs.~\cite{Bhat:1998qq,Creminelli:2001ij-hepph}.
}
(for a general introduction to Bayesian Probability Theory see
\cite{Jeffreys-book-39,Loredo-90,Loredo-92,Jaynes-book-95,D'Agostini-99}).
Another model independent analysis of solar neutrino data
was performed several years ago
\cite{Bilenky:1994ti}
using a different statistical approach.
Apart from the differences due to the statistical methods,
the main difference with the analysis performed in
Ref.~\cite{Bilenky:1994ti}
is the impressive increase of the quality and quantity of data,
that,
as we will show,
allow now to obtain much more stringent model independent information.

As discussed in Ref.~\cite{Garzelli:2001zu},
the Bayesian analysis of solar neutrino data
assuming the neutrino fluxes predicted by the Standard Solar Model
takes into account the uncertainty of the fluxes
in the covariance matrix of the least-squares function
of the data,
assuming a multi-normal distribution
for the errors of the fluxes.
This approach cannot be extended to the model independent analysis
of solar neutrino data,
in which the fluxes must be considered as free parameters
that may be determined from the data
in a similar way as the neutrino mixing parameters.
Therefore,
we have developed a new approach
in which the neutrino fluxes are considered as unknown quantities
with a flat prior probability distribution.
The model independent information on neutrino mixing
is calculated
through marginalization of the posterior probability distribution
with respect to the neutrino fluxes.
We also derive information on each neutrino flux
through marginalization of the posterior probability distribution
with respect to the other neutrino fluxes and
the oscillation parameters.

In the framework under consideration,
the theoretical calculation
of the
rates measured in solar neutrino experiments
depends on the neutrino mixing parameters
$\tan^2\vartheta$
and
$\Delta{m}^2$,
on the neutrino fluxes produced in the core of the sun
by the thermonuclear reaction of the $pp$ and CNO cycles,
and on the energy-averaged
cross sections of neutrino interaction in the detectors.
In order to have a uniform treatment of these
uncertain theoretical quantities,
we consider all of them as parameters
that have some uncertainty,
which is taken into account in the prior probability distribution.
This approach is different from the standard one
(see \cite{Garzelli:2001zu}),
in which the uncertainties of the fluxes and cross sections
are taken into account in the covariance matrix
that determines the likelihood function
(sampling distribution).
However,
we think that the new approach
is equally, if not more,
reasonable than the standard one,
because the values of the fluxes and cross sections
are part of our prior knowledge.
In any case,
the two approaches should give approximately the same result
in the model dependent analysis
in which the neutrino fluxes are assumed to be those
predicted by the Standard Solar Model.
We check this consistency in Section~\ref{Model Dependent Analysis}.

In the following it is convenient to
group the solar neutrino fluxes in the array
\begin{equation}
\mathbf{\Phi}
=
(\Phi_{1}, \ldots, \Phi_{8})
\,.
\label{fluxes}
\end{equation}
The neutrino flux
$\Phi_{i}$
is produced in the $i^{\mathrm{th}}$
thermonuclear reaction,
with
$i=1,\ldots,8$
for the
$pp$,
$pep$,
$hep$,
${^7\mathrm{Be}}$,
${^8\mathrm{B}}$,
${^{13}\mathrm{N}}$,
${^{15}\mathrm{O}}$,
${^{17}\mathrm{F}}$
thermonuclear reactions,
respectively.

In both the model dependent and independent analyses
we perform a Rates Analysis
that takes into account the total rates
measured in the
Homestake,
GALLEX+GNO+SAGE,
SNO,
Super-Kamiokande experiments,
and a Global Analysis
in which the Super-Kamiokande total rate is replaced by
the 38 bins of the Super-Kamiokande day and night electron energy spectra.
Hence,
calling $N_{\mathrm{S}}$ the number of solar data points,
we have $N_{\mathrm{S}}=4$ in the Rates Analysis
and
$N_{\mathrm{S}}=41$ in the Global Analysis.
We call $C_{ij}$ the
energy-averaged cross section of detection
of the $i^{\mathrm{th}}$ neutrino flux
for the $j^{\mathrm{th}}$ solar data point,
with $j=1,\ldots,N_{\mathrm{S}}$.
We group these cross sections in the
$8{\times}N_{\mathrm{S}}$
matrix
\begin{equation}
\mathbf{C}
=
\begin{pmatrix}
C_{11} & \cdots & C_{1N_{\mathrm{S}}}
\\
\vdots & \vdots & \vdots
\\
C_{81} & \cdots & C_{8N_{\mathrm{S}}}
\end{pmatrix}
\,.
\label{C}
\end{equation}

Bayesian Probability Theory
allows to calculate the posterior probability distribution
$p(\tan^2\!\vartheta,\Delta{m}^2,\mathbf{\Phi},\mathbf{C}|\mathrm{D},\mathrm{I})$
for the neutrino mixing parameters,
the neutrino fluxes
and the detection cross sections
through Bayes Theorem:
\begin{equation}
p(\tan^2\!\vartheta,\Delta{m}^2,\mathbf{\Phi},\mathbf{C}|\mathrm{D},\mathrm{I})
=
\frac{
p(\mathrm{D}|\tan^2\!\vartheta,\Delta{m}^2,\mathbf{\Phi},\mathbf{C},\mathrm{I})
\,
p(\tan^2\!\vartheta,\Delta{m}^2|\mathrm{I})
\,
p(\mathbf{\Phi}|\mathrm{I})
\,
p(\mathbf{C}|\mathrm{I})
}{
p(\mathrm{D}|\mathrm{I})
}
\,,
\label{Bayes1}
\end{equation}
where
$\mathrm{D}$
represents the data
and
$\mathrm{I}$
represents all the background information and assumptions
on solar physics, neutrino physics,
etc.
The \emph{sampling distribution}
$p(\mathrm{D}|\tan^2\!\vartheta,\Delta{m}^2,\mathbf{\Phi},\mathbf{C},\mathrm{I})$
is also known as \emph{likelihood function}.
The \emph{prior distributions}
$p(\tan^2\!\vartheta,\Delta{m}^2|\mathrm{I})$,
$p(\mathbf{\Phi}|\mathrm{I})$
and
$p(\mathbf{C}|\mathrm{I})$
quantify our prior knowledge,
or lack of knowledge,
on the values of the
mixing parameters, neutrino fluxes and detection cross sections,
respectively.
The probability distribution
$p(\mathrm{D}|\mathrm{I})$
is known as \emph{global likelihood}
and acts as a normalization constant.

The prior probability distribution of mixing parameters,
$p(\tan^2\!\vartheta,\Delta{m}^2|\mathrm{I})$,
must be chosen in order to quantify appropriately
the prior lack of knowledge on the values of
the parameters
$\tan^2\!\vartheta$ and $\Delta{m}^2$.
Assuming $\nu_e\to\nu_{\mu,\tau}$ oscillations
(included in $\mathrm{I}$),
we know that solar neutrino data are sensitive to
several different order of magnitude of
$\tan^2\!\vartheta$ and $\Delta{m}^2$,
through vacuum oscillations
\cite{Gribov:1969kq}
or resonant MSW transitions
\cite{Wolfenstein:1978ue,Mikheev:1985gs}.
Therefore,
the most reasonable non-informative
prior probability distribution function,
that we will use in the following,
is a flat distribution in the
$\log(\tan^2\!\vartheta)$--$\log(\Delta{m}^2)$ plane.
Indeed,
the authors of Ref.~\cite{Creminelli:2001ij}
have chosen, independently,
the same prior,
which is consistent with the calculation of the
credible regions for the neutrino oscillation parameters
in the
$\log(\tan^2\!\vartheta)$--$\log(\Delta{m}^2)$ plane.

Using a flat prior probability distribution in the
$\log(\tan^2\!\vartheta)$--$\log(\Delta{m}^2)$ plane,
Eq.~(\ref{Bayes1})
becomes
\begin{equation}
p(\tan^2\!\vartheta,\Delta{m}^2,\mathbf{\Phi},\mathbf{C}|\mathrm{D},\mathrm{I})
=
\frac{
p(\mathrm{D}|\tan^2\!\vartheta,\Delta{m}^2,\mathbf{\Phi},\mathbf{C},\mathrm{I})
\,
p(\mathbf{\Phi}|\mathrm{I})
\,
p(\mathbf{C}|\mathrm{I})
}{
\int\!\mathrm{d}\!\log(\tan^2\!\vartheta) \, \mathrm{d}\!\log(\Delta{m}^2) \,
p(\mathrm{D}|\tan^2\!\vartheta,\Delta{m}^2,\mathbf{\Phi},\mathbf{C},\mathrm{I})
\,
p(\mathbf{\Phi}|\mathrm{I})
\,
p(\mathbf{C}|\mathrm{I})
}
\,,
\label{Bayes2}
\end{equation}
where we have expressed the global likelihood
$p(\mathrm{D}|\mathrm{I})$
as the appropriate normalization constant.
Integral probabilities must be calculated integrating
$p(\tan^2\!\vartheta,\Delta{m}^2,\mathbf{\Phi},\mathbf{C}|\mathrm{D},\mathrm{I})$
over
$
\mathrm{d}\!\log(\tan^2\!\vartheta)
\,
\mathrm{d}\!\log(\Delta{m}^2)
\,
\mathrm{d}\mathbf{\Phi}
\,
\mathrm{d}\mathbf{C}
$.

The prior probability distribution of cross sections,
$p(\mathbf{C}|\mathrm{I})$,
represents our knowledge of the cross sections,
which is reasonably accurate.
Therefore,
we take this prior probability distribution
as a multinormal distribution as explained in Section~\ref{Cross Sections}.

As explained in Subsection~\ref{SSM Prior Distribution of Fluxes},
in the Standard Solar Model dependent analysis
the prior probability distribution of neutrino fluxes,
$p(\mathbf{\Phi}|\mathrm{I})$,
is a multinormal distribution.
In the model independent analysis
we consider a flat prior probability distribution of neutrino fluxes,
as explained in Subsection~\ref{Model Independent Prior Distribution of Fluxes}.

The posterior distribution for the oscillation
parameters $\tan^2\vartheta$, $\Delta{m}^2$
is obtained through marginalization
of the posterior distribution (\ref{Bayes2})
with respect to the neutrino fluxes
and cross sections:
\begin{equation}
p(\tan^2\!\vartheta,\Delta{m}^2|\mathrm{D},\mathrm{I})
=
\int\!\mathrm{d}\mathbf{\Phi}
\int\!\mathrm{d}\mathbf{C}
\,
p(\tan^2\!\vartheta,\Delta{m}^2,\mathbf{\Phi},\mathbf{C}|\mathrm{D},\mathrm{I})
\,.
\label{posterior1}
\end{equation}
Using this posterior distribution,
we calculate Bayesian
\emph{credible regions}\footnote{
Credible regions,
also known as
\emph{highest posterior density regions},
contain a specified fraction of
the posterior probability and all values of
the parameters inside the credible regions
have higher probability density than those outside
(see, for example, Ref.~\cite{Loredo-90}).
}
in the plane of the oscillation parameters
$\tan^2\!\vartheta$ and $\Delta{m}^2$.

In both the Rates Analysis and the Global Analysis
we consider also the data of the CHOOZ experiment,
that are important
because they exclude large mixing for
$\Delta{m}^2 \gtrsim 10^{-3} \, \mathrm{eV}^2$
\cite{Apollonio:1999ae}.

For convenience,
in the discussion we use the standard
SMA, LMA, LOW and VO
names
for the regions in the
$\tan^2\!\vartheta$--$\Delta{m}^2$ plane.
We consider
these regions as enlarged with respect to the usual ones
(see \cite{Garzelli:2001zu}),
because the model independent credible regions that we obtain
are slightly larger than the corresponding
model dependent ones:
Small Mixing Angle (SMA) for
\begin{equation}
\tan^2\!\vartheta
\lesssim
10^{-2}
\,,
\quad
10^{-8} \, \mathrm{eV}^2
\lesssim
\Delta{m}^2
\lesssim
10^{-3} \, \mathrm{eV}^2
\,,
\label{SMA}
\end{equation}
Large Mixing Angle (LMA) for
\begin{equation}
10^{-2}
\lesssim
\tan^2\!\vartheta
\lesssim
10^2
\,,
\quad
3 \times 10^{-6} \, \mathrm{eV}^2
\lesssim
\Delta{m}^2
\lesssim
10^{-3} \, \mathrm{eV}^2
\,,
\label{LMA}
\end{equation}
LOW for
\begin{equation}
10^{-2}
\lesssim
\tan^2\!\vartheta
\lesssim
10^2
\,,
\quad
10^{-8} \, \mathrm{eV}^2
\lesssim
\Delta{m}^2
\lesssim
3 \times 10^{-6} \, \mathrm{eV}^2
\,,
\label{LOW}
\end{equation}
Vacuum Oscillation (VO) for
\begin{equation}
10^{-2}
\lesssim
\tan^2\!\vartheta
\lesssim
10^2
\,,
\quad
\Delta{m}^2
\lesssim
10^{-8} \, \mathrm{eV}^2
\,.
\label{VO}
\end{equation}

The plan of this paper is as follows.
In Section~\ref{Sampling Distribution} we present the sampling distributions
in the two types of analysis that we perform:
the Rates Analysis of
Homestake,
GALLEX+GNO+SAGE,
SNO and
Super-Kamiokande total rates
and the Global Analysis of
Homestake,
GALLEX+GNO+SAGE and
SNO rates
and
Super-Kamiokande day and night spectra.
In Section~\ref{Cross Sections}
we present the prior distributions for the cross sections
in the Rates and Global Analyses.
In Section~\ref{Model Dependent Analysis}
we perform a SSM model dependent analysis in order to check
the validity of our approach
by comparing the results with those
obtained with a standard Bayesian analysis in Ref.~\cite{Garzelli:2001zu}.
In Section~\ref{Model Independent Analysis}
we present and discuss the results of the model independent analysis
of solar neutrino data.
In the concluding Section~\ref{Conclusions} we summarize our results.

\section{Sampling Distribution}
\label{Sampling Distribution}

Following the tradition,
we perform two types of analysis,
with different data sets.
In the Rates Analysis (Subsection~\ref{Sampling: Rates Analysis})
we consider the total rates
measured by solar neutrino experiments
(Homestake,
GALLEX+GNO+SAGE,
SNO,
Super-Kamiokande).
In the Global Analysis (Subsection~\ref{Sampling: Global Analysis})
we consider
the total rates of the
Homestake,
GALLEX+GNO+SAGE,
SNO experiments
and
the Super-Kamiokande day and night electron energy spectra.
Both analyses take into account also
the data of the CHOOZ experiment
\cite{Apollonio:1999ae}.

Since the solar and CHOOZ data are independent,
the sampling probability distribution can be written as
\begin{equation}
p(\mathrm{D}|\tan^2\!\vartheta,\Delta{m}^2,\mathbf{\Phi},\mathbf{C},\mathrm{I})
=
p(\mathrm{D}_{\mathrm{S}}|\tan^2\!\vartheta,\Delta{m}^2,\mathbf{\Phi},\mathbf{C},\mathrm{I})
\,
p(\mathrm{D}_{\mathrm{C}}|\tan^2\!\vartheta,\Delta{m}^2,\mathrm{I})
\,,
\label{sampling1}
\end{equation}
where
$\mathrm{D}_{\mathrm{S}}$
represents the solar data
and
$\mathrm{D}_{\mathrm{C}}$
represents the positron spectra measured in the CHOOZ experiment,
that are obviously independent from the solar neutrino fluxes
and detection cross sections.

Assuming a normal distribution of experimental errors,
the sampling probability distribution of CHOOZ data
is given by
\begin{equation}
p(\mathrm{D}_{\mathrm{C}}|\tan^2\!\vartheta,\Delta{m}^2,\mathrm{I})
=
\int
\mathrm{d} \alpha_{\mathrm{C}}
\,
\frac{
e^{-X^2_{\mathrm{C}}/2}
}{
(2\pi)^{N_{\mathrm{C}}/2}\sqrt{|V_{\mathrm{C}}|}
}
\,,
\label{sampling2}
\end{equation}
where
we have marginalized over the nuisance parameter
$\alpha_{\mathrm{C}}$,
that is the absolute normalization constant of CHOOZ positron spectra
\cite{Apollonio:1999ae}.
Here
$N_{\mathrm{C}}=14$
is the number of CHOOZ data points,
$X^2_{\mathrm{C}}$
is the CHOOZ least-squares function
and $V_{\mathrm{C}}$ is the corresponding covariance matrix of uncertainties.

The CHOOZ least-squares function
$X^2_{\mathrm{C}}$
is calculated as in the analysis A of Ref.~\cite{Apollonio:1999ae},
with the following approximations.
Since we do not know the
antineutrino spectrum,
the spatial distribution functions of the reactor cores and detector
and the detector response function linking the real and visible
positron energies,
for each energy bin we calculated the oscillation
probability at the average energy
of the bin and at the average distance of
the detector from each of the two reactors
\cite{Apollonio:1998xe}.
This approximation is quite good
because we are interested in small values of
$\Delta{m}^2$,
for which the energy and distance dependence of
the survival probability of the $\bar\nu_e$'s
in the CHOOZ experiment is very weak.
We calculate $X^2_{\mathrm{C}}$
as in Eq.~(13) of Ref.~\cite{Apollonio:1999ae},
with the only difference that we neglect the energy-scale
calibration factor,
whose small uncertainty (1.1\%) is practically negligible:
\begin{equation}
X^2_{\mathrm{C}}
=
\sum_{j_1,j_2=1}^{N_{\mathrm{C}}}
\left(
R^{\mathrm{(th)}}_{j_1}
-
\alpha_{\mathrm{C}}
\,
R^{\mathrm{(ex)}}_{j_1}
\right)
(V^{-1}_{\mathrm{C}})_{j_1j_2}
\left(
R^{\mathrm{(th)}}_{j_2}
-
\alpha_{\mathrm{C}}
\,
R^{\mathrm{(ex)}}_{j_2}
\right)
+
\left(
\frac{\alpha_{\mathrm{C}}-1}{\sigma_{\alpha_{\mathrm{C}}}}
\right)^2
\,,
\label{X2CHOOZ}
\end{equation}
where
$\sigma_{\alpha_{\mathrm{C}}} = 2.7 \times 10^{-2}$
is the uncertainty
of the absolute normalization constant
$\alpha_{\mathrm{C}}$
\cite{Apollonio:1999ae}.
We calculate the CHOOZ covariance matrix
$V_{\mathrm{C}}$
as described in Eq.~(12) of Ref.~\cite{Apollonio:1999ae}.
The only missing information in Ref.~\cite{Apollonio:1999ae}
is the value of the systematic uncertainties of the positron energy bins,
for which only the values for positron energy
2 and 6 MeV are given.
For the other bins we take systematic uncertainties
interpolated linearly between these two values.

\subsection{Rates Analysis}
\label{Sampling: Rates Analysis}

Assuming a normal distribution of experimental errors,
the sampling probability distribution of solar data
in the Rates Analysis
is given by
\begin{equation}
p(\mathrm{D}_{\mathrm{S}}|\tan^2\!\vartheta,\Delta{m}^2,\mathbf{\Phi},\mathbf{C},\mathrm{I})
=
\frac{
e^{-X^2_{\mathrm{S}}/2}
}{
(2\pi)^{N_{\mathrm{S}}/2}\sqrt{|V_{\mathrm{S}}|}
}
\,,
\label{sampling3}
\end{equation}
where
$N_{\mathrm{S}}=4$
is the number of solar data points.
$X^2_{\mathrm{S}}$
is the solar least-squares function
and $V_{\mathrm{S}}$ is the corresponding covariance matrix.

The solar least squares function $X^2_{\mathrm{S}}$
is given by
\begin{equation}
X^2_{\mathrm{S}}
=
\sum_{j=1}^{N_{\mathrm{S}}}
\frac
{ \left( R^{\mathrm{(ex)}}_{j} - R^{\mathrm{(th)}}_{j} \right)^2 }
{ \sigma^2_{j} }
\,,
\label{X2sunRates}
\end{equation}
where
$R^{\mathrm{(ex)}}_{j}$
is the event rate measured in the $j^{\mathrm{th}}$ experiment
and
$R^{\mathrm{(th)}}_{j}$
is the corresponding theoretical event rate,
that depends on
$\Delta{m}^2$, $\tan^2\theta$, the neutrino fluxes
and the neutrino interaction cross sections in the detectors.
The index $j=1,\ldots,4$
indicate the four types of solar neutrino experiments
listed in Table~\ref{rates}
together with the corresponding event rates
and experimental uncertainties
$\sigma_{j}$,
calculated by adding in quadrature the statistical and systematic
uncertainties for each experiment.

Since the uncertainties of the rates of different experiments are uncorrelated,
the determinant $|V_{\mathrm{S}}|$ of the solar covariance
matrix in Eq.~(\ref{sampling3}) is simply given by
\begin{equation}
|V_{\mathrm{S}}|
=
\prod_{j=1}^{N_{\mathrm{S}}}
\sigma^2_{j}
\,.
\label{DetVSrates}
\end{equation}

The theoretical event rate $R_{j}^{\mathrm{(th)}}$
can be written as
\begin{equation}
R_{j}^{\mathrm{(th)}}
=
\sum_{i=1}^{8}
R_{ij}^{\mathrm{(th)}}
\,,
\label{Rj}
\end{equation}
where
\begin{equation}
R_{ij}^{\mathrm{(th)}}
=
\Phi_{i}
\
C_{ij}
\
P_{ij}(\Delta{m}^2,\tan^2\theta)
\,,
\label{Rij}
\end{equation}
is the theoretical event rate in the $j^{\mathrm{th}}$
experiment
due to the neutrino flux
$\Phi_{i}$
produced in the $i^{\mathrm{th}}$
thermonuclear reaction of the $pp$
and CNO cycles in the sun
($i=1,\ldots,8$
for the
$pp$,
$pep$,
$hep$,
${^7\mathrm{Be}}$,
${^8\mathrm{B}}$,
${^{13}\mathrm{N}}$,
${^{15}\mathrm{O}}$,
${^{17}\mathrm{F}}$
thermonuclear reactions,
respectively).
$C_{ij}$
is the corresponding energy-averaged cross section
and
$P_{ij}(\Delta{m}^2,\tan^2\theta)$
is the corresponding averaged survival probability of solar $\nu_e$'s,
that depends on $\Delta{m}^2$ and $\tan^2\theta$
(in the case of the Super-Kamiokande experiment, $j=4$,
also the averaged $\nu_e\to\nu_a$ transition probability
must be properly taken into account).

For the calculation of the probabilities
$P_{ij}(\Delta{m}^2,\tan^2\theta)$
we have used the tables of neutrino spectra, solar 
density and radiochemical detector cross sections available in
Bahcall's web pages \cite{Bahcall-WWW},
BP2000 Standard Solar Model \cite{BP2000}.
For the calculation of the theoretical rate of the SNO experiment
we used the charged-current cross section
given in Refs.~\cite{Kubodera-deuteron-01,Kubodera-www,Butler:2000zp}.
We did not consider the contribution of radiative corrections
discussed in Refs.~\cite{Beacom:2001hr,Kurylov:2001av},
which increase the neutrino-deuteron cross section by a few percents,
that is smaller than the experimental uncertainty of the SNO event rate
\cite{SNO-01,Poon:2001ee}.

The probability of neutrino oscillations
is calculated with an unified approach that allows to
pass continuously from the vacuum oscillation regime to MSW transitions
\cite{Wolfenstein:1978ue,Mikheev:1985gs} 
through the quasi-vacuum regime
\cite{Friedland-vo-00,Fogli-Lisi-Montanino-Palazzo-Quasi-vacuum-00},
using the quasi-vacuum analytical
prescription given in Ref.~\cite{Lisi:2000su}
(see also Refs.~\cite{Petcov:1988wv,Petcov:1989du}),
the usual prescription
for the MSW survival probability
(see
\cite{Fogli-Lisi-Montanino-Palazzo-Quasi-vacuum-00,%
Gonzalez-Garcia:2000sk})
and the level crossing probability 
appropriate for an exponential density profile
\cite{Petcov-analytic-87,Kuo-Pantaleone-RMP-89}.
We calculate the regeneration in the Earth
using a two-step model of the Earth density profile
\cite{Liu-Maris-Petcov-earth1-97,Petcov-diffractive-98,%
Akhmedov-parametric-99,%
Chizhov-Petcov-earth-1-99,Chizhov-Petcov-earth-2-99},
that is known to produce results that do not differ appreciably
from those obtained with
a less approximate
density profile.

\subsection{Global Analysis}
\label{Sampling: Global Analysis}

In the Global Analysis,
instead of the total rate of the Super-Kamiokande experiment
we consider the data on the
Super-Kamiokande day and night electron energy spectra
presented in Ref.~\cite{SK-sun-hep-ex-0103032},
that contain information on the total rate
plus the shape of the energy spectrum.
Assuming a normal distribution of experimental errors,
the sampling probability distribution of solar data
in the Global Analysis
is given by
\begin{equation}
p(\mathrm{D}_{\mathrm{S}}|\tan^2\!\vartheta,\Delta{m}^2,\mathbf{\Phi},\mathbf{C},\mathrm{I})
=
\frac{
e^{-X^2_{\mathrm{S}}/2}
}{
(2\pi)^{N_{\mathrm{S}}/2}\sqrt{|V_{\mathrm{S}}|}
}
\frac{
e^{-X^2_{\mathrm{SK}}/2}
}{
(2\pi)^{N_{\mathrm{SK}}/2}\sqrt{|V_{\mathrm{SK}}|}
}
\,,
\label{sampling4}
\end{equation}
where
the first factor takes into account the rates measured in the
Homestake,
GALLEX+GNO+SAGE
and
SNO experiments.
Therefore, the first factor in Eq.~(\ref{sampling4})
is calculated as in the Rates Analysis,
with the only difference that
$N_{\mathrm{S}}=3$.

The second factor in Eq.~(\ref{sampling4})
takes into account the Super-Kamiokande data on the
day and night electron energy spectra.

The Super-Kamiokande least-squares function
$X^2_{\mathrm{SK}}$
is given by
\begin{equation}
X^2_{\mathrm{SK}}
=
\sum_{\bin_1,\bin_2=1}^{38}
\left( R^{\mathrm{(ex)}}_{\bin_1} - R^{\mathrm{(th)}}_{\bin_1} \right)
(V^{-1}_{\mathrm{SK}})_{\bin_1\bin_2}
\left( R^{\mathrm{(ex)}}_{\bin_2} - R^{\mathrm{(th)}}_{\bin_2} \right)
\,,
\label{X2sunGlobalSK}
\end{equation}
with
the indexes $\bin_1$, $\bin_2$
run from 1 to 38
for the bins of the Super-Kamiokande day and night electron energy spectra
given in Table~III of Ref.~\cite{SK-sun-hep-ex-0103032}.
The indexes $\bin_1, \bin_2 = 1, \ldots, 19$
refer to the day spectrum
and
the indexes $\bin_1, \bin_2 = 20, \ldots, 38$
refer to the night spectrum.
The covariance matrix $V_{\mathrm{SK}}$ is written as
\begin{equation}
(V_{\mathrm{SK}})_{\bin_1 \bin_2}
=
\delta_{\bin_1 \bin_2}
\left[
(\sigma^{\mathrm{(sta)}})_{\bin_1}^2
+
(\sigma^{\mathrm{(sys)}}_{\mathrm{unc}})_{\bin_1}^2
\right]
+
(\sigma^{\mathrm{(sys)}}_{\mathrm{cor}})_{\bin_1}
\,
(\sigma^{\mathrm{(sys)}}_{\mathrm{cor}})_{\bin_2}
+
(\sigma^{\mathrm{(sys)}}_{\mathrm{unc}})_{\bin_1}^2
\left[
\delta_{\bin_1(\bin_2-19)}
+
\delta_{\bin_1(\bin_2+19)}
\right]
\,.
\label{VSK}
\end{equation}
The statistical uncertainties
$(\sigma^{\mathrm{(sta)}})_\bin$
are given
in the third (for $\bin=1,\ldots,19$, day spectrum)
and fourth (for $\bin=20,\ldots,38$, night spectrum)
columns of Table~III in Ref.~\cite{SK-sun-hep-ex-0103032}.
The experimental uncorrelated systematic uncertainties
$(\sigma^{\mathrm{(sys)}}_{\mathrm{unc}})_\bin$
(for $\bin=1,\ldots,38$)
are given by
$
R^{\mathrm{(ex)}}_{\bin}
\,
\delta^{\mathrm{(ex)}}_{\bin,\mathrm{unc}}
$,
with $R^{\mathrm{(ex)}}_{\bin}$
listed in the third (for $\bin=1,\ldots,19$, day spectrum)
and fourth (for $\bin=20,\ldots,38$, night spectrum)
columns
of Table~III in Ref.~\cite{SK-sun-hep-ex-0103032}.
The last term in Eq.~(\ref{VSK}) is due to the assumption
of a full correlation of the
systematic uncertainties of the day and night bins
with the same energy.
The values of
$\delta^{\mathrm{(ex)}}_{\bin,\mathrm{unc}}=\delta^{\mathrm{(ex)}}_{\bin+19,\mathrm{unc}}$
for $\bin=1,\ldots,19$
are listed in the sixth column of Table~III in
Ref.~\cite{SK-sun-hep-ex-0103032}
(we took the bigger between the positive and negative values).

The experimental correlated systematic uncertainties are given by
$
(\sigma^{\mathrm{(sys)}}_{\mathrm{cor}})_\bin
=
R^{\mathrm{(ex)}}_{\bin}
\,
\delta^{\mathrm{(sys)}}_{\bin,\mathrm{cor}}
$.
Unfortunately,
the relative correlated systematic uncertainties
presented by the Super-Kamiokande collaboration
in the fifth column of Table~III in
Ref.~\cite{SK-sun-hep-ex-0103032}
contain the contribution from the energy-dependent part of
the theoretical uncertainty
on the ${^8\mathrm{B}}$ neutrino spectrum given in Ref.~\cite{Ortiz:2000nf},
that is taken into account by the prior probability distribution
discussed in Subsection~\ref{Cross: Global Analysis}.
We list in Table~\ref{delta}
the values of relative theoretical uncertainty
$\delta^{\mathrm{(th)}}_{\bin,\mathrm{cor}}$
for the rates in the Super-Kamiokande spectral bins ($\bin=1,\ldots,38$)
due to the theoretical uncertainty
on the ${^8\mathrm{B}}$ neutrino spectrum
extracted
from Fig.~5 of Ref.~\cite{Ortiz:2000nf}.
In order to avoid double-counting,
we subtracted quadratically
the energy-dependent part of $\delta^{\mathrm{(th)}}_{\bin,\mathrm{cor}}$
(\textit{i.e.}
$
\delta^{\mathrm{(th)}}_{\bin,\mathrm{cor}}
-
\delta^{\mathrm{(th)}}_{1,\mathrm{cor}}
$,
where $\bin=1$ is the index of the first Super-Kamiokande energy bin)
from the relative correlated systematic uncertainties
in the fifth column of Table~III in
Ref.~\cite{SK-sun-hep-ex-0103032}
(for which we took the bigger between the positive and negative values).
The resulting values of
$\delta^{\mathrm{(sys)}}_{\bin,\mathrm{cor}}$
are listed in Table~\ref{delta}.

\section{Cross Sections}
\label{Cross Sections}

In this section we discuss the prior distribution
$p(\mathbf{C}|\mathrm{I})$
for the
energy-averaged cross sections
of neutrino interaction in the detectors.
In the case of the Super-Kamiokande experiment
there is practically no uncertainty on the value of the neutrino-electron
elastic cross section,
but there is an uncertainty of the shape of the ${^8\mathrm{B}}$
neutrino spectrum
\cite{Bahcall:1996qv,Ortiz:2000nf}.
We take this uncertainty into account
both in the Rates Analysis and Global Analysis,
as described in the following subsections.

\subsection{Rates Analysis}
\label{Cross: Rates Analysis}

In the Rates Analysis
the number of experimental data points is
$N_{\mathrm{S}}=4$
(see Table~\ref{rates}),
leading to a $8\times4$
matrix $\mathbf{C}$
of energy averaged cross sections.

Assuming
a complete correlation of the errors of the
energy-averaged cross sections of different neutrino fluxes in each experiment
\cite{Garzelli-Giunti-cs-00},
the cross section $C_{ij}$
of neutrino detection in the $j^{\mathrm{th}}$
experiment
averaged over the energy spectrum of the
$i^{\mathrm{th}}$
neutrino flux
can be written as
\begin{equation}
C_{ij}
=
\left( 1 + \xi_j \, \Delta\ln\!C_{ij} \right) C_{ij}^0
\,,
\label{cross2}
\end{equation}
where
$C_{ij}^0$
is the standard averaged cross section,
$\Delta\ln\!C_{ij}$
is the relative uncertainty of $C_{ij}^0$,
and $\xi_j$ is an unknown parameter that quantifies the deviation of
the averaged cross section from its standard value.
Assuming a normal distribution of errors,
for each $\xi_j$ we adopt the prior distribution
\begin{equation}
p(\xi_j|\mathrm{I})
=
\frac{e^{-\xi_j^2/2}}{\sqrt{2\pi}}
\,.
\label{cross1}
\end{equation}
The prior distributions for different
$\xi_j$'s are independent because
the theoretical errors
of the energy-averaged cross sections
of neutrino interaction in different experiments
are uncorrelated.
The prior distribution $p(\mathbf{C}|\mathrm{I})$
of energy averaged cross section is obtained from Eqs.~(\ref{cross2})
and (\ref{cross1}).
In practice,
since all quantities that we calculate are integrated over
$\mathrm{d}\mathbf{C}$ using a Monte Carlo,
the prior distribution $p(\mathbf{C}|\mathrm{I})$
is generated by the Monte Carlo
from the prior distribution (\ref{cross1}) of $\xi_j$,
using Eq.~(\ref{cross2}).

This procedure may seem arbitrarily complicated,
but one must notice that
it is not possible to write a multi-normal
distribution for fully correlated quantities
as $C_{ij}$ for $i=1,\ldots,8$,
because the covariance matrix would be singular.
Indeed,
complete correlation means that the errors are not independent:
the deviation of one $C_{ij}$
with respect to $C_{ij}^0$
determines the deviations of all the others
$C_{i'j}$ from $C_{i'j}^0$,
with $i' \neq i$,
as quantified in Eq.~(\ref{cross2}).

The values of $C_{ij}^0$ that we use
for the Chlorine ($j=1$)
and Gallium ($j=2$)
experiments
are given in Table~V of Ref.~\cite{Fogli:1999zg}.
For the SNO experiment ($j=3$)
with kinetic energy threshold
$T_e^{\mathrm{SNO}} = 6.75 \, \mathrm{MeV}$
for the detected electrons,
we calculated
\begin{equation}
C_{33}^0 = 1.82 \times 10^{-7} \, \mathrm{cm}^2 \, \mathrm{s}
\,,
\qquad
C_{53}^0 = 1.98 \times 10^{-7} \, \mathrm{cm}^2 \, \mathrm{s}
\,,
\label{cross3}
\end{equation}
and all the other
$C_{i3}^0$ are zero.
For the Super-Kamiokande experiment ($j=4$)
with kinetic energy threshold
$T_e^{\mathrm{SK}} = 4.50 \, \mathrm{MeV}$
for the detected electrons,
we calculated
\begin{equation}
C_{34}^0 = 1.04 \times 10^{-7} \, \mathrm{cm}^2 \, \mathrm{s}
\,,
\qquad
C_{54}^0 = 1.98 \times 10^{-7} \, \mathrm{cm}^2 \, \mathrm{s}
\,,
\label{cross4}
\end{equation}
and all the other
$C_{i4}^0$ are zero.

The values of the relative uncertainties
$
\Delta\ln\!C_{ij}
=
\Delta{C}_{ij} / C_{ij}
$
for $j=1,2$
(${}^{37}\mathrm{Cl}$,
${}^{71}\mathrm{Ga}$
experiments)
are given in Table~VI of Ref.~\cite{Fogli:1999zg}.
For the SNO experiment
($j=3$),
$
\Delta\ln\!C_{33}
=
\Delta\ln\!C_{53}
=
3.0 \times 10^{-2}
$
\cite{Kubodera-deuteron-01,Butler:2000zp,SNO-01}
and all the other
$\Delta\ln\!C_{i3}$ are zero.
For the Super-Kamiokande experiment
($j=4$) we take
$
\Delta\ln\!C_{54}
=
2.0 \times 10^{-2}
$
in order to take into account the
uncertainty of the shape of the ${^8\mathrm{B}}$ neutrino spectrum
calculated in Ref.~\cite{Ortiz:2000nf}.
All the other $\Delta\ln\!C_{i4}$ are zero.

\subsection{Global Analysis}
\label{Cross: Global Analysis}

In the Global Analysis the prior distribution of the averaged cross sections
in the Chlorine ($j=1$), Gallium ($j=2$) and SNO ($j=3$) experiments
is the same as in the Rates Analysis.
For the Super-Kamiokande experiment
instead of the total rate,
we consider the data on the 38 bins of the
Super-Kamiokande day and night electron energy spectra
(see Table~\ref{delta}).
Hence,
the number of experimental data points in the Global Analysis is
$N_{\mathrm{S}}=41$,
leading to a $8\times41$
matrix $\mathbf{C}$
of energy averaged cross sections.

The energy-averaged cross sections $C_{ij}$ for $j\leq3$
are given by Eq.~(\ref{cross2}),
with the distribution of the normalized deviations $\xi_j$
given in Eq.~(\ref{cross1}).

The energy-averaged cross sections
$C_{5j}$, with $j=4,\ldots,41$,
of ${^8\mathrm{B}}$ neutrino detection
for the Super-Kamiokande day and night spectral bins
that take into account the fully correlated uncertainty
due to the uncertainty of
the shape of the ${^8\mathrm{B}}$ neutrino spectrum
\cite{Bahcall:1996qv,Ortiz:2000nf}
are given by 
\begin{equation}
C_{5(\bin+3)}
=
\left( 1 + \xi_4 \, \delta^{\mathrm{(th)}}_{\bin,\mathrm{cor}} \right)
C_{5(\bin+3)}^0
\qquad
(\bin=1,\ldots,38)
\,,
\label{cross5}
\end{equation}
with
the normalized deviation $\xi_4$
distributed as $p(\xi_4|\mathrm{I})$
given in Eq.~(\ref{cross1}) with $j=4$.
In Eq.~(\ref{cross5}),
$C_{5(\bin+3)}^0$
is the standard energy-averaged cross section of
${^8\mathrm{B}}$ neutrino detection in the $\bin^{\mathrm{th}}$
Super-Kamiokande bin
given in Table~\ref{delta},
together with $C_{3(\bin+3)}^0$,
relative to the $hep$ neutrino flux
(we take $C_{3(\bin+3)}=C_{3(\bin+3)}^0$
and all the other $C_{i(\bin+3)}=0$).
The values of relative theoretical uncertainty
$\delta^{\mathrm{(th)}}_{\bin,\mathrm{cor}}$
given in Fig.~5 of Ref.~\cite{Ortiz:2000nf}
for the rates in the Super-Kamiokande spectral bins ($\bin=1,\ldots,38$)
due to the theoretical uncertainty
on the ${^8\mathrm{B}}$ neutrino spectrum
are listed in Table~\ref{delta}.

\section{Model Dependent Analysis}
\label{Model Dependent Analysis}

In this Section we perform a Bayesian model dependent analysis
of solar neutrino data assuming the
solar neutrino fluxes
predicted by the BP2000 Standard Solar Model \cite{BP2000}.
In this analysis we take into account of all
the theoretical uncertainties through the prior distributions.
We check the validity
of this approach comparing the results with those
obtained with a standard Bayesian analysis in Ref.~\cite{Garzelli:2001zu}.

The sampling distributions and the prior distributions of detection
cross sections are described,
respectively, in Sections~\ref{Sampling Distribution}
and \ref{Cross Sections}.
In the following Subsection~\ref{SSM Prior Distribution of Fluxes}
we describe the SSM prior distribution of fluxes
used in the model dependent analysis.
In Section~\ref{Credible Regions}
we discuss our results for the calculation of
the Bayesian credible regions
in the plane of the oscillation parameters
$\tan^2\!\vartheta$ and $\Delta{m}^2$,
to be compared with the corresponding ones
presented in Ref.~\cite{Garzelli:2001zu}.

\subsection{SSM Prior Distribution of Fluxes}
\label{SSM Prior Distribution of Fluxes}

In the model dependent analysis we assume
a multi-normal prior distribution for the neutrino fluxes
centered on the BP2000 SSM values of the neutrino fluxes,
$\Phi_{i}^{\mathrm{SSM}}$ \cite{BP2000},
with the standard correlated uncertainties
(see \cite{Fogli-Lisi-correlations-95,Fogli:1999zg}):
\begin{equation}
p(\mathbf{\Phi}|\mathrm{I})
=
N
\,
\exp\left[
- \frac{1}{2}
\sum_{i_1,i_2=1}^{8}
\left( \Phi_{i_1} - \Phi_{i_1}^{\mathrm{SSM}} \right)
(V^{-1}_{\Phi})_{i_1 i_2}
\left( \Phi_{i_2} - \Phi_{i_2}^{\mathrm{SSM}} \right)
\right]
\prod_{i=1}^{8}
\theta(\Phi_{i})
\,,
\label{prior-fluxes}
\end{equation}
where
$N$
is a normalization constant and
$\theta(\Phi_{i})=1$
for
$\Phi_{i}\geq0$
and
$\theta(\Phi_{i})=0$
for
$\Phi_{i}<0$.
$V_{\Phi}$
is the covariance matrix of the fluxes,
given by
\cite{Fogli-Lisi-correlations-95,Fogli:1999zg}
\begin{equation}
(V_{\Phi})_{i_1 i_2}
=
\Phi_{i_1}^{\mathrm{SSM}}
\Phi_{i_2}^{\mathrm{SSM}}
\sum_{k=1}^{12}
\alpha_{i_1,k}
\alpha_{i_2,k}
\left( \Delta\!\ln\!X_k \right)^2
\,,
\label{VPhi}
\end{equation}
Here $X_k$,
with $k=1,\ldots,12$,
are the twelve input astrophysical parameters 
in the SSM
($S_{1,1}$,
$S_{3,3}$,
$S_{3,4}$,
$S_{1,14}$,
$S_{1,7}$,
Luminosity,
$Z/X$,
Age,
Opacity,
Diffusion,
$C_{{^7\mathrm{Be}}}$,
$S_{1,16}$),
whose relative uncertainties
$\Delta\!\ln\!X_k$
determine the correlated uncertainties of the neutrino fluxes
$\Phi_{i}^{\mathrm{SSM}}$
through the logarithmic derivatives
\begin{equation}
\alpha_{i,k}
=
\frac{\partial\ln\Phi_{i}^{\mathrm{SSM}}}{\partial\ln\!X_k}
\,.
\label{alpha}
\end{equation}
We adopt the values of
$\alpha_{i,k}$
and
$\Delta\!\ln\!X_k$
given in Ref.~\cite{Fogli:1999zg},
except for $\Delta\!\ln\!X_7$ (relative to $Z/X$),
whose value has been updated in the BP2000 SSM \cite{BP2000}
from 0.033 to 0.061,
and
$\alpha_{i,12}=\delta_{i8}$,
$\Delta\!\ln\!X_{12}=0.181$,
that have been introduced for the first time in
the BP2000 SSM \cite{BP2000}.
$X_{12}$
is the astrophysical factor $S_{1,16}$ for the reaction
${^{16}\mathrm{O}}(p,\gamma){^{17}\mathrm{F}}$
that determines the small $^{17}\mathrm{F}$ neutrino flux ($i=8$).

\subsection{Credible Regions}
\label{Credible Regions}

The credible regions
with
90\%,
95\% and
99\%
posterior probability
calculated through Eq.~(\ref{posterior1})
are shown in Fig.~\ref{ssmra} for the Rates Analysis
and in Fig.~\ref{ssmga} for the Global Analysis.
One can see that they are almost identical to the
corresponding regions in
Figs.~3 and 4 of Ref.~\cite{Garzelli:2001zu},
obtained with the standard method,
in which the uncertainties of the fluxes and cross sections
are taken into account in the covariance matrix
that determines the likelihood function.

The integral probabilities of the LMA, VO, SMA and LOW
regions presented in Table~\ref{prob}
almost coincide with those obtained in Ref.~\cite{Garzelli:2001zu}.
As noted in Ref.~\cite{Garzelli:2001zu},
these probabilities favor the LMA solution of the solar neutrino problem.

Table~\ref{prob} shows also the values of the integral
probabilities of large and small values of
$\tan^2\!\vartheta$
and
$\Delta{m}^2$,
that favor large mixing and large values of
$\Delta{m}^2$.

In conclusion of this section,
we have shown that
in the model dependent analysis of solar neutrino data
the new approach
that we follow here,
in which all the theoretical uncertainties
are taken into accounts in the priors,
gives approximately the same results as the standard one
followed in Ref.~\cite{Garzelli:2001zu},
in which the theoretical uncertainties
are taken into account in the covariance matrix
that determines the likelihood function.
Therefore,
we conclude that the new approach
is compatible with the standard one
and we can apply it to the model independent analysis
of solar neutrino data.

Let us notice,
however,
that the new approach is not convenient for
the model dependent analysis of solar neutrino data,
because it needs the integrations
over the neutrino fluxes and cross sections
in Eq.~(\ref{posterior1}).

\section{Model Independent Analysis}
\label{Model Independent Analysis}

This is the main Section of the paper,
in which we present the results of the model independent analysis
of solar neutrino data assuming
a flat prior probability distribution of neutrino fluxes,
that is explained in the following
Subsection~\ref{Model Independent Prior Distribution of Fluxes}.

In Subsection~\ref{Oscillation Parameters}
we present the model independent allowed ranges
for the neutrino mixing parameters
$\tan^2\!\vartheta$, $\Delta{m}^2$.
In Subsection~\ref{Neutrino Fluxes}
we discuss the information on the solar neutrino fluxes
obtained from the model independent posterior distribution.
Finally,
in Subsection~\ref{Oscillations versus no oscillations}
we compare the probabilities
of the hypotheses of no neutrino oscillations and neutrino oscillations.

\subsection{Model Independent Prior Distribution of Fluxes}
\label{Model Independent Prior Distribution of Fluxes}

The Standard Solar Model has reached a reasonable
degree of credibility in recent years
as a consequence of its very good agreement with helioseismological
measurements
\cite{BP2000,Bahcall:2000ue}.
Therefore,
we think that
the true values of the neutrino fluxes cannot be too different
from those of the SSM.

In our model independent analysis
we assume a flat prior distribution
from zero to twice the BP2000 SSM value \cite{BP2000}
for all the neutrino fluxes except $hep$ and $pp$.

Since the astrophysical $S$-factor of the $hep$
reaction is very difficult to calculate
(see Ref.~\cite{Bahcall:1998se}),
the value of the $hep$ flux is the most uncertain one
and no uncertainty is given in the BP2000 SSM.
However, the Super-Kamiokande Collaboration measured
\cite{SK-sun-01}
a 90\% CL upper limit for the $hep$ neutrino flux
of 4.3 times the BP2000 SSM prediction.
Hence,
in the model independent analysis
we take for the $hep$ flux
a flat prior distribution
from zero to ten times the BP2000 SSM value \cite{BP2000}.

The prior distribution of the
$pp$ flux is obtained from the
prior distributions of the other fluxes
and the luminosity constraint
(see Ref.~\cite{Bahcall:2001pf})
\begin{equation}
\sum_i
\alpha_i
\,
\Phi_i
=
K_\odot
\,,
\label{luminosity}
\end{equation}
where
$ \alpha_i $
is the average thermal energy released together
with a neutrino from the source $i$ given in Ref.~\cite{Bahcall:2001pf}
and
$
K_\odot
=
\mathcal{L}_\odot / 4 \pi R^2
=
8.527 \times 10^{11} \, \mathrm{MeV} \, \mathrm{cm}^{-2} \, \mathrm{s}^{-1}
$
is the solar constant
($ \mathcal{L}_\odot = 2.398 \times 10^{39} \, \mathrm{MeV} \, \mathrm{s}^{-1} $
is the luminosity of the sun
and
$ R = 1.496 \times 10^{13} \, \mathrm{cm} $
is the sun-earth distance).
The choice of the $pp$ flux
as the dependent flux
to be determined by the luminosity constraint (\ref{luminosity})
is mandatory,
because
the $pp$ flux,
being the largest one in the SSM,
is the only flux that can exhaust by itself
the luminosity constraint without assuming an unreasonably large value.

The resulting prior distribution of the ratio
$\Phi_{pp}/\Phi_{pp}^{\mathrm{SSM}}$
is almost flat in the range
$[0.904,1.094]$,
as can be seen from Figs.~\ref{misra_f1} and \ref{misga_f1}
(thick dashed curve).
The upper value of this range corresponds to the maximum
$pp$ flux allowed by the luminosity constraint (\ref{luminosity}).
The lower value
depends on the choice of
the range of the prior distributions of the other fluxes.
As shown in Figs.~\ref{misra_f1} and \ref{misga_f1},
the prior distribution of the $pp$ flux
is almost exactly symmetric with respect to
$\Phi_{pp}/\Phi_{pp}^{\mathrm{SSM}}=1$
and is exactly flat
in the range $[0.94,1.06]$
(the small fluctuations of the prior distribution in
Figs.~\ref{misra_f1} and \ref{misga_f1}
is due to the fact that
the prior distributions of the neutrino fluxes are
generated through a Monte Carlo,
in order to perform the integrals over the neutrino fluxes
in Eq.~(\ref{posterior1})).

In Table~\ref{dfl} we list the relative prior uncertainties
$\Delta\!\Phi_{i}/\Phi_{i}^{\mathrm{SSM}}$
of the neutrino fluxes in the model independent analysis
and we confront them with
the relative uncertainties
$(\Delta\!\Phi_{i}^{\mathrm{SSM}}/\Phi_{i}^{\mathrm{SSM}})_{(1\sigma)}$
in the BP2000 Standard Solar Model
and
with the relative
differences between the fluxes predicted by several old and new models
and the BP2000 SSM.
One can see that the relative prior uncertainties
in the model independent analysis
are much larger than
the relative uncertainties
in the BP2000 Standard Solar Model
and
they are also much larger than
the relative differences between the fluxes predicted by other models
and the BP2000 SSM.
Therefore,
our choice of prior distributions
for the neutrino fluxes is indeed appropriate for a model independent
analysis of solar neutrino data.

\subsection{Oscillation Parameters}
\label{Oscillation Parameters}

In this Subsection we present our results
for the model independent credible regions in the
$\tan^2\!\vartheta$--$\Delta{m}^2$ plane
obtained with the posterior distribution (\ref{posterior1})
marginalized
with respect to the neutrino fluxes
and cross sections.

The credible regions
with
90\%,
95\% and
99\%
posterior probability
obtained with the Rates Analysis
are shown in Fig.~\ref{misra}.
One can see that they are significantly larger than the model dependent
credible regions in Fig.~\ref{ssmra},
but still restricted in the parameter space:
a large area of the parameter space shown in
Fig.~\ref{misra} is excluded\footnote{
For convenience,
we call ``excluded'' the areas outside of the credible regions,
although they are not strictly speaking excluded,
since they have a non-zero, albeit small, posterior probability.
Similarly,
we sometimes call ``allowed'' the areas within the credible regions.
}
in a model independent way.
In particular,
the absence of neutrino oscillations,
corresponding to
$\left( \Delta{m}^2 \, \tan^2\vartheta \right) \lesssim 10^{-12} \mathrm{eV}^2$,
is excluded.

In Fig.~\ref{misra}
one can distinguish clearly the 90\% and 95\% allowed
LMA, SMA, LOW and VO regions
that are obviously larger than the corresponding regions
obtained in the model dependent analysis
(Fig.~\ref{ssmra}).
The 99\% allowed regions
are less defined,
but large values of
$\Delta{m}^2$,
corresponding to the LMA and SMA regions with
$\Delta{m}^2 \gtrsim 3 \times 10^{-6} \mathrm{eV}^2$,
are separated from small values of
$\Delta{m}^2$,
corresponding to the LOW and VO regions with
$\Delta{m}^2 \lesssim 3 \times 10^{-6} \mathrm{eV}^2$.
The LMA and SMA 99\% allowed regions are connected
through a channel at
$\Delta{m}^2 \simeq 1.3 \times 10^{-4} \mathrm{eV}^2$.
The LOW 99\% region is connected with the upper
VO 99\% region,
but there are still two separated
VO 99\% regions
at
$\Delta{m}^2 \simeq 3 \times 10^{-10} \mathrm{eV}^2$
and
$\Delta{m}^2 \simeq 1 \times 10^{-10} \mathrm{eV}^2$.

All experiments are important in order to
restrict the model independent allowed regions,
as one can see confronting Fig.~\ref{misra}
with the four figures \ref{nogal_misra}--\ref{nokam_misra},
each one obtained neglecting the data
of one of the four experiments in Table~\ref{rates}.
In particular,
one can see that the Gallium data of GALLEX+GNO+SAGE
and the rate measured in the Super-Kamiokande experiment
are crucial in order to separate the 90\% LMA and LOW regions,
and all experiment are important
for the separation of the 95\% and 99\% LMA and LOW regions.
The SNO and Super-Kamiokande rates are important for the exclusion
of a large part of the ``dark-side'' region
($\tan^2\vartheta > 1$).

Figure~\ref{misra_fit}
illustrates the quality of the fit of solar neutrino data
in the Rates Analysis
for some selected values of $\tan^2\!\vartheta$ and $\Delta{m}^2$.
The corresponding best-fit values of the neutrino fluxes,
normalized to the BP2000 SSM prediction,
are given in Table~\ref{misra_fit_flu}.

The solid, long-dashed, short-dashed and dotted
lines in Fig.~\ref{misra_fit}A
correspond, respectively, to the best-fit points
in the SMA, LMA, LOW and VO regions
(we call ``best-fit point''
the point with highest posterior density of probability).
From Fig.~\ref{misra_fit}A
one can see that the best-fit LMA solution fits very well
the data,
the best-fit SMA solution fits well
the data, but it is slightly off the SNO rate,
the best-fit VO solution
do not fit well the Gallium rate,
and
the best-fit LOW solution fit
of the Gallium and Chlorine data is rather bad.

The solid line in Fig.~\ref{misra_fit}B
corresponds to
$\tan^2\!\vartheta = \protect\input{fig/misra_fit_par_5_1.tab}$
and
$\Delta{m}^2 = \protect\input{fig/misra_fit_par_5_2.tab} \, \mathrm{eV}^2$,
a point near the center of a quasi-triangular excluded region
in Figs.~\ref{misra}--\ref{nokam_misra}.
From Fig.~\ref{misra_fit}B
one can see that in this case the energy dependence of the
oscillation probability is very strong,
typical of small mixing angle MSW transitions
(see \cite{Mikheev:1985gs,%
Bilenky-Petcov-RMP-87,%
CWKim-book-93,%
BGG-review-98}),
with
small suppression of low-energy $pp$ neutrinos
that do not cross the resonance
and
a large suppression of intermediate and high energy neutrinos.
The fit of the Super-Kamiokande rate is
less disastrous than the fit of the SNO rate because
of the high flux of $\nu_{\mu,\tau}$
due to a large $^8\mathrm{B}$ flux at the limit of
the maximum value
$\Phi_{^8\mathrm{B}}/\Phi_{^8\mathrm{B}}^{\mathrm{SSM}}=2$,
as shown in Table~\ref{misra_fit_flu}.
From Fig.~\ref{misra_fit}B one can infer that
the quasi-triangular excluded region
centered at
$\tan^2\!\vartheta \simeq \protect\input{fig/misra_fit_par_5_1.tab}$
and
$\Delta{m}^2 \simeq \protect\input{fig/misra_fit_par_5_2.tab} \, \mathrm{eV}^2$
is still excluded eliminating one data point,
in agreement with Figs.~\ref{nogal_misra}--\ref{nokam_misra}.

The long-dashed line in Fig.~\ref{misra_fit}B
corresponds to
$\tan^2\!\vartheta = \protect\input{fig/misra_fit_par_6_1.tab}$
and
$\Delta{m}^2 = \protect\input{fig/misra_fit_par_6_2.tab} \, \mathrm{eV}^2$,
a point in a quasi-triangular excluded region
at the center of Figs.~\ref{misra}--\ref{nokam_misra}.
One can see that in this region it is not possible
to fit the Gallium, Chlorine and SNO data.
The reason is a strong suppression of all energy neutrinos
due to the crossing of the MSW resonance in the sun.
Only the Super-Kamiokande rate can be fitted well
taking a large $^8\mathrm{B}$ flux at the limit of
the maximum value of the prior distribution,
as shown in Table~\ref{misra_fit_flu}.

The short-dashed line in Fig.~\ref{misra_fit}B
corresponds to
$\tan^2\!\vartheta = \protect\input{fig/misra_fit_par_7_1.tab}$
and
$\Delta{m}^2 = \protect\input{fig/misra_fit_par_7_2.tab} \, \mathrm{eV}^2$,
a point near the center of an excluded area
in the VO region in Figs.~\ref{misra}--\ref{noclo_misra} and \ref{nokam_misra}.
One can see that this region is excluded by the SNO rate,
in agreement with the fact that it is not excluded in
Fig.~\ref{nosno_misra},
obtained with the model independent Rates Analysis
of
Gallium,
Chlorine and
Super-Kamiokande data only.

The dotted line in Fig.~\ref{misra_fit}B
corresponds to
$\tan^2\!\vartheta = \protect\input{fig/misra_fit_par_8_1.tab}$
and
$\Delta{m}^2 = \protect\input{fig/misra_fit_par_8_2.tab} \, \mathrm{eV}^2$,
a point in the
``Just So$^2$''
region found in Ref.~\cite{Bahcall:2001hv}
in the SSM model dependent analysis of pre-SNO solar neutrino data.
One can see that the SNO rate disfavors this region,
that is indeed allowed in
Fig.~\ref{nosno_misra},
obtained with the model independent Rates Analysis
of
Gallium,
Chlorine and
Super-Kamiokande data only.

Finally,
the dash-dotted line in Fig.~\ref{misra_fit}B
corresponds to the case of no oscillations.
Obviously,
as shown in Table~\ref{misra_fit_flu},
in this case the flux of intermediate energy neutrinos
must be suppressed
and the $^8\mathrm{B}$
neutrino flux must have an intermediate value between those
measured in the Super-Kamiokande \cite{SK-sun-01}
and SNO \cite{SNO-01} experiments
assuming no neutrino oscillations.
The smaller experimental uncertainty of the Super-Kamiokande rate
forces the $^8\mathrm{B}$ neutrino flux to be close to the one
measured in the Super-Kamiokande experiment in the case of no-oscillations,
leading to a bad fit of the SNO rate.
This $^8\mathrm{B}$ neutrino flux adds to the $pp$ flux to
give a rate in Gallium experiments that is too high,
leading to a bad fit of Gallium data,
as shown in Fig.~\ref{misra_fit}B.

Let us consider now the Global Analysis.
Figure~\ref{misga} shows the credible regions
with
90\%,
95\% and
99\%
posterior probability
obtained with the Global Analysis.
Again,
the allowed regions are larger than those obtained with the
model dependent analysis
(see Fig.~\ref{ssmga}),
but they are still clearly distinguishable
and a large region of the parameter space is excluded.
One can notice that in the model independent Global Analysis
there is a very small SMA credible region
at the 99\% probability level,
that is not present in the model dependent analysis.

In Fig.~\ref{misga}
the
LMA regions with
90\%,
95\% and
99\%
probability are clearly separated from the other regions,
whereas
the LOW and VO credible regions
are connected through the quasi-vacuum region
at $\Delta{m}^2 \sim 6 \times 10^{-9} \, \mathrm{eV}^2$.
Several disconnected VO credible regions
are present for
$
8 \times 10^{-11} \, \mathrm{eV}^2
\lesssim
\Delta{m}^2
\lesssim
3 \times 10^{-9} \, \mathrm{eV}^2
$.

Figure~\ref{misga_fit}
illustrates the quality of the fit of solar neutrino data
in the Global Analysis
for some selected values of $\tan^2\!\vartheta$ and $\Delta{m}^2$,
with the corresponding best values of the neutrino fluxes
normalized to the BP2000 SSM prediction
given in Table~\ref{misga_fit_flu}.

The lines in Fig.~\ref{misga_fit}A
correspond to the same values of
$\tan^2\!\vartheta$ and $\Delta{m}^2$
as in Fig.~\ref{misra_fit}A,
\textit{i.e.}
to the best-fit point
in the SMA, LMA, LOW and VO regions
obtained in the Rates Analysis.
The solid, long-dashed, short-dashed and dotted
lines in Fig.~\ref{misga_fit}B
correspond, respectively, to the best-fit point
in the SMA, LMA, LOW and VO regions
obtained in the Global Analysis.
The dash-dotted line
in Fig.~\ref{misga_fit}B
corresponds to the case of no oscillations.

The best-fit SMA solution in the Rates Analysis
(solid line in Fig.~\ref{misga_fit}A)
is disfavored by the Super-Kamiokande energy spectrum,
because the corresponding electron-neutrino survival probability
increases with energy too steeply.
The best-fit SMA solution in the Global Analysis
(solid line in Fig.~\ref{misga_fit}B)
has a smaller mixing angle,
leading to a flatter energy dependence
of the electron-neutrino survival probability,
in agreement with the Super-Kamiokande spectrum.
The survival probability of high-energy electron-neutrinos
is close to one and the Super-Kamiokande data are fitted
with a small $^8\mathrm{B}$ neutrino flux,
as shown in Table~\ref{misga_fit_flu}
($ \Phi_{^8\mathrm{B}}/\Phi_{^8\mathrm{B}}^{\mathrm{SSM}} \simeq \protect\input{fig/misga_fit_f5_5.tab} $).
In this case
the theoretical rate for the SNO experiment
($j=3$) is too high,
as shown in Fig.~\ref{misga_fit}B,
disfavoring the SMA solution.

Figure~\ref{misga_fit}
shows that the data are well fitted in the LMA and LOW regions,
whereas the VO solution is slightly disfavored
by the highest-energy bins of the Super-Kamiokande spectrum.
It is interesting to notice the difference
of the LMA prediction for the day and night Super-Kamiokande
energy bins,
that is especially evident in Fig.~\ref{misga_fit}A.
This difference is due to the regeneration of
electron neutrinos in their passage through the Earth.

The case of no oscillations (dash-dotted line in Fig.~\ref{misga_fit}B)
is compatible with the Super-Kamiokande energy spectrum
through a small $^8\mathrm{B}$ neutrino flux,
as shown in Table~\ref{misga_fit_flu}
($ \Phi_{^8\mathrm{B}}/\Phi_{^8\mathrm{B}}^{\mathrm{SSM}} \simeq \protect\input{fig/misga_fit_f5_9.tab} $).
However,
this flux is too high to fit the Gallium ($j=1$), Chlorine ($j=2$)
and SNO ($j=3$) rates.

Let us now determine
the probabilities of the
SMA, LMA, LOW and VO regions,
that are given by the integrals
of the posterior density
over the appropriate
ranges of the parameters given in Eqs.~(\ref{SMA})--(\ref{VO}):
\begin{equation}
p(\mathrm{R}|\mathrm{D},\mathrm{I})
=
\int_{\mathrm{R}}
\mathrm{d}\!\log(\tan^2\!\vartheta) \, \mathrm{d}\!\log(\Delta{m}^2) \,
p(\tan^2\!\vartheta,\Delta{m}^2|\mathrm{D},\mathrm{I})
\,,
\label{pR}
\end{equation}
with
$\mathrm{R} = \mathrm{SMA}, \mathrm{LMA}, \mathrm{LOW}, \mathrm{VO}$.
The values of these probabilities
in both the Rates Analysis and Global Analysis
are listed in Table~\ref{prob},
together with
those obtained in the
model dependent analysis
presented in Section~\ref{Model Dependent Analysis}
(see also \cite{Garzelli:2001zu}).
One can see that the model independent and model dependent
integral probabilities of the four regions are similar
and the LMA region is favored in both cases,
especially with the Global Analysis that,
on the other hand,
strongly disfavors the SMA region.

Bayesian Probability Theory allows also to
calculate the separate posterior probability distributions
for
$\tan^2\!\vartheta$
and
$\Delta{m}^2$
through the marginalizations
\begin{align}
\null & \null
p(\tan^2\!\vartheta|\mathrm{D},\mathrm{I})
=
\int\!\mathrm{d}\!\log(\Delta{m}^2) \,
p(\tan^2\!\vartheta,\Delta{m}^2|\mathrm{D},\mathrm{I})
\,,
\label{margial1}
\\
\null & \null
p(\Delta{m}^2|\mathrm{D},\mathrm{I})
=
\int\!\mathrm{d}\!\log(\tan^2\!\vartheta) \,
p(\tan^2\!\vartheta,\Delta{m}^2|\mathrm{D},\mathrm{I})
\,.
\label{marginal2}
\end{align}

Figures~\ref{misra_t2t} and \ref{misga_t2t}
show the model independent posterior distribution
of $\tan^2\!\vartheta$ in the Rates Analysis and Global Analysis,
respectively.
One can see that
in both cases there is a wide and high peak at large mixing angles
($\tan^2\!\vartheta \sim 0.3$).
In the Rates Analysis there is also
a small peak at small mixing angles
($\tan^2\!\vartheta \sim 10^{-3}$).
The integral probabilities of large and small mixing
are given in Table~\ref{prob},
together
with those obtained in the
model dependent analysis of Section~\ref{Model Dependent Analysis}.
One can see that the values are similar
and both
the model independent and the model dependent analyses
favor large mixing
($0.1<\tan^2\!\vartheta<10$).
This indication is very strong
in both the model independent and the model dependent
Global Analyses.

Small mixing is obtained either through small
($<0.1$) or large ($>10$)
values of
$\tan^2\!\vartheta$,
that correspond, respectively,
to
$\nu_e \simeq \nu_1$
and
$\nu_e \simeq \nu_2$
(see Eq.~(\ref{mixing})).
Table~\ref{prob} shows that both small
and large values of
$\tan^2\!\vartheta$
are disfavored,
especially large values,
in what could be called ``extreme dark side''
($\tan^2\!\vartheta > 10$),
following the ``dark side'' denomination of the region
$\tan^2\!\vartheta > 1$
\cite{deGouvea:2000cq}.

Figures~\ref{misra_dm2} and \ref{misga_dm2}
show the posterior distribution
of $\Delta{m}^2$ in the Rates Analysis and Global Analysis,
respectively.
The high and wide peak at
$\Delta{m}^2 \sim 10^{-5}-10^{-4} \, \mathrm{eV}^2$
in both figures
implies that large values of $\Delta{m}^2$
are favored.
This is confirmed by the values of the integral
probabilities of large
($>10^{-5} \, \mathrm{eV}^2$)
and small
($<10^{-5} \, \mathrm{eV}^2$)
values of
$\Delta{m}^2$ listed in Table~\ref{prob},
that are comparable to those obtained in the
model dependent analysis,
also reported in Table~\ref{prob}.

The model independent indications in favor of
large mixing and large values of
$\Delta{m}^2$
are very important for future terrestrial experiments
that could explore this region of the parameter space
with reactor neutrinos
\cite{KAMLAND,BOREXINO}.

\subsection{Neutrino Fluxes}
\label{Neutrino Fluxes}

In this section we present the information
on neutrino fluxes obtained through the marginal
posterior distributions
\begin{equation}
p(\Phi_i|\mathrm{D},\mathrm{I})
=
\int\!\prod_{i' \neq i} \mathrm{d}\Phi_{i'}
\int\!\mathrm{d}\mathbf{C}
\int\!\mathrm{d}\!\log(\tan^2\!\vartheta) \, \mathrm{d}\!\log(\Delta{m}^2) \,
p(\tan^2\!\vartheta,\Delta{m}^2,\mathbf{\Phi},\mathbf{C}|\mathrm{D},\mathrm{I})
\,.
\label{flux}
\end{equation}

As could have been expected,
we found that only
the posterior distributions of the
$pp$, ${^7\mathrm{Be}}$ and ${^8\mathrm{B}}$ fluxes,
to which the existing experiments are most sensitive,
are significantly different from the corresponding prior distributions.
The posterior distributions of all the other
fluxes, including $hep$,
is practically flat
in the same range as the prior distribution.

These distributions
of the $pp$, ${^7\mathrm{Be}}$ and ${^8\mathrm{B}}$ fluxes
are shown in Figs.~\ref{misra_f1}--\ref{misga_f5}.
In all these figures
the thick solid line is the model independent posterior distribution,
the thick dashed line is the prior distribution
and
the thick dotted line is the Gaussian BP2000 Standard Solar Model distribution.
The intervals in which
the thick solid line
lies above the
thin horizontal dotted, dashed and solid lines have,
respectively,
90\%, 95\% and 99\% probability.
The thick solid lines representing the posterior distributions
in Figs.~\ref{misra_f1}--\ref{misga_f5}
have small fluctuations
because the integrals over the fluxes and cross sections in Eq.~(\ref{flux})
have been performed through a Monte Carlo.
The fluctuations of the
thick dashed lines
is due to the fact that the prior distribution is generated by the Monte Carlo.

Table~\ref{fluxint} gives
the credible intervals of the $pp$, ${^7\mathrm{Be}}$ and ${^8\mathrm{B}}$ fluxes
with 90\%, 95\% and 99\%
probability
in the Rates and Global Analyses.

Figures~\ref{misra_f1} and \ref{misga_f1}
show that the posterior distribution of the $pp$ flux
is in good agreement with the SSM distribution,
with a slight preference for high values.
The upper limit of the credible intervals
with 90\%, 95\% and 99\%
probability
reported in Table~\ref{fluxint}
does not give a significant information,
since it
almost coincides with the upper limit of
the prior distribution,
that is determined by the luminosity constraint (\ref{luminosity}),
as explained in Section~\ref{Model Independent Prior Distribution of Fluxes}.
On the other hand,
the lower limit of the credible intervals reported in Table~\ref{fluxint}
is interesting,
because it is significantly larger than the lower limit of
the prior distribution,
that follows from the prior distributions of the other fluxes and
the luminosity constraint (\ref{luminosity}).
In other words,
the existing solar neutrino data allow to improve our knowledge
of the $pp$ flux,
increasing its allowed lower limit.

Also the posterior distributions of the $^7\mathrm{Be}$ flux
shown in
Figures~\ref{misra_f4} and \ref{misga_f4}
are in good agreement with the SSM distribution,
with a slight preference for low values.
One can see from Table~\ref{fluxint}
that,
according to the model independent analysis
of solar neutrino data,
the $^7\mathrm{Be}$ flux could be very small,
but it is limited from above.
Especially the Global Analysis
restricts the allowed upper value for the $^7\mathrm{Be}$ flux
significantly below the upper value of the prior distribution,
as one can see from Fig.~\ref{misga_f4}
and
Table~\ref{fluxint}.
Let us remember that the $^7\mathrm{Be}$ neutrino flux
has also been constrained in Ref.~\cite{Ricci:2000jd} within
$
\pm 25 \%
$
($1\sigma$ error)
of the SSM value from helioseismological measurements.

Finally,
Figures~\ref{misra_f5} and \ref{misga_f5}
show that the posterior distribution of the $^8\mathrm{B}$ flux
is in excellent agreement with the SSM distribution
and it is bounded both from above and from below
in an interval smaller than the prior distribution.
This is a very interesting result,
that agrees with the ranges for the $^8\mathrm{B}$ flux
obtained in
Refs.~\cite{SNO-01,Fogli:2001vr,Giunti-aoe-01,%
Bahcall:2001zu,Barbieri:2000sv-hepph,Krastev:2001tv,%
Smy:2001yn,%
Fiorentini:2001jt,Bahcall:2001cb}
from the comparison of the SNO and Super-Kamiokande total rates only,
and in Ref.~\cite{Krastev:2001tv,Bahcall:2001cb}
from the least-squares analysis of all solar neutrino data
with the $^8\mathrm{B}$ flux considered as a free parameter.
For example,
the authors of Ref.~\cite{Fogli:2001vr} found
$
0.44
\lesssim
\Phi_{^8\mathrm{B}}/\Phi_{^8\mathrm{B}}^{\mathrm{SSM}}
\lesssim
1.62
$
at $3\sigma$,
\textit{i.e.}
99.73\% confidence level.
We find that the credible interval of
$\Phi_{^8\mathrm{B}}/\Phi_{^8\mathrm{B}}^{\mathrm{SSM}}$
with 99.73\% probability is
\begin{equation}
\protect\input{fig/misra_f5_1.tab}
<
\Phi_{^8\mathrm{B}}/\Phi_{^8\mathrm{B}}^{\mathrm{SSM}}
<
\protect\input{fig/misra_f5_2.tab}
\label{8B9973rates}
\end{equation}
in the Rates Analysis and
\begin{equation}
\protect\input{fig/misga_f5_1.tab}
<
\Phi_{^8\mathrm{B}}/\Phi_{^8\mathrm{B}}^{\mathrm{SSM}}
<
\protect\input{fig/misga_f5_2.tab}
\label{8B9973global}
\end{equation}
in the Global Analysis.
Although the meaning of confidence level intervals
and credible intervals is different,
it is fair to say that
the Global Analysis allows to restrict
the allowed range of the $^8\mathrm{B}$ neutrino flux
more than the comparison of the SNO and Super-Kamiokande total rates only.

\subsection{Oscillations versus no oscillations}
\label{Oscillations versus no oscillations}

Bayesian Probability Theory
allows to compare the compatibility of different hypotheses
with the observed data calculating
the ratio of their probabilities,
that is usually called ``odds''.
Hence,
we can compare in a model independent way
the oscillation hypothesis with the
no oscillation hypothesis,
\textit{i.e.} the Standard Model of neutrino physics.

In order to perform this task we must reconsider Bayes Theorem
in the form (\ref{Bayes1}).
The posterior probability of the oscillation hypothesis is given by
\begin{align}
\null & \null
p(\mathrm{OSC}|\mathrm{D},\mathrm{I})
=
\int\!\mathrm{d}\!\log(\tan^2\!\vartheta) \, \mathrm{d}\!\log(\Delta{m}^2)
\int\!\mathrm{d}\mathbf{\Phi}
\int\!\mathrm{d}\mathbf{C} \,
p(\tan^2\!\vartheta,\Delta{m}^2,\mathbf{\Phi},\mathbf{C}|\mathrm{D},\mathrm{I})
\nonumber
\\
\null & \null
=
\frac{
\int\!\mathrm{d}\!\log(\tan^2\!\vartheta) \, \mathrm{d}\!\log(\Delta{m}^2)
\int\!\mathrm{d}\mathbf{\Phi}
\int\!\mathrm{d}\mathbf{C}
\,
p(\mathrm{D}|\tan^2\!\vartheta,\Delta{m}^2,\mathbf{\Phi},\mathbf{C},\mathrm{I})
\,
p(\tan^2\!\vartheta,\Delta{m}^2|\mathrm{I})
\,
p(\mathbf{\Phi}|\mathrm{I})
\,
p(\mathbf{C}|\mathrm{I})
}{
p(\mathrm{D}|\mathrm{I})
}
\,.
\label{osc}
\end{align}
The flat prior distribution in the
$\log(\tan^2\!\vartheta)$--$\log(\Delta{m}^2)$ plane
of Figs.~\ref{ssmra}--\ref{misga}
is given by
\begin{equation}
p(\tan^2\!\vartheta,\Delta{m}^2|\mathrm{I})
=
\frac{1}{\Delta\!\log(\tan^2\!\vartheta) \, \Delta\!\log(\Delta{m}^2)}
\,,
\label{flat}
\end{equation}
where
$\Delta\!\log(\tan^2\!\vartheta)$
and
$\Delta\!\log(\Delta{m}^2)$
are the logarithmic ranges of
$\tan^2\!\vartheta$
and
$\Delta{m}^2$:
$\Delta\!\log(\tan^2\!\vartheta)=7$
and
$\Delta\!\log(\Delta{m}^2)=11$.

We compare the probability of neutrino oscillations
(\ref{osc})
with the probability of no oscillations
\begin{equation}
p(\mathrm{NO-OSC}|\mathrm{D},\mathrm{I})
=
\frac{
\int\!\mathrm{d}\mathbf{\Phi}
\int\!\mathrm{d}\mathbf{C}
\,
p(\mathrm{D}|\tan^2\!\vartheta=0,\Delta{m}^2=0,\mathbf{\Phi},\mathbf{C},\mathrm{I})
\,
p(\mathbf{\Phi}|\mathrm{I})
\,
p(\mathbf{C}|\mathrm{I})
}{
p(\mathrm{D}|\mathrm{I})
}
\,.
\label{nos}
\end{equation}
Notice that in the ratio of Eqs.~(\ref{nos}) and (\ref{osc})
the unknown probability
$p(\mathrm{D}|\mathrm{I})$
cancels.

Our result for the comparison
of no neutrino oscillations
versus neutrino oscillations is
\begin{equation}
\frac
{ p(\mathrm{NO-OSC}|\mathrm{D},\mathrm{I}) }
{ p(\mathrm{OSC}|\mathrm{D},\mathrm{I}) }
=
\protect\input{fig/misra_odds.tab}
\label{rat-rates}
\end{equation}
in the Rates Analysis and
\begin{equation}
\frac
{ p(\mathrm{NO-OSC}|\mathrm{D},\mathrm{I}) }
{ p(\mathrm{OSC}|\mathrm{D},\mathrm{I}) }
=
\protect\input{fig/misga_odds.tab}
\,.
\label{rat-global}
\end{equation}
in the Global Analysis.
Therefore,
there is a rather strong model independent indication in favor of
neutrino oscillations,
in agreement with those obtained in
Refs.~\cite{SNO-01,Fogli:2001vr,Giunti-aoe-01,Fiorentini:2001jt,Bahcall:2001pe}.
The odds in favor of neutrino oscillations
are slightly stronger in the Rates Analysis
than in the Global Analysis
because the Super-Kamiokande energy spectrum is rather flat,
which would be in agreement with the hypothesis of no oscillations
if the Super-Kamiokande rate were equal to the SNO rate.

Finally,
let us notice that
the prior distribution (\ref{flat})
and the result of the comparison of
the oscillation and no-oscillation
hypotheses that we have calculated
depends on the ranges
$\Delta\!\log(\tan^2\!\vartheta)$
and
$\Delta\!\log(\Delta{m}^2)$.
However,
one must always keep in mind that
these ranges have not been chosen randomly,
but follow from our prior knowledge that
there can be a measurable oscillation effect
in solar neutrino experiments
only if the neutrino mixing parameters lie within them.
It is possible that somebody may argue in favor of a slight
expansion or contraction of
the prior ranges of
$\Delta\!\log(\tan^2\!\vartheta)$
and
$\Delta\!\log(\Delta{m}^2)$
that we have considered.
However,
the effect of such expansion or contraction
on the result of the comparison of
the oscillation and no-oscillation
hypotheses is small,
not changing its order of magnitude.

\section{Conclusions}
\label{Conclusions}

In this paper we have analyzed the data of solar neutrino experiment
in terms of $\nu_e\to\nu_{\mu,\tau}$ oscillations
without assuming the Standard Solar Model values of
the neutrino fluxes.
Working in the framework of Bayesian Probability Theory,
the assumed lack of knowledge of
the neutrino fluxes
has been quantified through a flat prior probability distribution
for the neutrino fluxes,
that have been considered as unknown quantities to be determined,
if possible,
from the analysis of the data.

In Section~\ref{Model Dependent Analysis}
we have checked the reliability of our approach
using a multi-normal
prior probability distribution
for the neutrino fluxes corresponding to the
prediction of the BP2000 Standard Solar Model.
We have shown that the resulting allowed regions
for the neutrino mixing parameters
$\tan^2\!\vartheta$, $\Delta{m}^2$
are practically equal with those obtained with
the standard method
in which the uncertainties of the fluxes
are taken into account in the covariance matrix
that determines the likelihood function.

The results of the model independent analysis
presented in Section~\ref{Model Independent Analysis}
show that the present solar neutrino data
allow to derive rather stringent model independent information
on the neutrino mixing parameters,
improving dramatically
the conclusions obtained several years ago
\cite{Bilenky:1994ti}
in another model independent analysis of solar neutrino data.

Figures~\ref{misra} and \ref{misga},
obtained with the Rates Analysis and Global Analysis,
respectively,
show that the allowed regions of the neutrino mixing parameters
are rather restricted
and not much larger than those
obtained in the SSM model dependent analysis
(shown in Figs.~\ref{ssmra} and \ref{ssmga}).
One can distinguish clearly the usual
SMA, LMA, LOW and VO regions.
Therefore,
we conclude that the indication that
the true values of the neutrino mixing parameters
lie in one of these regions is robust.

We have also shown in Section~\ref{Oscillation Parameters}
that there is a strong model independent
indication in favor of
large neutrino mixing and large values of $\Delta{m}^2$.
This is very important for future terrestrial experiments
that could explore this part of the parameter space of neutrino mixing
\cite{KAMLAND,BOREXINO}.
In particular,
the model independent Global Analysis strongly disfavors the SMA solution
of the solar neutrino problem.
This is an impressive consequence of the
high quality of the existing data.

In Section~\ref{Neutrino Fluxes}
we have presented the posterior distributions of the
$pp$, $^7\mathrm{Be}$ and $^8\mathrm{B}$
neutrino fluxes,
to which the existing data are most sensitive
(the posterior distribution of the other fluxes is similar
to the prior distribution).
We have shown that the posterior distributions of these fluxes
are in excellent agreement with the BP2000 Standard Solar Model distribution,
with a slight preference of a high $pp$ flux
and a low $^7\mathrm{Be}$ flux.
The $^8\mathrm{B}$
is severely constrained in an interval around the
SSM value
(see Table~\ref{fluxint} and Eqs.~(\ref{8B9973rates}), (\ref{8B9973global})),
in agreement with the limits
found in
Refs.~\cite{SNO-01,Fogli:2001vr,Giunti-aoe-01,Krastev:2001tv,
Smy:2001yn,Fiorentini:2001jt,Bahcall:2001cb}.

Finally,
in Section~\ref{Oscillations versus no oscillations}
we have shown that the hypothesis of no neutrino oscillations
is strongly disfavored with respect to the hypothesis
of $\nu_e\to\nu_{\mu,\tau}$ oscillations,
in agreement with the results obtained in
Refs.~\cite{SNO-01,Fogli:2001vr,Giunti-aoe-01,Fiorentini:2001jt,Bahcall:2001pe}.

In conclusion, we would like to emphasize the usefulness of
the quantity and quality of the existing solar neutrino data,
that nowadays allow to obtain stringent model independent information
on neutrino physics and solar physics.
We enthusiastically look forward towards the realization
of new more sensitive
solar neutrino experiments in order to improve our knowledge.

\begin{acknowledgments}
We would like to express our gratitude to
M.C. Gonzalez-Garcia and
C. Pena-Garay for useful discussions on the analysis of solar neutrino data.
M.V. Garzelli thanks A. Palazzo for very useful
discussions and suggestions
and
A. Bottino for encouragments and support.
C. Giunti would like to thank the
Korea Institute for Advanced Study (KIAS)
for warm hospitality during the final part of this work.
\end{acknowledgments}

\input{mis.bbl}
\clearpage

\input{mis_tab.tex}

\clearpage

\input{mis_fig.tex}

\end{document}

%% file: fig/misga_f5_1.tab
0.45

%% file: fig/misga_f5_2.tab
1.42

%% file: fig/misra_fit_par_5_1.tab
1.0 \times 10^{-2}

%% file: fig/misra_fit_par_5_2.tab
3.0 \times 10^{-5}

%% file: fig/misra_fit_par_6_1.tab
5.0 \times 10^{-2}

%% file: fig/misra_fit_par_6_2.tab
1.0 \times 10^{-6}

%% file: fig/misra_fit_par_7_1.tab
1.00

%% file: fig/misra_fit_par_7_2.tab
9.0 \times 10^{-11}

%% file: fig/misra_fit_par_8_1.tab
1.00

%% file: fig/misra_fit_par_8_2.tab
5.5 \times 10^{-12}

%% file: fig/misga_fit_f5_5.tab
0.53

%% file: fig/misga_fit_f5_9.tab
0.44

%% file: fig/misra_f5_1.tab
0.42

%% file: fig/misra_f5_2.tab
1.75

%% file: fig/misra_odds.tab
8.7 \times 10^{-4}

%% file: fig/misga_odds.tab
7.4 \times 10^{-4}

%% file: mis_tab.tex
\begin{table}
\begin{center}
\begin{tabular}{|c|c|c|}
\hline
\vphantom{\bigg|}
$j$
&
Detection Material and Process
&
Data
\\
\hline
\vphantom{\bigg|}
1
&
\begin{tabular}{c}
${}^{37}\mathrm{Cl}$:
\quad
$\nu_e + {}^{37}\mathrm{Cl} \to {}^{37}\mathrm{Ar} + e^-$
\\
(Homestake \protect\cite{Homestake-98})
\end{tabular}
&
$ 2.56 \pm 0.23 \, \mathrm{SNU} $
\\
\hline
\vphantom{\bigg|}
2
&
\begin{tabular}{c}
${}^{71}\mathrm{Ga}$:
\quad
$\nu_e + {}^{71}\mathrm{Ga} \to {}^{71}\mathrm{Ge} + e^-$
\\
(GALLEX \protect\cite{GALLEX-99}
+
GNO \protect\cite{GNO-00}
+
SAGE \protect\cite{SAGE-nu00})
\end{tabular}
&
$ 74.7 \pm 5.1 \, \mathrm{SNU} $
\\
\hline
\vphantom{\bigg|}
3
&
\begin{tabular}{c}
$\mathrm{D}_2\mathrm{O}$:
\quad
$\nu_e + d \to p + p + e^-$
\\
(SNO \protect\cite{SNO-01})
\end{tabular}
&
$ 0.347 \pm 0.028 $
\\
\hline
\vphantom{\bigg|}
4
&
\begin{tabular}{c}
$\mathrm{H}_2\mathrm{O}$:
\quad
$\nu + e^- \to \nu + e^-$
\\
(Super-Kamiokande \protect\cite{SK-sun-01})
\end{tabular}
&
$ 0.459 \pm 0.017 $
\\
\hline
\end{tabular}
\end{center}
\caption{
\label{rates}
The rates measured in solar neutrino experiments.
The rates of the Homestake and GALLEX+SAGE+GNO
experiments are expressed in SNU units
($ 1 \, \mathrm{SNU} \equiv 10^{-36} \,
\mathrm{events} \, \mathrm{atom}^{-1} \, \mathrm{s}^{-1} $),
whereas the results of the Kamiokande and SNO experiments
are expressed in terms of the ratio of the
experimental rate and the BP2000 Standard Solar Model
prediction \protect\cite{BP2000}.
The statistical and systematic uncertainties have been added in quadrature.
The GALLEX+SAGE+GNO rate is a weighted average of the
GALLEX+GNO rate reported in Ref.~\protect\cite{GNO-00}
and the SAGE rate reported in Ref.~\protect\cite{SAGE-nu00}.
The rate of the SNO experiment is that measured through CC weak interactions.
}
\end{table}


\begin{table}
\begin{center}
\begin{tabular}{|c|c|c|c|c|c|}
\hline
$
\begin{array}{c}
\bin
\\
\text{day}
\end{array}
$
&
$
\begin{array}{c}
\bin
\\
\text{day}
\end{array}
$
&
$
\begin{array}{c}
K_{3\bin}^0
\\
(\mathrm{cm}^2 \, \mathrm{s})
\end{array}
$
&
$
\begin{array}{c}
K_{5\bin}^0
\\
(\mathrm{cm}^2 \, \mathrm{s})
\end{array}
$
&
$ \delta^{\mathrm{(th)}}_{\bin,\mathrm{cor}} $
&
$ \delta^{\mathrm{(sys)}}_{\bin,\mathrm{cor}} $
\\
\hline
 1 & 20 & $6.21\times10^{-8}$ & $1.98\times10^{-7}$ & 0.0039 & 0.0020 \\ \hline
 2 & 21 & $6.62\times10^{-8}$ & $1.98\times10^{-7}$ & 0.0039 & 0.0020 \\ \hline
 3 & 22 & $7.09\times10^{-8}$ & $1.98\times10^{-7}$ & 0.0047 & 0.0029 \\ \hline
 4 & 23 & $7.63\times10^{-8}$ & $1.98\times10^{-7}$ & 0.0055 & 0.0058 \\ \hline
 5 & 24 & $8.22\times10^{-8}$ & $1.98\times10^{-7}$ & 0.0062 & 0.0077 \\ \hline
 6 & 25 & $9.23\times10^{-8}$ & $1.98\times10^{-7}$ & 0.0070 & 0.0106 \\ \hline
 7 & 26 & $1.00\times10^{-7}$ & $1.98\times10^{-7}$ & 0.0078 & 0.0134 \\ \hline
 8 & 27 & $1.12\times10^{-7}$ & $1.98\times10^{-7}$ & 0.0086 & 0.0163 \\ \hline
 9 & 28 & $1.29\times10^{-7}$ & $1.98\times10^{-7}$ & 0.0093 & 0.0203 \\ \hline
10 & 29 & $1.47\times10^{-7}$ & $1.98\times10^{-7}$ & 0.0101 & 0.0242 \\ \hline
11 & 30 & $1.72\times10^{-7}$ & $1.98\times10^{-7}$ & 0.0109 & 0.0292 \\ \hline
12 & 31 & $2.03\times10^{-7}$ & $1.98\times10^{-7}$ & 0.0117 & 0.0331 \\ \hline
13 & 32 & $2.44\times10^{-7}$ & $1.98\times10^{-7}$ & 0.0124 & 0.0381 \\ \hline
14 & 33 & $2.91\times10^{-7}$ & $1.97\times10^{-7}$ & 0.0132 & 0.0440 \\ \hline
15 & 34 & $3.72\times10^{-7}$ & $1.97\times10^{-7}$ & 0.0140 & 0.0500 \\ \hline
16 & 35 & $4.72\times10^{-7}$ & $1.97\times10^{-7}$ & 0.0148 & 0.0570 \\ \hline
17 & 36 & $6.08\times10^{-7}$ & $1.97\times10^{-7}$ & 0.0155 & 0.0640 \\ \hline
18 & 37 & $8.21\times10^{-7}$ & $1.97\times10^{-7}$ & 0.0163 & 0.0730 \\ \hline
19 & 38 & $1.95\times10^{-6}$ & $1.94\times10^{-7}$ & 0.0205 & 0.1058 \\ \hline
\end{tabular}
\end{center}
\caption{ \label{delta}
Values of the relative theoretical uncertainties
$\delta^{\mathrm{(th)}}_{\bin,\mathrm{cor}}$
and experimental uncertainties
$\delta^{\mathrm{(sys)}}_{\bin,\mathrm{cor}}$
of the Super-Kamiokande energy bins
used in the Global Analysis
(see Section~\ref{Sampling: Global Analysis}).
}
\end{table}


\begin{table}
\begin{center}
\input{fig/intpro.tab}
\end{center}
\caption{ \label{prob}
Integral probabilities
of regions in the
$\tan^2\!\vartheta$--$\Delta{m}^2$ plane
and of intervals of
$\tan^2\!\vartheta$
and
$\Delta{m}^2$
obtained with the model dependent and model independent
Rates Analysis and Global Analysis.
The SMA, LMA, LOW and VO regions are defined in Eqs.~(\ref{SMA})--(\ref{VO}).
}
\end{table}


\begin{table}
\input{fig/dfl.tab}
\caption{ \label{dfl}
Relative uncertainties
$\Delta\!\Phi_{i}/\Phi_{i}^{\mathrm{SSM}}$
of the prior distributions of the neutrino fluxes
in the model independent analysis
confronted with the
relative uncertainties
$(\Delta\!\Phi_{i}^{\mathrm{SSM}}/\Phi_{i}^{\mathrm{SSM}})_{(1\sigma)}$
in the BP2000 Standard Solar Model
and the differences between the fluxes predicted by several models
and the BP2000 SSM.
}
\end{table}


\begin{table}
\begin{center}
Rates Analysis
\\
\begin{tabular}{|c|c|c|c|c|c|c|c|c|c|}
\hline
$\tan^2\!\vartheta$
&
$\Delta{m}^2 \, (\mathrm{eV}^2)$
&
$pp$
&
$pep$
&
$hep$
&
$^7\mathrm{Be}$
&
$^8\mathrm{B}$
&
$^{13}\mathrm{N}$
&
$^{15}\mathrm{O}$
&
$^{17}\mathrm{F}$
\vphantom{\bigg|}
\\
\hline
\input{fig/misra_fit_flu.tab}
\end{tabular}
\end{center}
\caption{ \label{misra_fit_flu}
Values of the neutrino fluxes
normalized to the BP2000 SSM fluxes
corresponding to the lines
in Fig.~\ref{misra_fit},
obtained in the Rates Analysis with the nine selected values of
$\tan^2\!\vartheta$ and $\Delta{m}^2$
listed in the first two columns.
The first four rows correspond,
respectively,
to the best-fit points in the SMA, LMA, LOW and VO regions
in the Rates Analysis.
The fifth to eighth rows correspond to points in
excluded areas in Fig.~\ref{misra}.
The last row corresponds to the case of no oscillations.
}
\end{table}


\begin{table}
\begin{center}
Global Analysis
\\
\begin{tabular}{|c|c|c|c|c|c|c|c|c|c|}
\hline
$\tan^2\!\vartheta$
&
$\Delta{m}^2 \, (\mathrm{eV}^2)$
&
$pp$
&
$pep$
&
$hep$
&
$^7\mathrm{Be}$
&
$^8\mathrm{B}$
&
$^{13}\mathrm{N}$
&
$^{15}\mathrm{O}$
&
$^{17}\mathrm{F}$
\vphantom{\bigg|}
\\
\hline
\input{fig/misga_fit_flu.tab}
\end{tabular}
\end{center}
\caption{ \label{misga_fit_flu}
Values of the neutrino fluxes
normalized to the BP2000 SSM fluxes
corresponding to the lines
in Fig.~\ref{misga_fit},
obtained in the Global Analysis with the nine selected values of
$\tan^2\!\vartheta$ and $\Delta{m}^2$
listed in the first two columns.
The first four rows correspond,
respectively,
to the best-fit points in the SMA, LMA, LOW and VO regions
in the Rates Analysis.
The fifth to eighth rows correspond,
respectively,
to the best-fit points in the SMA, LMA, LOW and VO regions
in the Global Analysis.
The last row corresponds to the case of no oscillations.
}
\end{table}


\begin{table}
\begin{center}
\input{fig/flulim.tab}
\end{center}
\caption{ \label{fluxint}
Credible intervals for the $pp$, ${^7\mathrm{Be}}$ and ${^8\mathrm{B}}$ fluxes
with 90\%, 95\% and 99\%
probability obtained
with the model independent Rates and Global Analyses.
}
\end{table}


%% file: fig/intpro.tab
\begin{tabular}{|c|c|c|c|c|}
\hline
\vphantom{\Big|}
&
\multicolumn{2}{|c|}{BP2000 SSM}
&
\multicolumn{2}{|c|}{Model Independent}
\\
\cline{2-5}
\vphantom{\Big|}
&
Rates
&
Global
&
Rates
&
Global
\\
\hline
\vphantom{\Big|}
$\mathrm{SMA}$
&
$9.9 \times 10^{-2}$
&
$7.2 \times 10^{-5}$
&
$0.15$
&
$1.0 \times 10^{-3}$
\\
\hline
\vphantom{\Big|}
$\mathrm{LMA}$
&
$0.72$
&
$0.86$
&
$0.67$
&
$0.80$
\\
\hline
\vphantom{\Big|}
$\mathrm{LOW}$
&
$6.1 \times 10^{-2}$
&
$0.12$
&
$3.8 \times 10^{-2}$
&
$0.13$
\\
\hline
\vphantom{\Big|}
$\mathrm{VO}$
&
$0.12$
&
$1.9 \times 10^{-2}$
&
$0.14$
&
$6.6 \times 10^{-2}$
\\
\hline
\vphantom{\Big|}
$\tan^2\!\vartheta<0.1$
&
$9.9 \times 10^{-2}$
&
$7.2 \times 10^{-5}$
&
$0.18$
&
$8.2 \times 10^{-3}$
\\
\hline
\vphantom{\Big|}
$0.1<\tan^2\!\vartheta<10$
&
$0.90$
&
$1.00$
&
$0.82$
&
$0.99$
\\
\hline
\vphantom{\Big|}
$\tan^2\!\vartheta>10$
&
$6.5 \times 10^{-8}$
&
$1.3 \times 10^{-8}$
&
$1.1 \times 10^{-3}$
&
$1.7 \times 10^{-3}$
\\
\hline
\vphantom{\Big|}
$\Delta{m}^2<10^{-5} \, \mathrm{eV}^2$
&
$0.32$
&
$0.14$
&
$0.41$
&
$0.20$
\\
\hline
\vphantom{\Big|}
$\Delta{m}^2>10^{-5} \, \mathrm{eV}^2$
&
$0.68$
&
$0.86$
&
$0.59$
&
$0.80$
\\
\hline
\end{tabular}

%% file: fig/dfl.tab
\begin{center}
\footnotesize
\begin{tabular}{|c|c|c|c|c|c|c|c|c|}
\hline
$i$&1&2&3&4&5&6&7&8
\\
&
$pp$
&
$pep$
&
$hep$
&
$^7\mathrm{Be}$
&
$^8\mathrm{B}$
&
$^{13}\mathrm{N}$
&
$^{15}\mathrm{O}$
&
$^{17}\mathrm{F}$
\vphantom{$\Big|$}
\\
\hline
$\Delta\!\Phi_{i}/\Phi_{i}^{\mathrm{SSM}}$
&
$
\begin{array}{l}
+ 0.094
\\
- 0.096
\end{array}
$
&
$\pm1$
&
$
\begin{array}{l}
+ 9
\\
- 1
\end{array}
$
&
$\pm1$
&
$\pm1$
&
$\pm1$
&
$\pm1$
&
$\pm1$
\vphantom{$\Big|$}
\\
\hline
$(\Delta\!\Phi_{i}^{\mathrm{SSM}}/\Phi_{i}^{\mathrm{SSM}})_{(1\sigma)}$
&
$\pm0.010$
&
$\pm0.015$
&
$-$
&
$\pm0.100$
&
$\pm0.200$
&
$\pm0.210$
&
$\pm0.250$
&
$\pm0.250$
\vphantom{$\Big|$}
\\
\hline
NACRE (3.842) \protect\cite{BP2000} 
&
$0.002$
&
$-0.007$
&
$0.011$
&
$0.008$
&
$0.077$
&
$-0.111$
&
$-0.129$
&
$-0.059$
\vphantom{$\Big|$}
\\
\hline
BP2000 (3.844) \protect\cite{BP2000} 
&
$0.002$
&
$0.000$
&
$0.000$
&
$0.010$
&
$0.020$
&
$0.015$
&
$0.017$
&
$0.018$
\vphantom{$\Big|$}
\\
\hline
NACRE (3.844) \protect\cite{BP2000} 
&
$0.003$
&
$-0.007$
&
$0.011$
&
$0.017$
&
$0.097$
&
$-0.100$
&
$-0.117$
&
$-0.043$
\vphantom{$\Big|$}
\\
\hline
AS00 \protect\cite{BP2000}
&
$0.007$
&
$0.007$
&
$0.011$
&
$-0.031$
&
$-0.069$
&
$-0.042$
&
$-0.050$
&
$-0.053$
\vphantom{$\Big|$}
\\
\hline
GN 93 \protect\cite{BP2000}
&
$-0.002$
&
$-0.007$
&
$-0.011$
&
$0.023$
&
$0.051$
&
$0.128$
&
$0.135$
&
$0.155$
\vphantom{$\Big|$}
\\
\hline
Pre-M.S. \protect\cite{BP2000}
&
$0.000$
&
$-0.007$
&
$-0.011$
&
$0.021$
&
$0.048$
&
$0.124$
&
$0.131$
&
$0.149$
\vphantom{$\Big|$}
\\
\hline
Rotation \protect\cite{BP2000}
&
$0.005$
&
$0.000$
&
$-0.011$
&
$-0.019$
&
$-0.028$
&
$0.016$
&
$0.015$
&
$0.028$
\vphantom{$\Big|$}
\\
\hline
Radius$_{78}$ \protect\cite{BP2000}
&
$-0.002$
&
$-0.007$
&
$-0.011$
&
$0.023$
&
$0.051$
&
$0.128$
&
$0.135$
&
$0.155$
\vphantom{$\Big|$}
\\
\hline
Radius$_{508}$ \protect\cite{BP2000}
&
$-0.002$
&
$-0.007$
&
$-0.011$
&
$0.023$
&
$0.051$
&
$0.128$
&
$0.135$
&
$0.155$
\vphantom{$\Big|$}
\\
\hline
No Diffusion \protect\cite{BP2000}
&
$0.017$
&
$0.021$
&
$0.032$
&
$-0.117$
&
$-0.234$
&
$-0.254$
&
$-0.279$
&
$-0.281$
\vphantom{$\Big|$}
\\
\hline
Old Physics \protect\cite{BP2000}
&
$0.000$
&
$0.007$
&
$-0.011$
&
$0.029$
&
$0.020$
&
$0.053$
&
$0.048$
&
$0.052$
\vphantom{$\Big|$}
\\
\hline
$S_{3,4}=0$ \protect\cite{BP2000}
&
$0.076$
&
$0.107$
&
$0.086$
&
$-1.000$
&
$-1.000$
&
$0.181$
&
$0.175$
&
$0.190$
\vphantom{$\Big|$}
\\
\hline
Mixed \protect\cite{BP2000}
&
$0.030$
&
$-0.093$
&
$-0.333$
&
$-0.252$
&
$-0.182$
&
$-0.445$
&
$-0.365$
&
$-0.359$
\vphantom{$\Big|$}
\\
\hline
BP98 \protect\cite{BP98}
&
$-0.002$
&
$-0.007$
&
$-0.774$
&
$0.006$
&
$0.020$
&
$0.104$
&
$0.108$
&
$0.124$
\vphantom{$\Big|$}
\\
\hline
BP95 \protect\cite{BP95}
&
$-0.007$
&
$0.000$
&
$-0.871$
&
$0.080$
&
$0.311$
&
$0.128$
&
$0.135$
&
$0.151$
\vphantom{$\Big|$}
\\
\hline
DS94 
&
$0.017$
&
$0.039$
&
$-0.866$
&
$-0.107$
&
$-0.298$
&
$-0.074$
&
$-0.083$
&
$-0.211$
\vphantom{$\Big|$}
\\
\hline
TL93 \protect\cite{TurckChieze-Lopes-93,TurckChieze-PR230-93}
&
$0.013$
&
$-0.007$
&
$-$
&
$-0.090$
&
$-0.123$
&
$-0.301$
&
$-0.337$
&
$-$
\vphantom{$\Big|$}
\\
\hline
BP92 \protect\cite{BP92}
&
$0.008$
&
$0.021$
&
$-0.868$
&
$0.025$
&
$0.127$
&
$-0.102$
&
$-0.113$
&
$-0.043$
\vphantom{$\Big|$}
\\
\hline
BU88 \protect\cite{BU88}
&
$0.008$
&
$0.000$
&
$-0.185$
&
$-0.017$
&
$0.141$
&
$0.111$
&
$0.088$
&
$-0.083$
\vphantom{$\Big|$}
\\
\hline
\end{tabular}
\end{center}

%% file: fig/misra_fit_flu.tab
$1.7 \times 10^{-3}$ & $8.0 \times 10^{-6}$ &
 $0.935$ & $0.220$ & $8.907$ & $1.992$ & $1.290$ & $1.090$ & $0.177$ & $0.063$\vphantom{\bigg|}
\\
\hline
$0.20$ & $1.2 \times 10^{-5}$ &
 $0.962$ & $0.960$ & $6.752$ & $1.508$ & $1.354$ & $1.362$ & $0.764$ & $1.253$\vphantom{\bigg|}
\\
\hline
$0.49$ & $1.4 \times 10^{-7}$ &
 $0.970$ & $0.813$ & $1.012$ & $1.437$ & $1.003$ & $0.805$ & $0.749$ & $1.250$\vphantom{\bigg|}
\\
\hline
$3.59$ & $8.0 \times 10^{-11}$ &
 $1.031$ & $1.779$ & $1.449$ & $0.739$ & $0.904$ & $0.800$ & $0.033$ & $1.032$\vphantom{\bigg|}
\\
\hline
$1.0 \times 10^{-2}$ & $3.0 \times 10^{-5}$ &
 $1.035$ & $1.344$ & $5.116$ & $0.349$ & $1.981$ & $1.967$ & $1.861$ & $1.632$\vphantom{\bigg|}
\\
\hline
$5.0 \times 10^{-2}$ & $1.0 \times 10^{-6}$ &
 $0.916$ & $1.885$ & $5.203$ & $1.996$ & $1.997$ & $1.758$ & $1.173$ & $1.844$\vphantom{\bigg|}
\\
\hline
$1.00$ & $9.0 \times 10^{-11}$ &
 $0.965$ & $0.841$ & $3.686$ & $1.464$ & $1.563$ & $0.454$ & $1.015$ & $0.202$\vphantom{\bigg|}
\\
\hline
$1.00$ & $5.5 \times 10^{-12}$ &
 $0.983$ & $0.757$ & $0.040$ & $1.212$ & $0.446$ & $1.822$ & $0.853$ & $0.025$\vphantom{\bigg|}
\\
\hline
$0$ & $0$ &
 $1.089$ & $0.679$ & $0.249$ & $0.021$ & $0.428$ & $0.233$ & $0.103$ & $0.481$\vphantom{\bigg|}
\\
\hline

%% file: fig/misga_fit_flu.tab
$1.7 \times 10^{-3}$ & $8.0 \times 10^{-6}$ &
 $0.929$ & $0.165$ & $3.922$ & $1.862$ & $1.260$ & $0.262$ & $1.523$ & $1.557$\vphantom{\bigg|}
\\
\hline
$0.20$ & $1.2 \times 10^{-5}$ &
 $0.966$ & $1.414$ & $0.335$ & $1.507$ & $1.286$ & $1.906$ & $0.301$ & $0.699$\vphantom{\bigg|}
\\
\hline
$0.49$ & $1.4 \times 10^{-7}$ &
 $0.968$ & $0.045$ & $1.561$ & $1.520$ & $0.992$ & $0.617$ & $0.503$ & $1.147$\vphantom{\bigg|}
\\
\hline
$3.59$ & $8.0 \times 10^{-11}$ &
 $1.014$ & $0.987$ & $1.004$ & $0.896$ & $0.931$ & $0.048$ & $0.651$ & $0.426$\vphantom{\bigg|}
\\
\hline
$3.8 \times 10^{-4}$ & $5.3 \times 10^{-6}$ &
 $1.035$ & $0.697$ & $2.764$ & $0.535$ & $0.526$ & $0.394$ & $1.186$ & $0.114$\vphantom{\bigg|}
\\
\hline
$0.34$ & $6.1 \times 10^{-5}$ &
 $1.004$ & $1.248$ & $5.544$ & $1.047$ & $1.104$ & $1.854$ & $0.186$ & $0.433$\vphantom{\bigg|}
\\
\hline
$0.60$ & $1.5 \times 10^{-7}$ &
 $0.994$ & $0.147$ & $5.910$ & $1.223$ & $0.915$ & $0.319$ & $0.342$ & $0.967$\vphantom{\bigg|}
\\
\hline
$3.16$ & $5.0 \times 10^{-10}$ &
 $1.050$ & $1.111$ & $4.100$ & $0.510$ & $0.620$ & $0.008$ & $0.141$ & $0.255$\vphantom{\bigg|}
\\
\hline
$0$ & $0$ &
 $1.088$ & $0.065$ & $0.664$ & $0.071$ & $0.438$ & $0.032$ & $0.031$ & $0.340$\vphantom{\bigg|}
\\
\hline

%% file: fig/flulim.tab
\begin{tabular}{|c|c|c|c|c|c|c|}
\hline
\vphantom{\Big|}
&
\multicolumn{3}{|c|}{Rates Analysis}
&
\multicolumn{3}{|c|}{Global Analysis}
\\
\cline{2-7}
\vphantom{\Big|}
&
90\%
&
95\%
&
99\%
&
90\%
&
95\%
&
99\%
\\
\hline
\vphantom{\Big|}
$\Phi_{pp}/\Phi_{pp}^{\mathrm{SSM}}$
&
$[0.95,1.08]$
&
$[0.93,1.08]$
&
$[0.92,1.09]$
&
$[0.99,1.08]$
&
$[0.98,1.08]$
&
$[0.95,1.09]$
\\
\hline
\vphantom{\Big|}
$\Phi_{^7\mathrm{Be}}/\Phi_{^7\mathrm{Be}}^{\mathrm{SSM}}$
&
$[0.02,1.65]$
&
$[0.02,1.82]$
&
$[0.02,1.95]$
&
$[0.02,1.15]$
&
$[0.02,1.28]$
&
$[0.02,1.62]$
\\
\hline
\vphantom{\Big|}
$\Phi_{^8\mathrm{B}}/\Phi_{^8\mathrm{B}}^{\mathrm{SSM}}$
&
$[0.62,1.48]$
&
$[0.58,1.58]$
&
$[0.48,1.68]$
&
$[0.62,1.22]$
&
$[0.58,1.28]$
&
$[0.48,1.35]$
\\
\hline
\end{tabular}

%% file: mis_fig.tex
\begin{figure}
\begin{center}
\includegraphics[bb=92 314 444 767, width=0.8\textwidth]{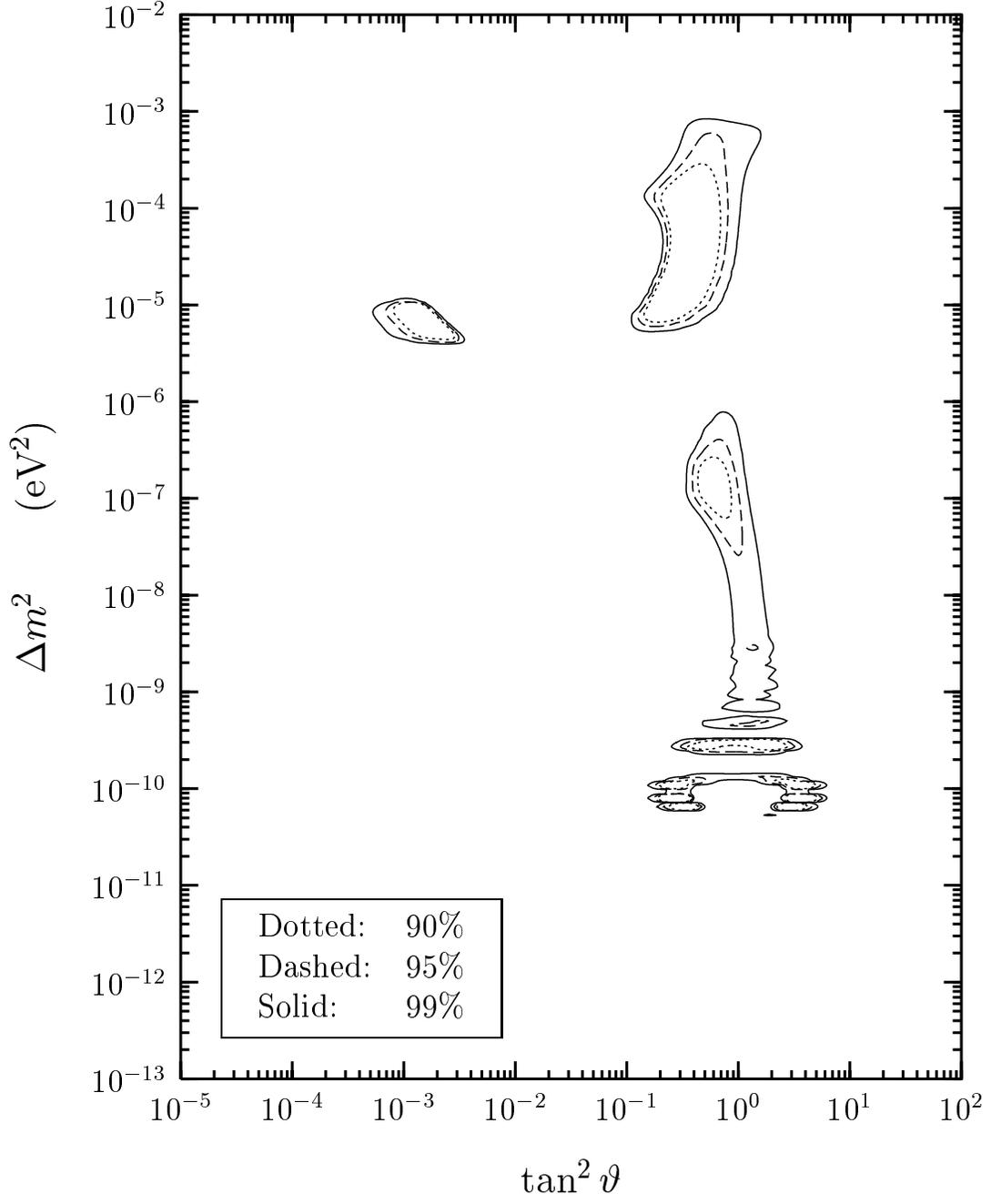}
\end{center}
\caption{ \label{ssmra}
Credible regions obtained with the Rates Analysis
of
Homestake,
GALLEX+GNO+SAGE,
SNO and
Super-Kamiokande total rates,
assuming the BP2000 Standard Solar Model neutrino fluxes.
}
\end{figure}

\clearpage

\begin{figure}
\begin{center}
\includegraphics[bb=92 314 444 767, width=0.8\textwidth]{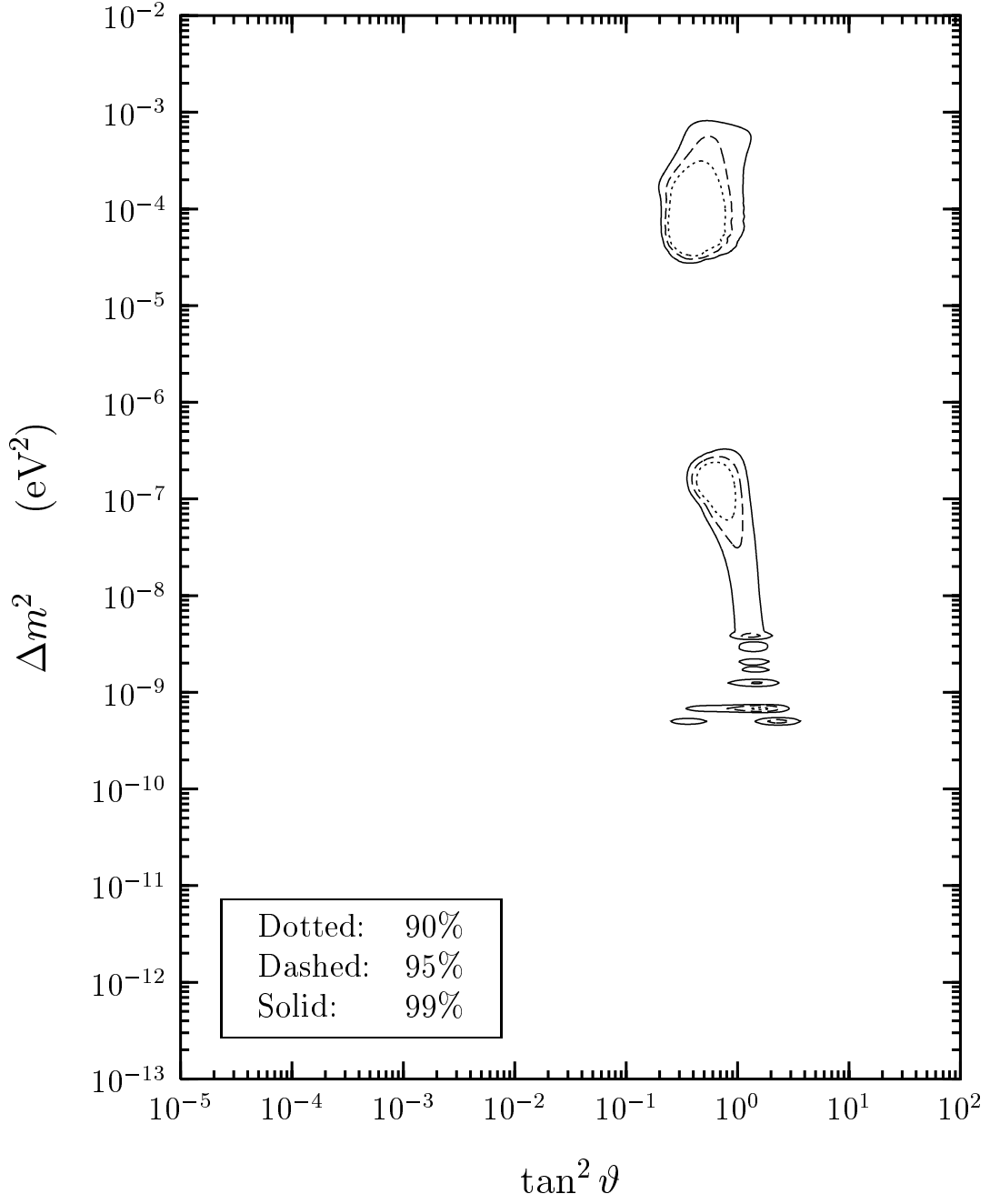}
\end{center}
\caption{ \label{ssmga}
Credible regions obtained with the Global Analysis
of
Homestake,
GALLEX+GNO+SAGE and
SNO rates
and
Super-Kamiokande day and night spectra,
assuming the BP2000 Standard Solar Model neutrino fluxes.
}
\end{figure}

\clearpage

\begin{figure}
\begin{center}
\includegraphics[bb=92 314 444 767, width=0.8\textwidth]{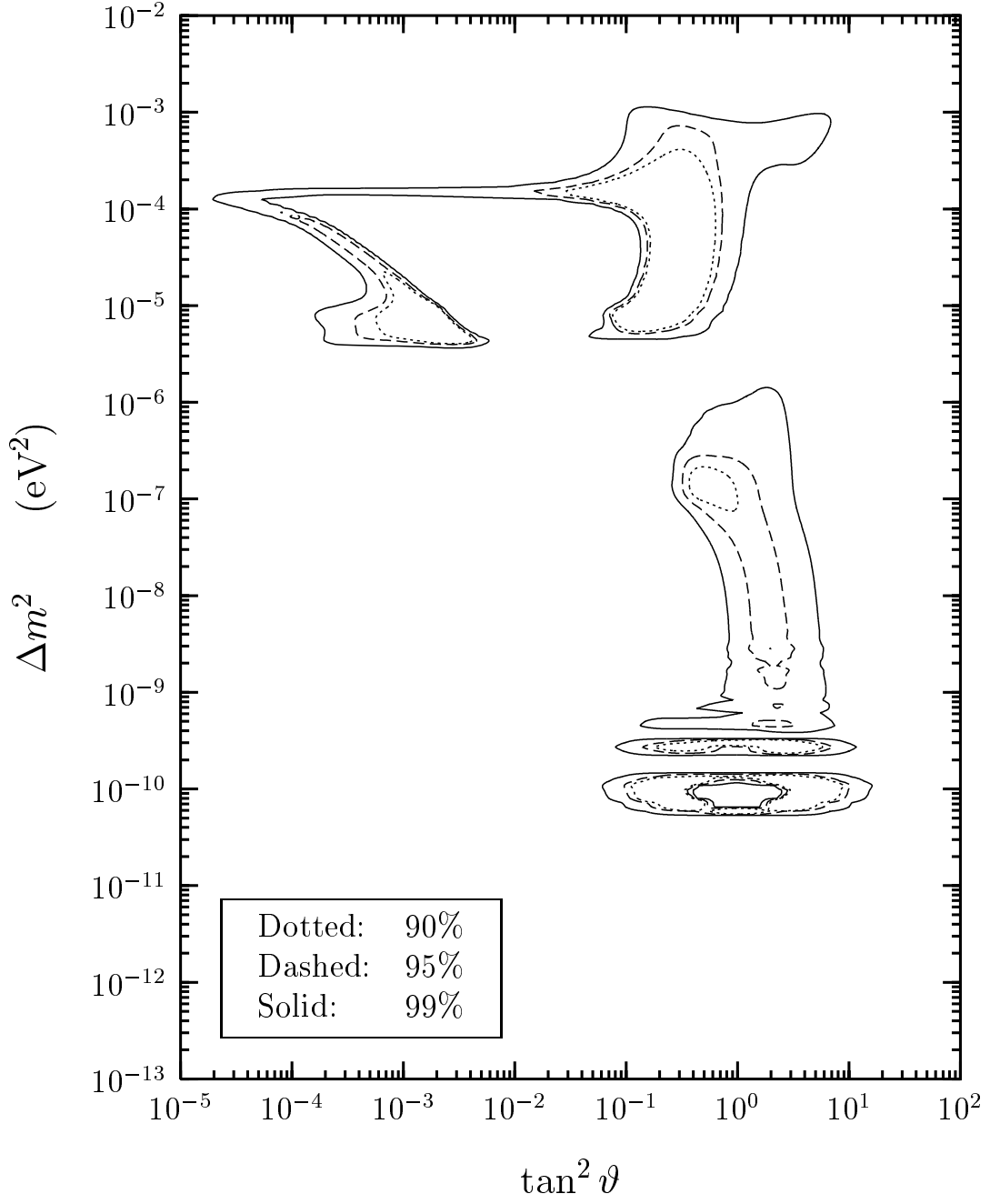}
\end{center}
\caption{ \label{misra}
Credible regions obtained with the model independent Rates Analysis
of
Homestake,
GALLEX+GNO+SAGE,
SNO and
Super-Kamiokande total rates.
}
\end{figure}

\clearpage

\begin{figure}
\begin{center}
\includegraphics[bb=92 314 444 767, width=0.8\textwidth]{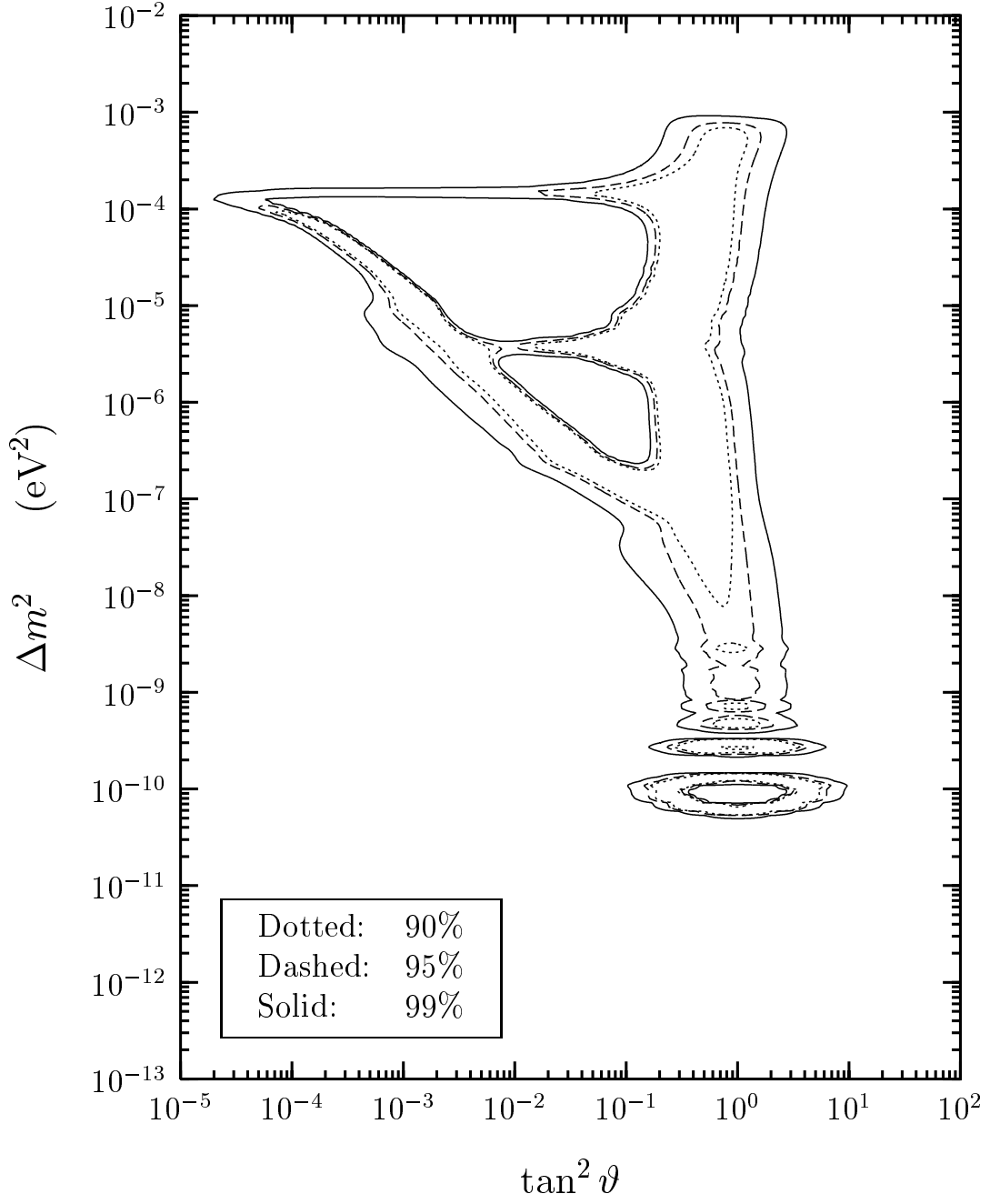}
\end{center}
\caption{ \label{nogal_misra}
Credible regions obtained with the model independent Rates Analysis
of
Homestake,
SNO and
Super-Kamiokande data,
neglecting the Gallium data of the GALLEX, GNO and SAGE experiment.
}
\end{figure}

\clearpage

\begin{figure}
\begin{center}
\includegraphics[bb=92 314 444 767, width=0.8\textwidth]{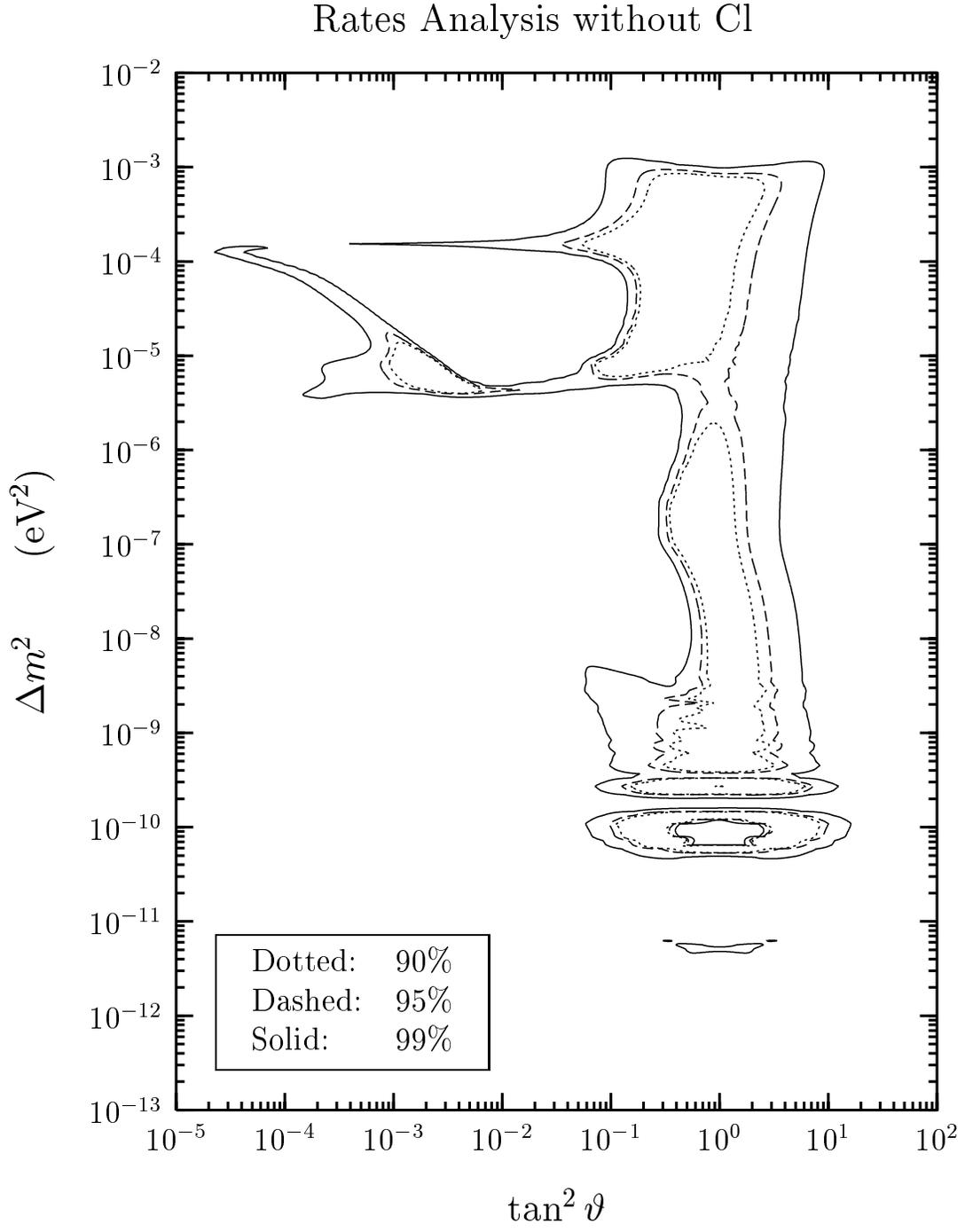}
\end{center}
\caption{ \label{noclo_misra}
Credible regions obtained with the model independent Rates Analysis
of
GALLEX+GNO+SAGE,
SNO and
Super-Kamiokande data,
neglecting the Chlorine data of the Homestake experiment.
}
\end{figure}

\clearpage

\begin{figure}
\begin{center}
\includegraphics[bb=92 314 444 767, width=0.8\textwidth]{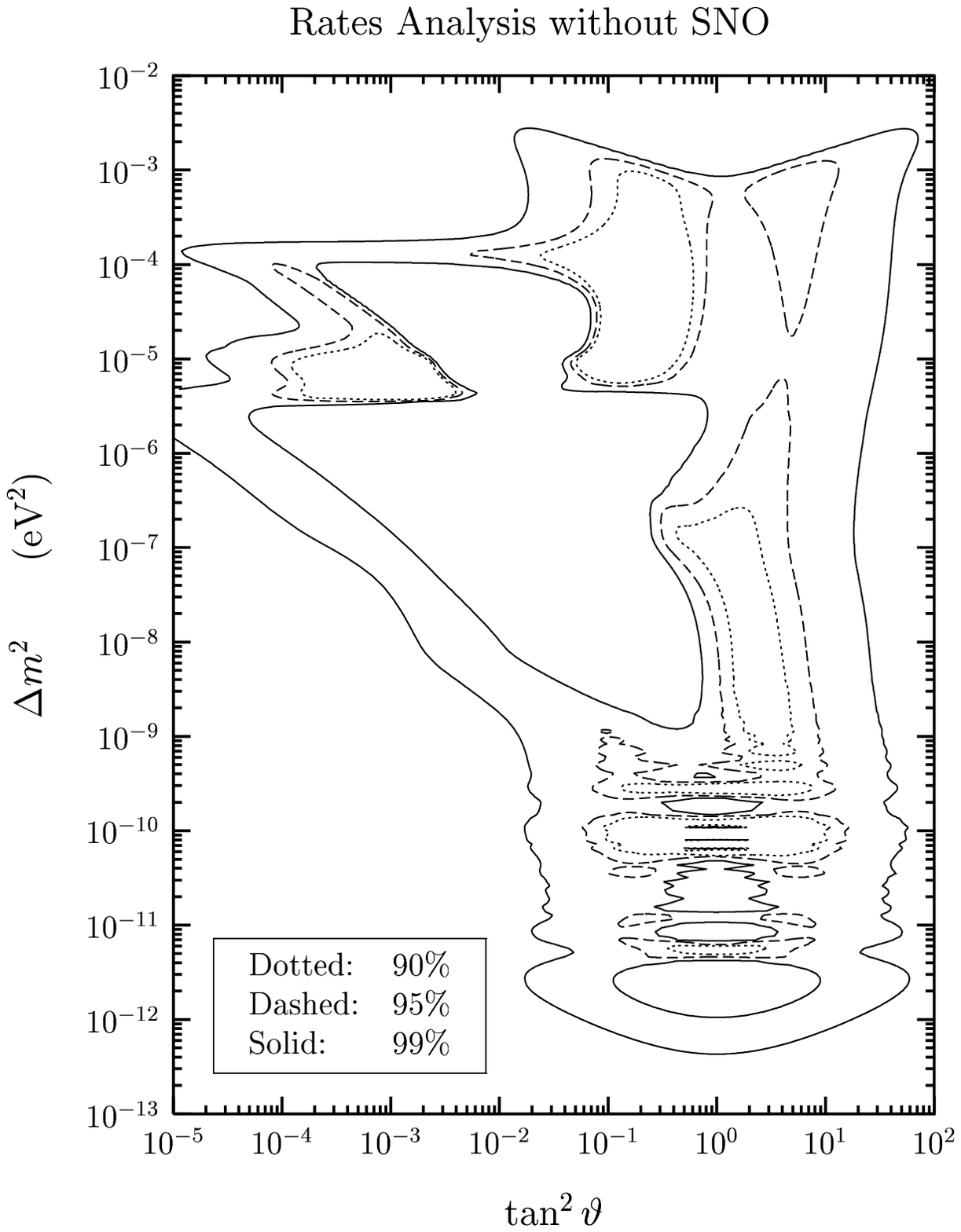}
\end{center}
\caption{ \label{nosno_misra}
Credible regions obtained with the model independent Rates Analysis
of
Homestake,
GALLEX+GNO+SAGE and
Super-Kamiokande data,
neglecting the rate measured in the SNO experiment.
}
\end{figure}

\clearpage

\begin{figure}
\begin{center}
\includegraphics[bb=92 314 444 767, width=0.8\textwidth]{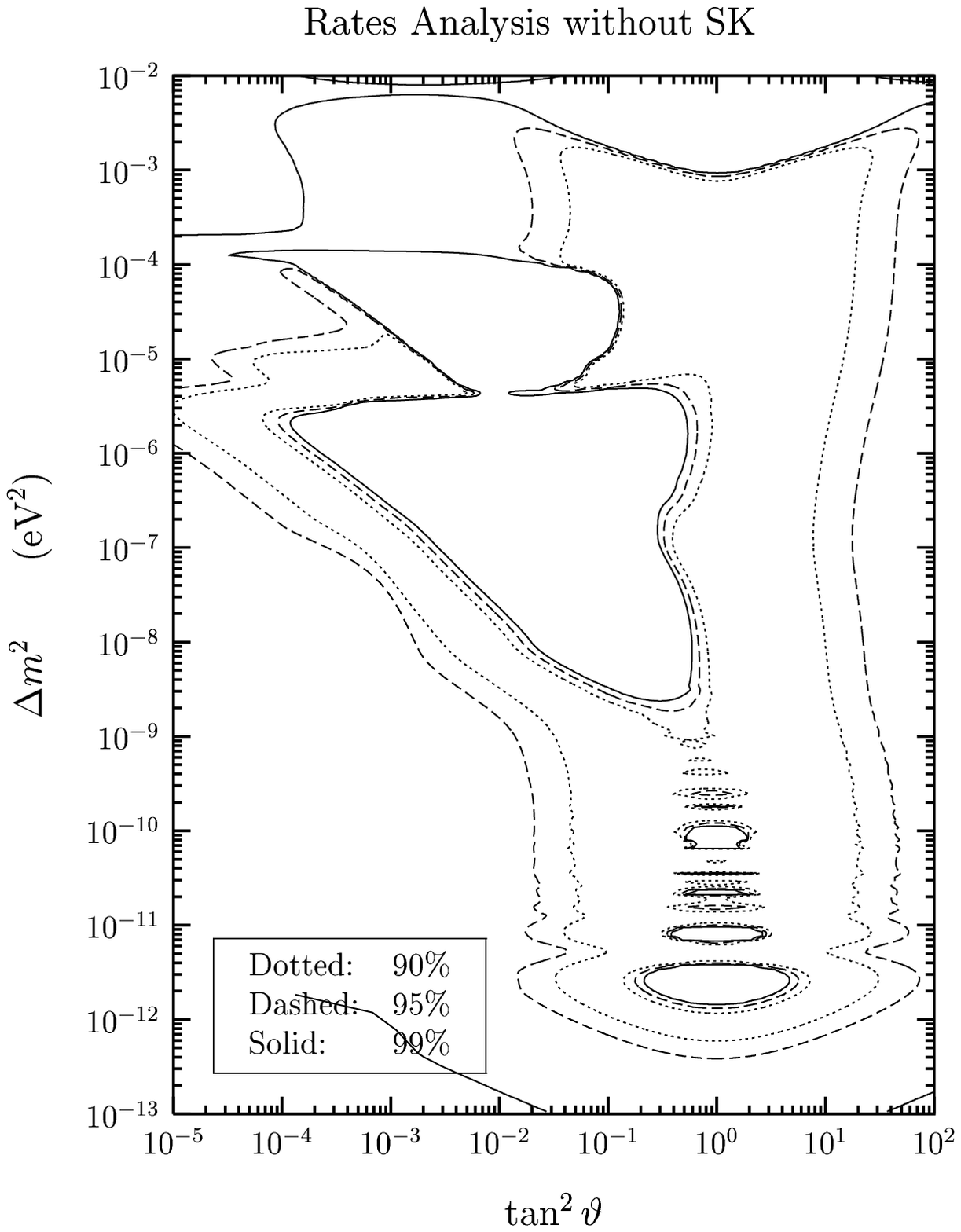}
\end{center}
\caption{ \label{nokam_misra}
Credible regions obtained with the model independent Rates Analysis
of
Homestake,
GALLEX+GNO+SAGE and
SNO data,
neglecting the rate measured in the Super-Kamiokande experiment.
}
\end{figure}

\clearpage

\begin{figure}
\begin{center}
\includegraphics[bb=80 425 437 642, width=0.7\textwidth]{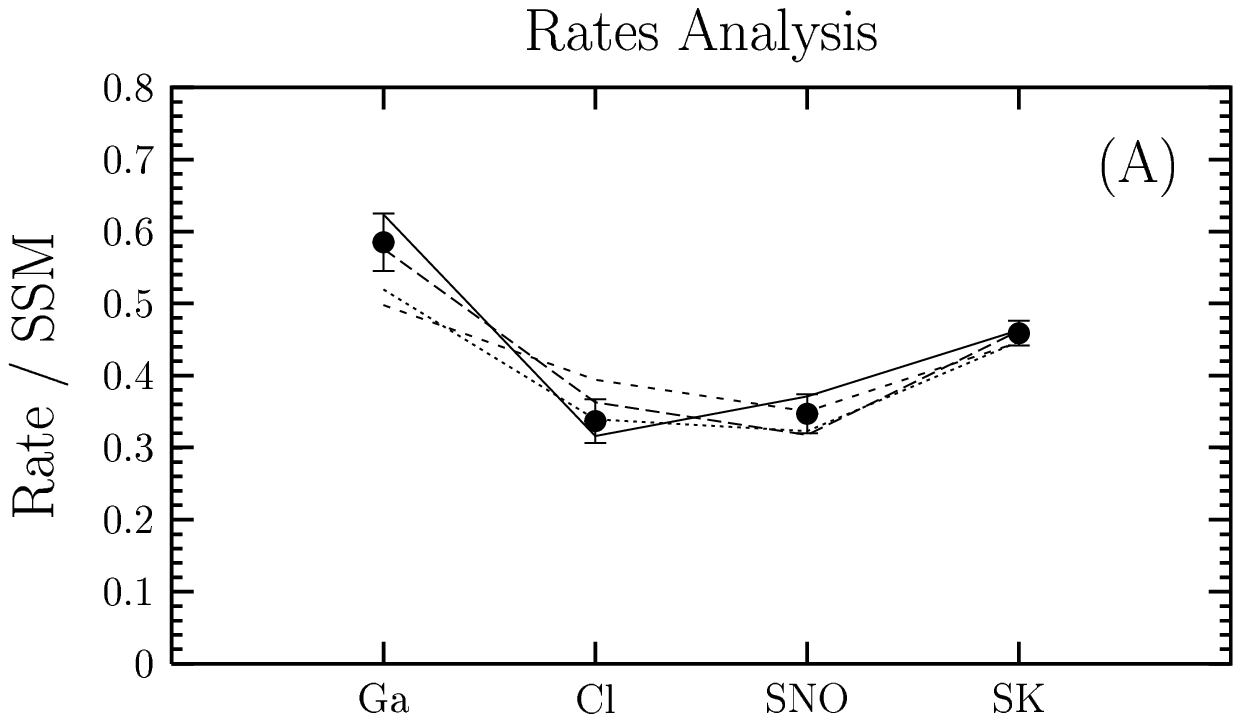}
\includegraphics[bb=80 425 437 642, width=0.7\textwidth]{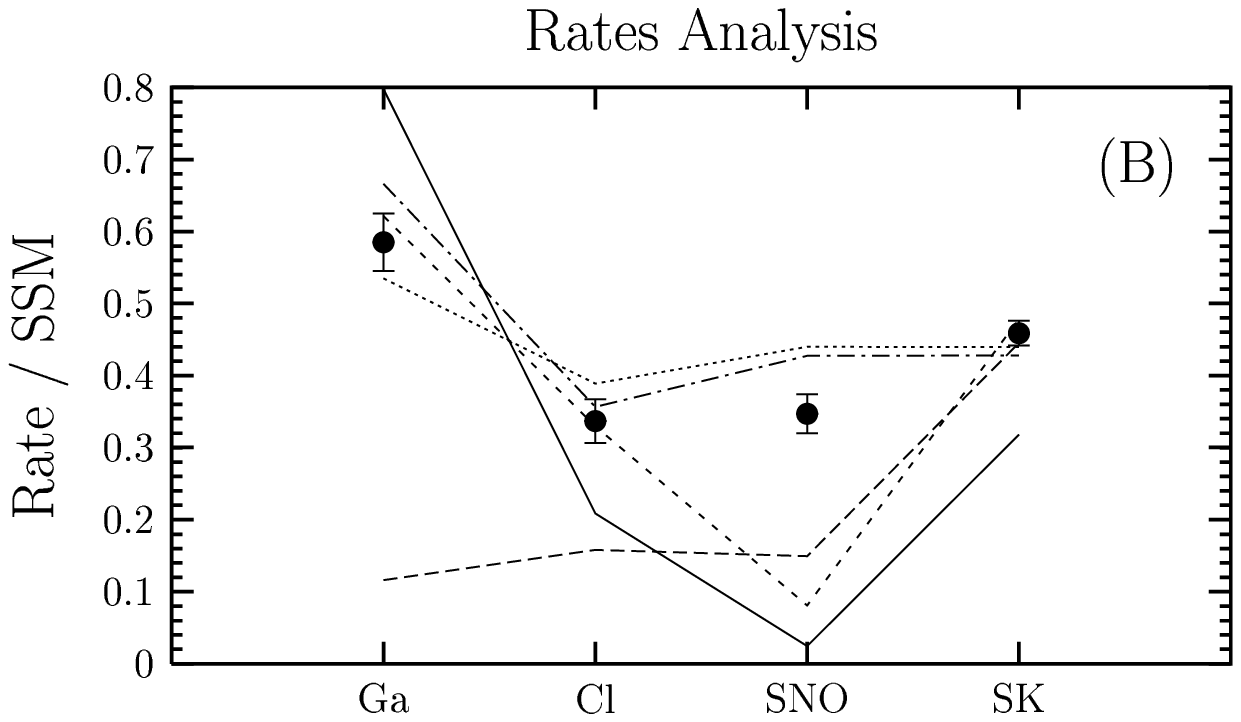}
\end{center}
\caption{ \label{misra_fit}
Experimental rates (points with errorbars) and
theoretical predictions
for nine selected values of
$\tan^2\!\vartheta$ and $\Delta{m}^2$
in the Rates Analysis.
\protect\input{fig/misra_fit_fig.tab}
dash-dotted line:
no oscillations.
}
\end{figure}

\clearpage

\begin{figure}
\begin{center}
\includegraphics[bb=92 314 444 767, width=0.8\textwidth]{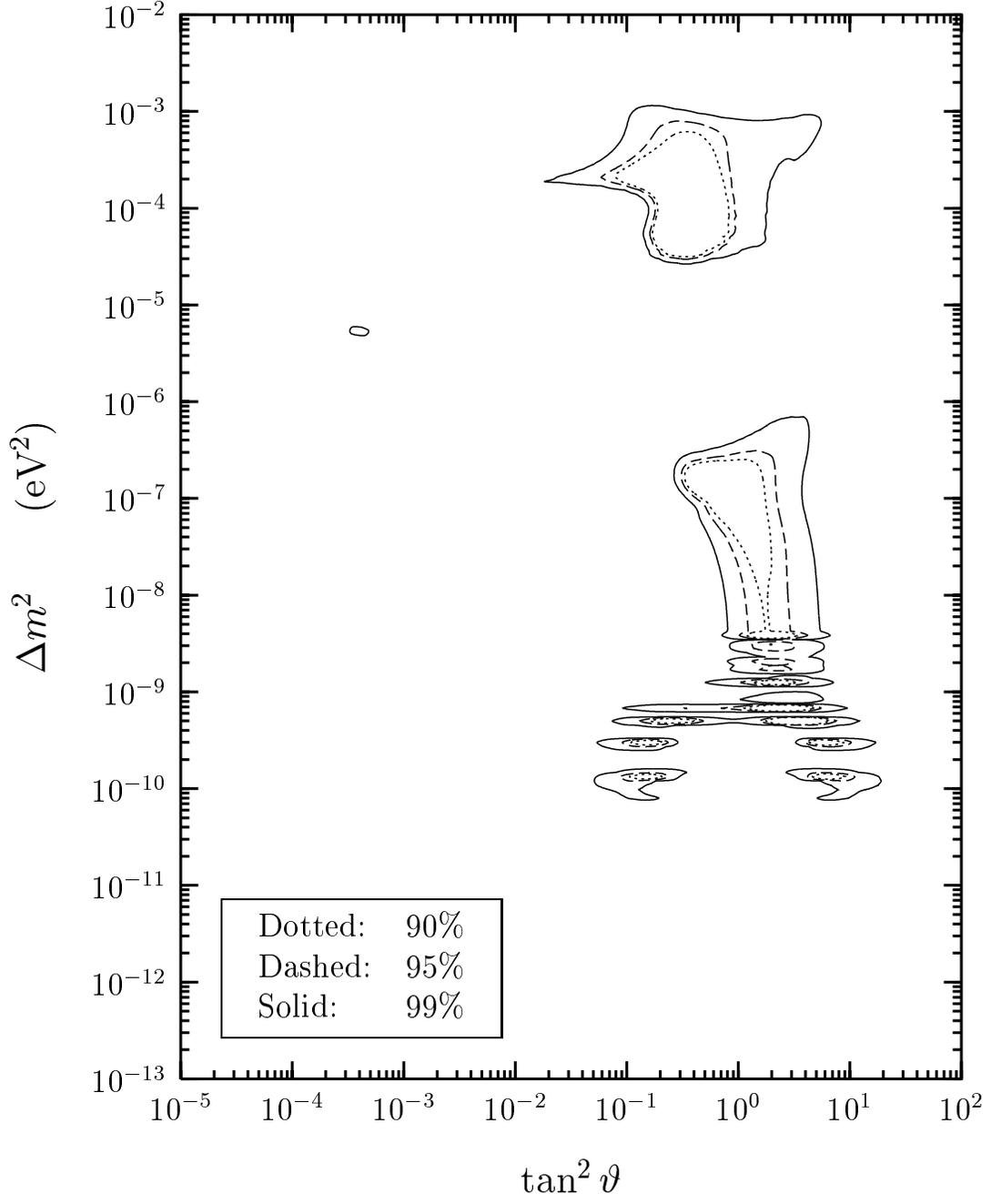}
\end{center}
\caption{ \label{misga}
Credible regions obtained with the model independent Global Analysis
of
Homestake,
GALLEX+GNO+SAGE and
SNO rates
and
Super-Kamiokande day and night spectra.
}
\end{figure}

\clearpage

\begin{figure}
\begin{center}
\includegraphics[bb=80 425 437 642, width=0.7\textwidth]{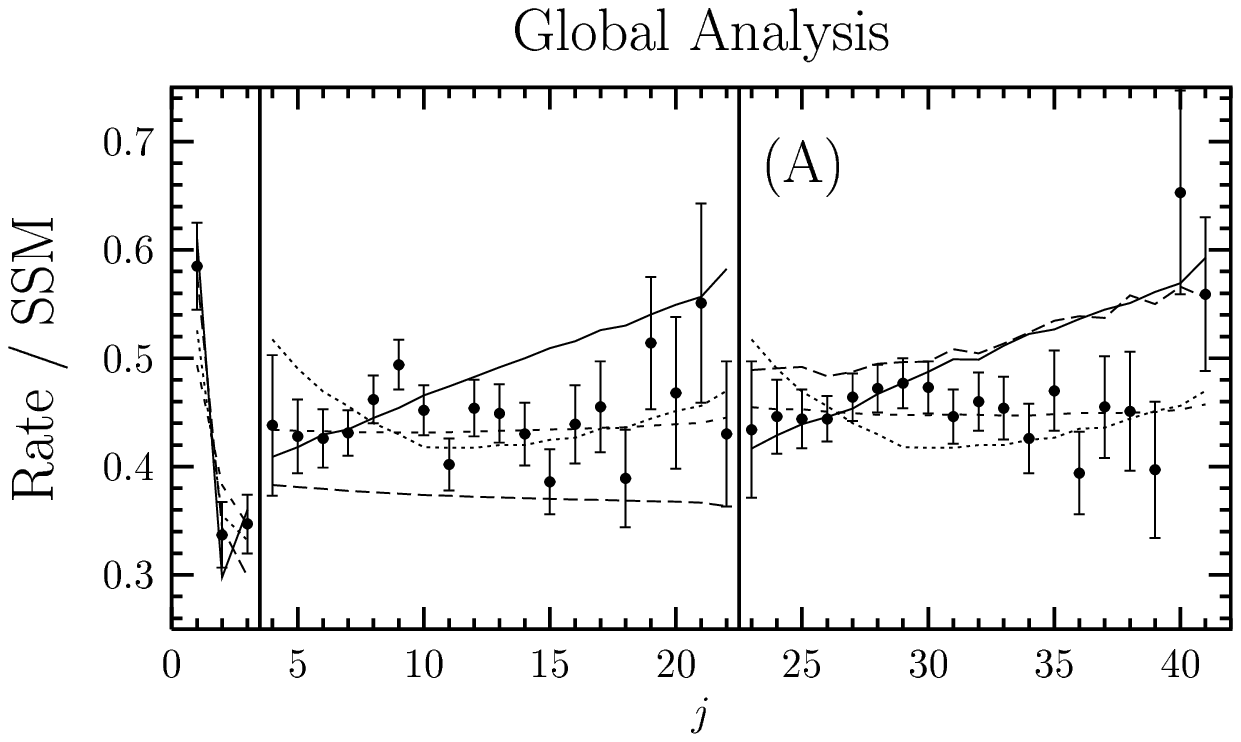}
\includegraphics[bb=80 425 437 642, width=0.7\textwidth]{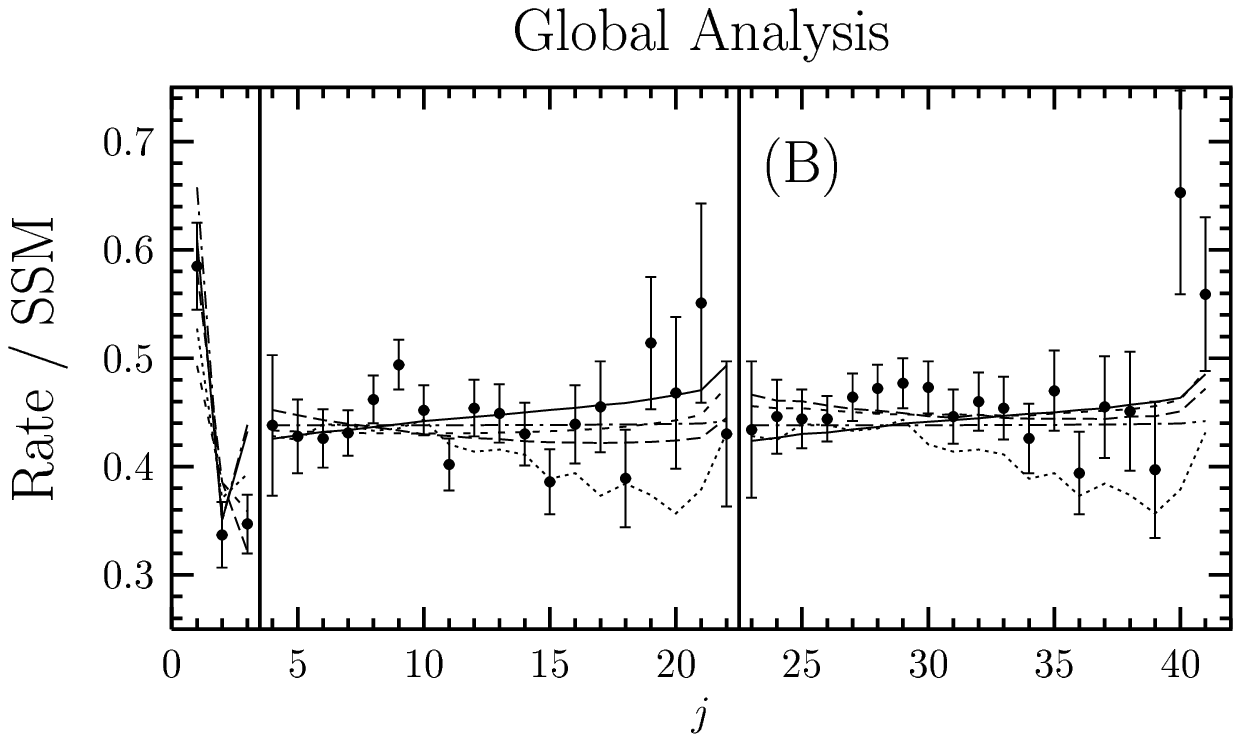}
\end{center}
\caption{ \label{misga_fit}
Experimental rates (points with errorbars) and
theoretical predictions
for nine selected values of
$\tan^2\!\vartheta$ and $\Delta{m}^2$
in the Global Analysis.
The index $j$
labels the solar data points:
$j=1,2,3$
for the Gallium, Chlorine and SNO event rates,
respectively;
$j=4,\ldots,22$
for the Super-Kamiokande energy spectrum in the day;
$j=23,\ldots,41$
for the Super-Kamiokande energy spectrum in the night.
\protect\input{fig/misga_fit_fig.tab}
dash-dotted line:
no oscillations.
}
\end{figure}

\clearpage

\begin{figure}
\begin{center}
\includegraphics[bb=90 427 431 767, width=0.8\textwidth]{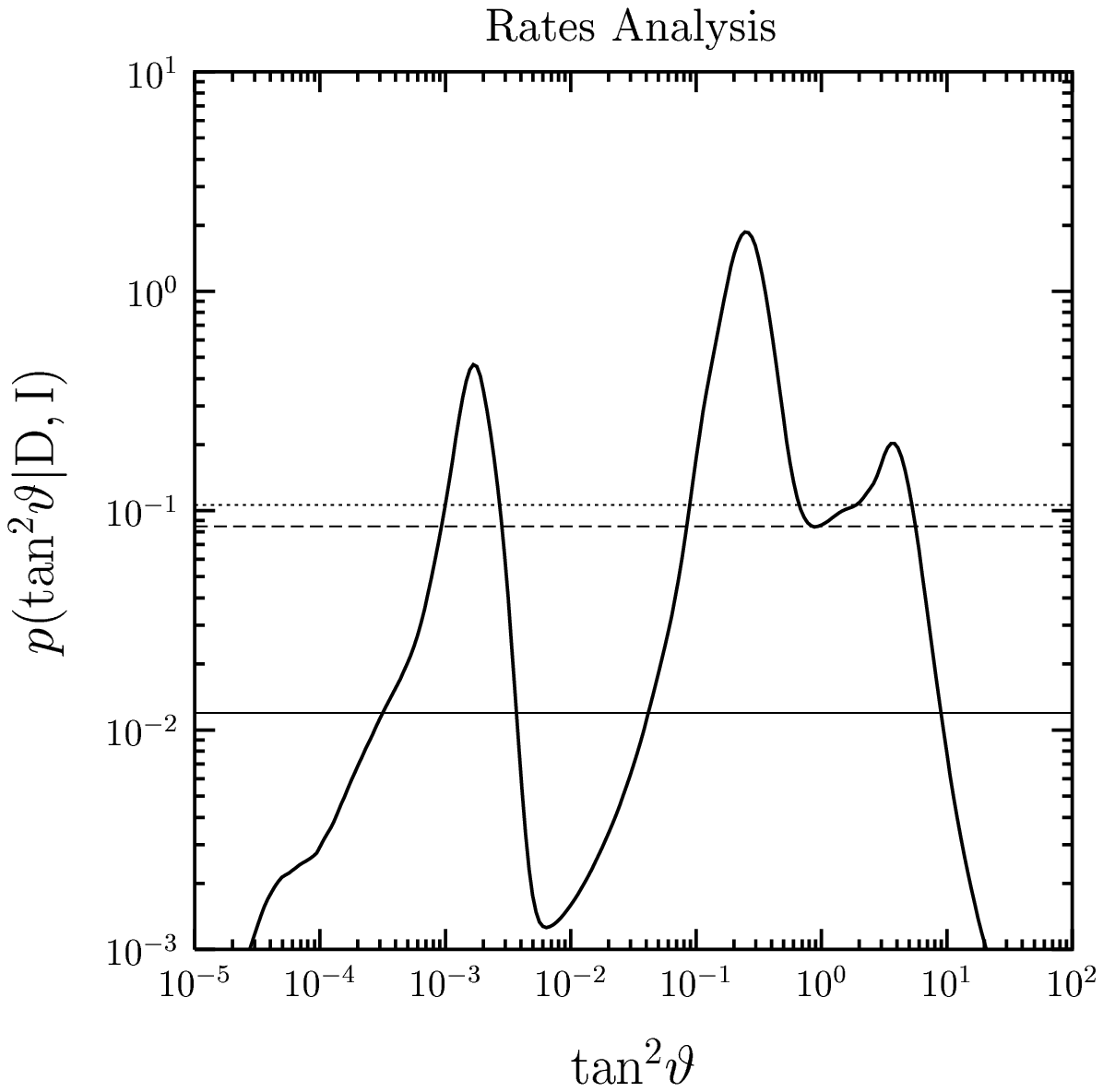}
\end{center}
\caption{ \label{misra_t2t}
Posterior distribution
of $\tan^2\!\vartheta$ obtained with the model independent Rates Analysis.
The intervals in which
the thick solid curve lies above the horizontal
dotted, dashed, and solid lines
have, respectively,
90\%,
95\% and
99\%
posterior probability to contain the true value of
$\tan^2\!\vartheta$.
}
\end{figure}

\clearpage

\begin{figure}
\begin{center}
\includegraphics[bb=90 427 431 767, width=0.8\textwidth]{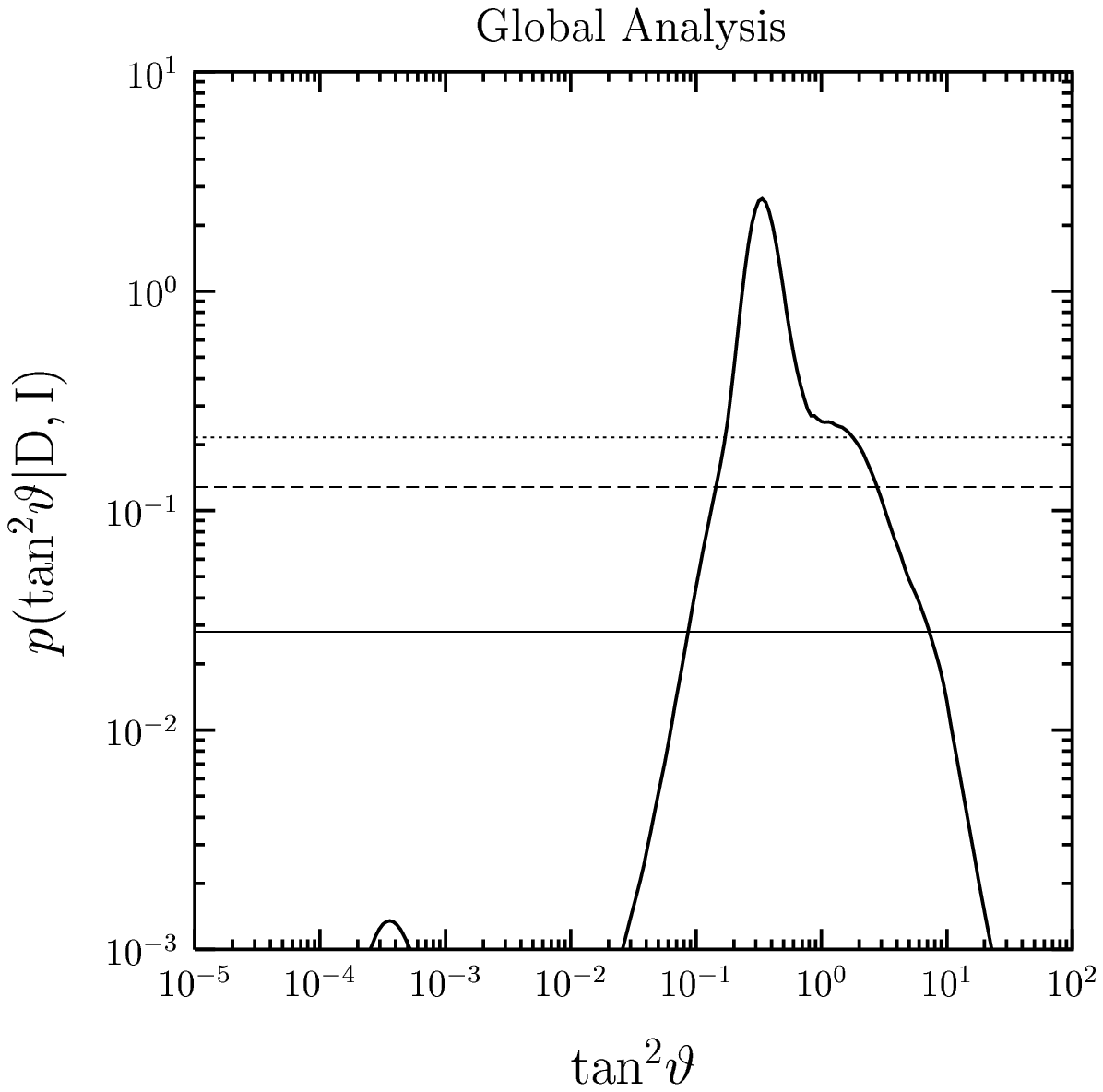}
\end{center}
\caption{ \label{misga_t2t}
Posterior distribution
of $\tan^2\!\vartheta$ obtained with the model independent Global Analysis.
The intervals in which
the thick solid curve lies above the horizontal
dotted, dashed, and solid lines
have, respectively,
90\%,
95\% and
99\%
posterior probability to contain the true value of
$\tan^2\!\vartheta$.
}
\end{figure}

\clearpage

\begin{figure}
\begin{center}
\includegraphics[bb=90 424 565 767, width=0.99\textwidth]{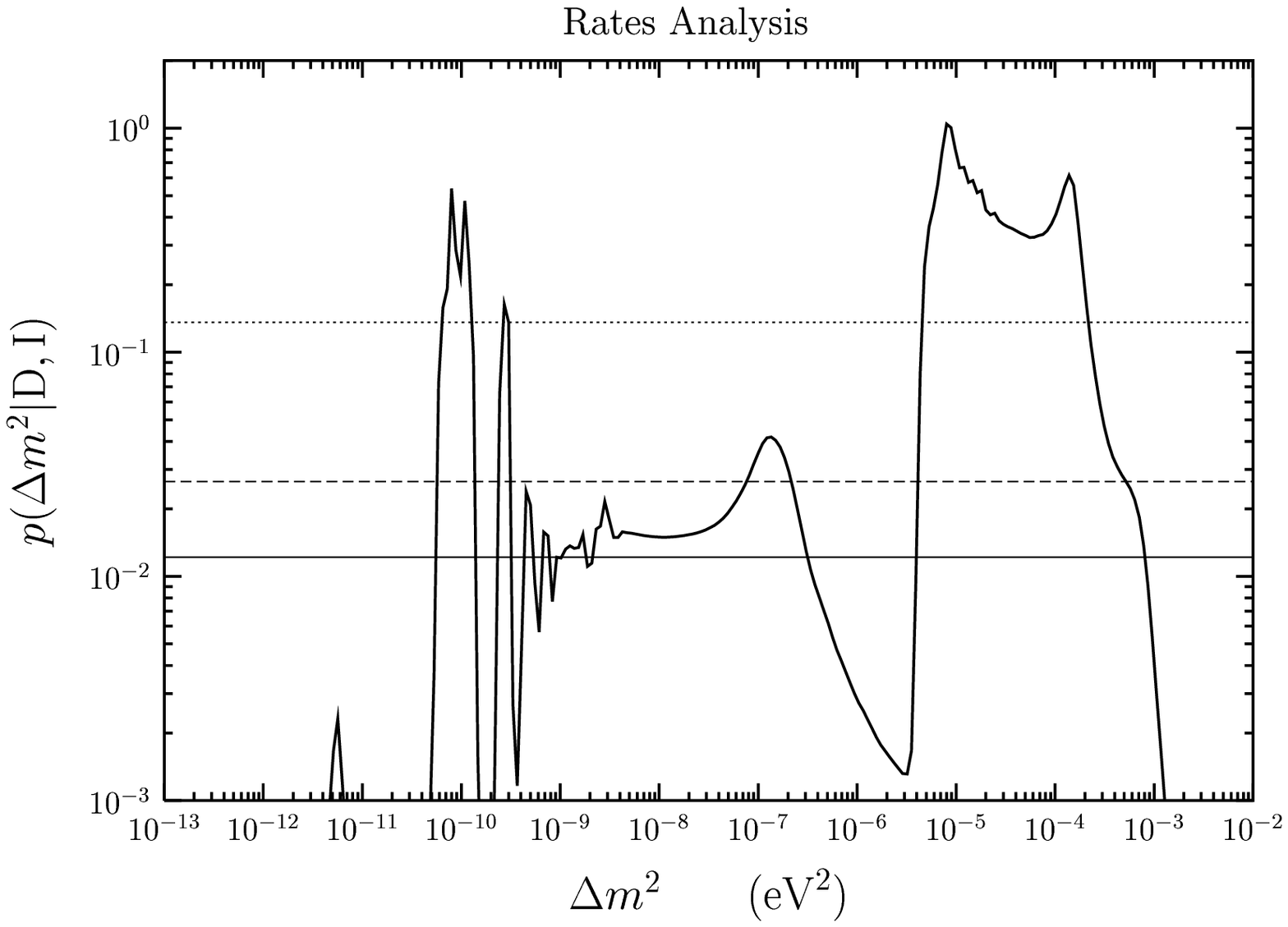}
\end{center}
\caption{ \label{misra_dm2}
Posterior distribution
of $\Delta{m}^2$ obtained with the model independent Rates Analysis.
The intervals in which
the thick solid curve lies above the horizontal
dotted, dashed, and solid lines
have, respectively,
90\%,
95\% and
99\%
posterior probability to contain the true value of
$\Delta{m}^2$.
}
\end{figure}

\clearpage

\begin{figure}
\begin{center}
\includegraphics[bb=90 424 565 767, width=0.99\textwidth]{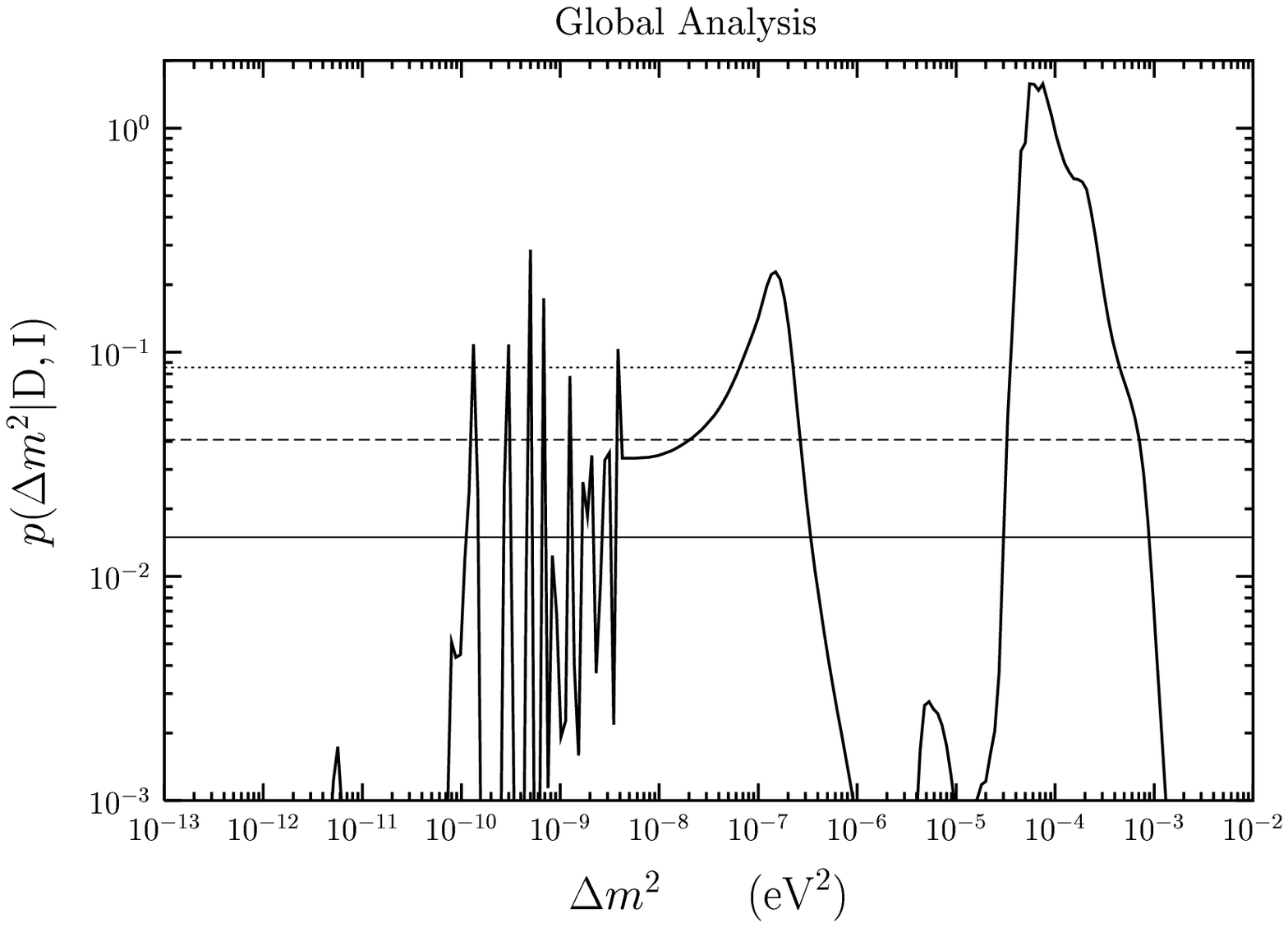}
\end{center}
\caption{ \label{misga_dm2}
Posterior distribution
of $\Delta{m}^2$ obtained with the model independent Global Analysis.
The intervals in which
the thick solid curve lies above the horizontal
dotted, dashed, and solid lines
have, respectively,
90\%,
95\% and
99\%
posterior probability to contain the true value of
$\Delta{m}^2$.
}
\end{figure}

\clearpage

\begin{figure}
\begin{center}
\includegraphics[bb=79 423 400 767, width=0.8\textwidth]{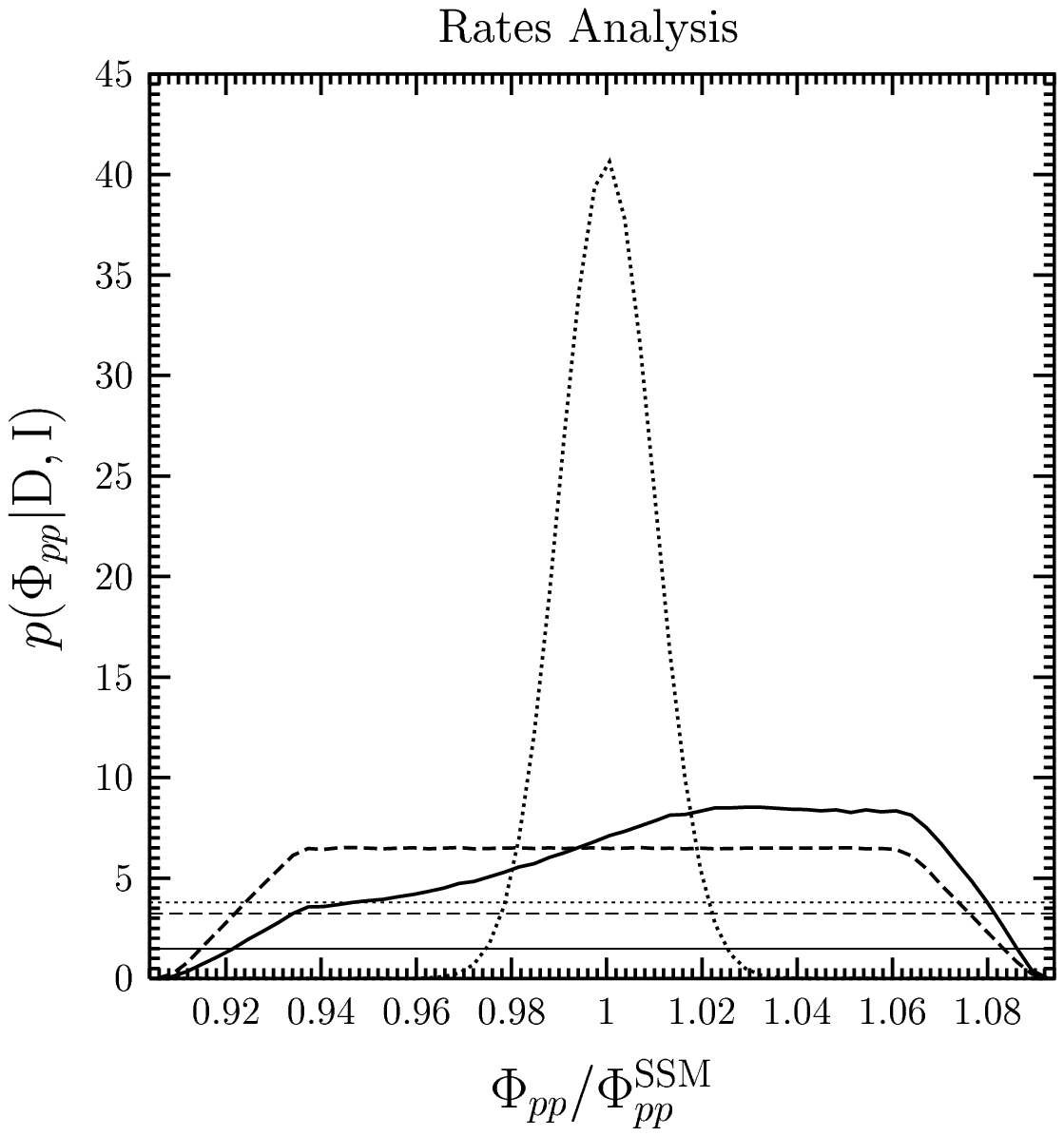}
\end{center}
\caption{ \label{misra_f1}
The thick solid line is the
posterior distribution
of the $pp$ flux obtained with the model independent Rates Analysis.
The intervals in which
the thick solid line lies above the horizontal thin
dotted, dashed, and solid lines
have, respectively,
90\%,
95\% and
99\%
posterior probability to contain the true value of
the $pp$ flux.
The gaussian thick dotted curve
is the BP2000 SSM distribution
and the thick dashed curve is the prior distribution
in the model independent analysis.
}
\end{figure}

\clearpage

\begin{figure}
\begin{center}
\includegraphics[bb=79 423 400 767, width=0.8\textwidth]{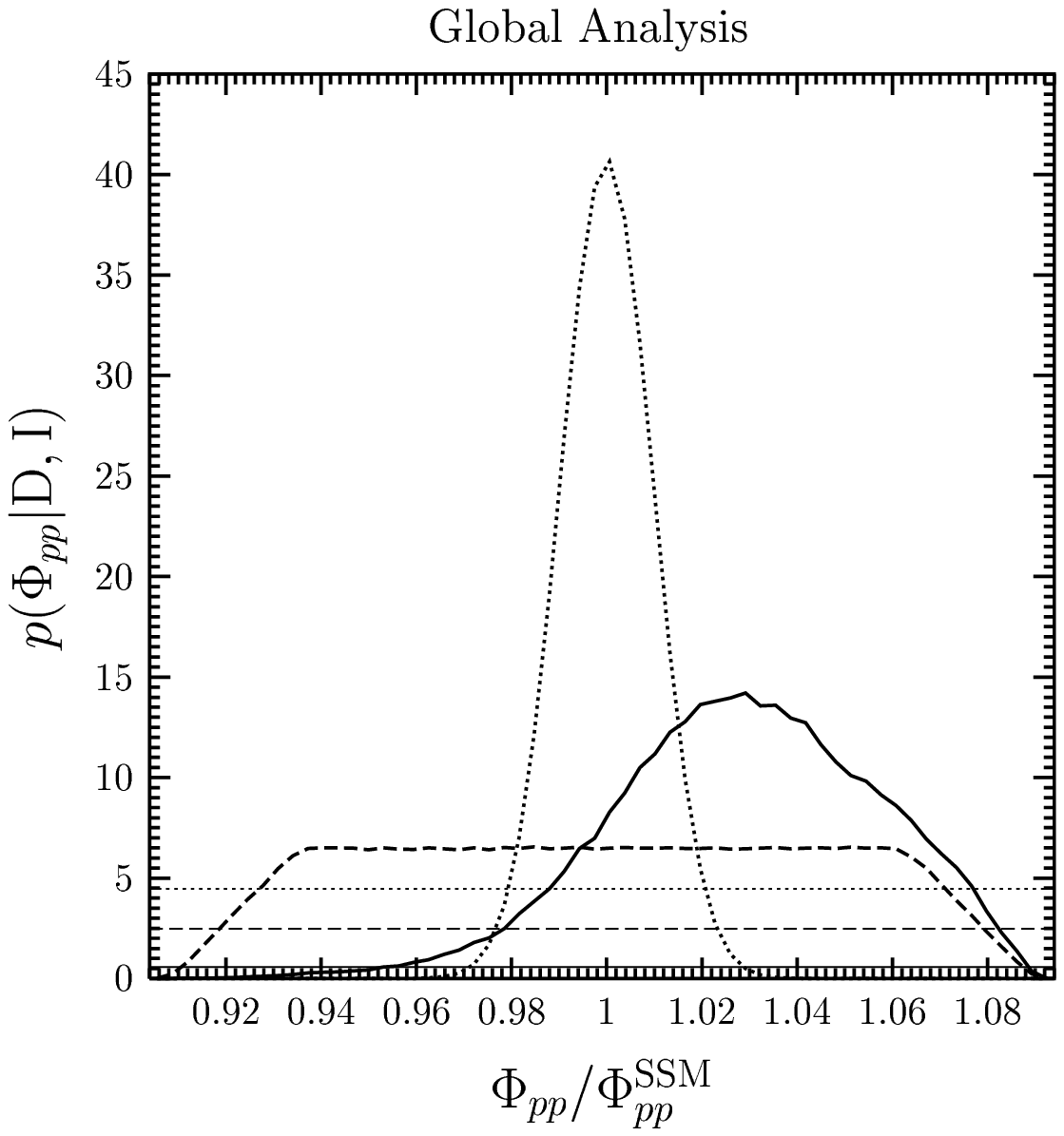}
\end{center}
\caption{ \label{misga_f1}
The thick solid line is the
posterior distribution
of the $pp$ flux obtained with the model independent Global Analysis.
The intervals in which
the thick solid line lies above the horizontal thin
dotted, dashed, and solid lines
have, respectively,
90\%,
95\% and
99\%
posterior probability to contain the true value of
the $pp$ flux.
The gaussian thick dotted curve
is the BP2000 SSM distribution
and the thick dashed curve is the prior distribution
in the model independent analysis.
}
\end{figure}

\clearpage

\begin{figure}
\begin{center}
\includegraphics[bb=79 423 400 767, width=0.8\textwidth]{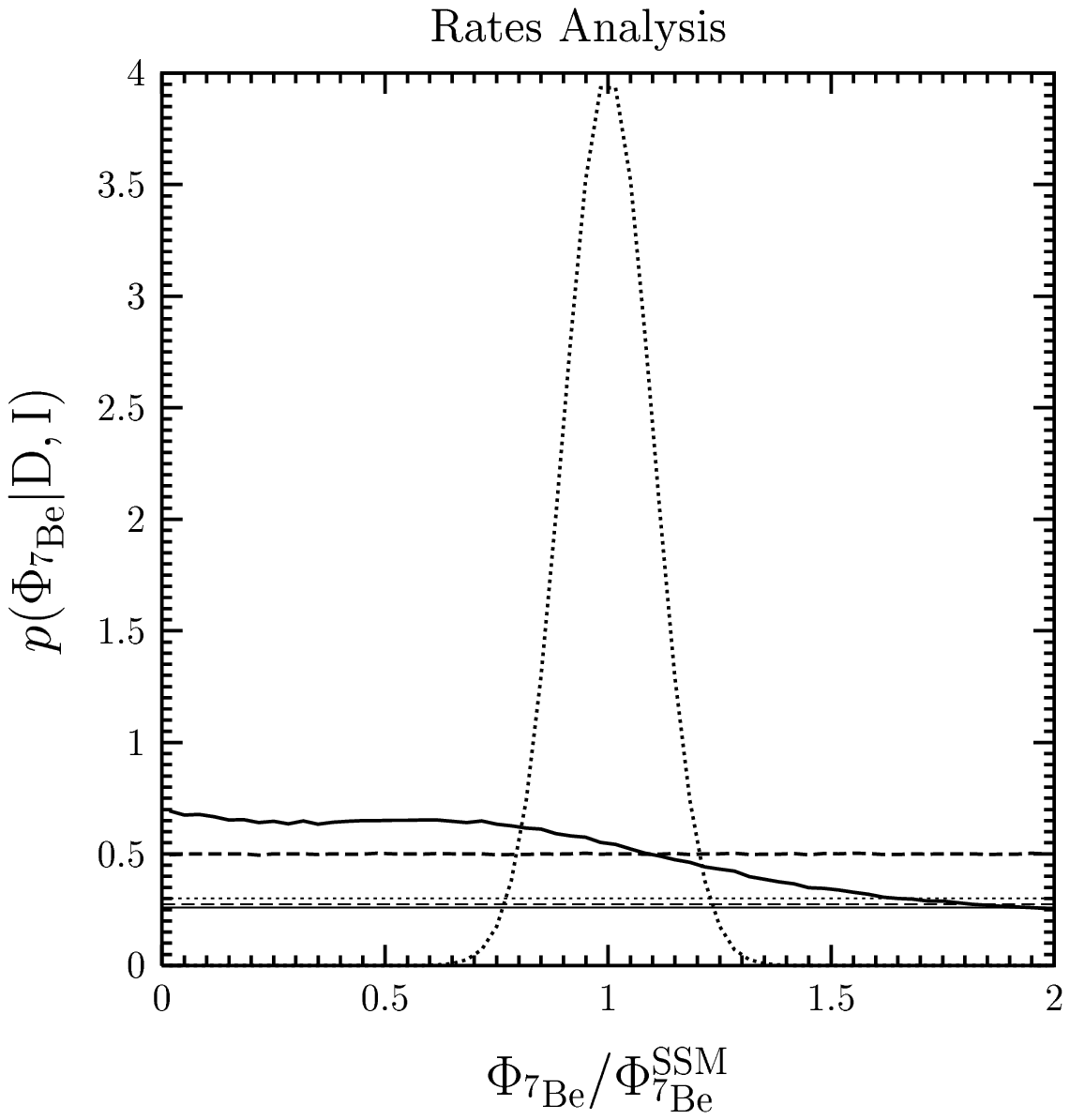}
\end{center}
\caption{ \label{misra_f4}
The thick solid line is the
posterior distribution
of the ${^7\mathrm{Be}}$ flux obtained with the model independent Rates Analysis.
The intervals in which
the thick solid line lies above the horizontal thin
dotted, dashed, and solid lines
have, respectively,
90\%,
95\% and
99\%
posterior probability to contain the true value of
the $pp$ flux.
The gaussian thick dotted curve
is the BP2000 SSM distribution
and the thick dashed curve is the prior distribution
in the model independent analysis.
}
\end{figure}

\clearpage

\begin{figure}
\begin{center}
\includegraphics[bb=79 423 400 767, width=0.8\textwidth]{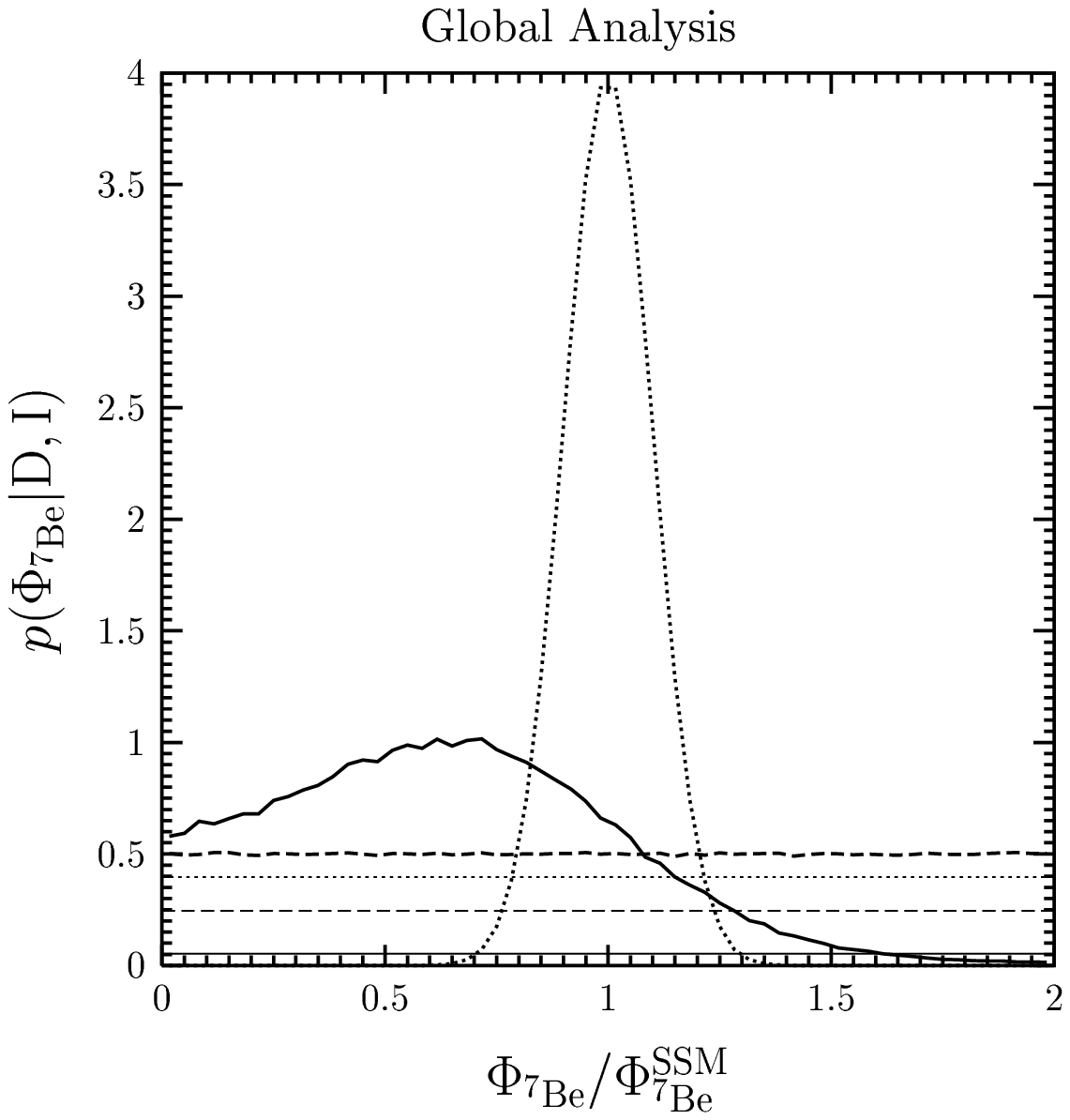}
\end{center}
\caption{ \label{misga_f4}
The thick solid line is the
posterior distribution
of the ${^7\mathrm{Be}}$ flux obtained with the model independent Global Analysis.
The intervals in which
the thick solid line lies above the horizontal thin
dotted, dashed, and solid lines
have, respectively,
90\%,
95\% and
99\%
posterior probability to contain the true value of
the $pp$ flux.
The gaussian thick dotted curve
is the BP2000 SSM distribution
and the thick dashed curve is the prior distribution
in the model independent analysis.
}
\end{figure}

\clearpage

\begin{figure}
\begin{center}
\includegraphics[bb=79 423 400 767, width=0.8\textwidth]{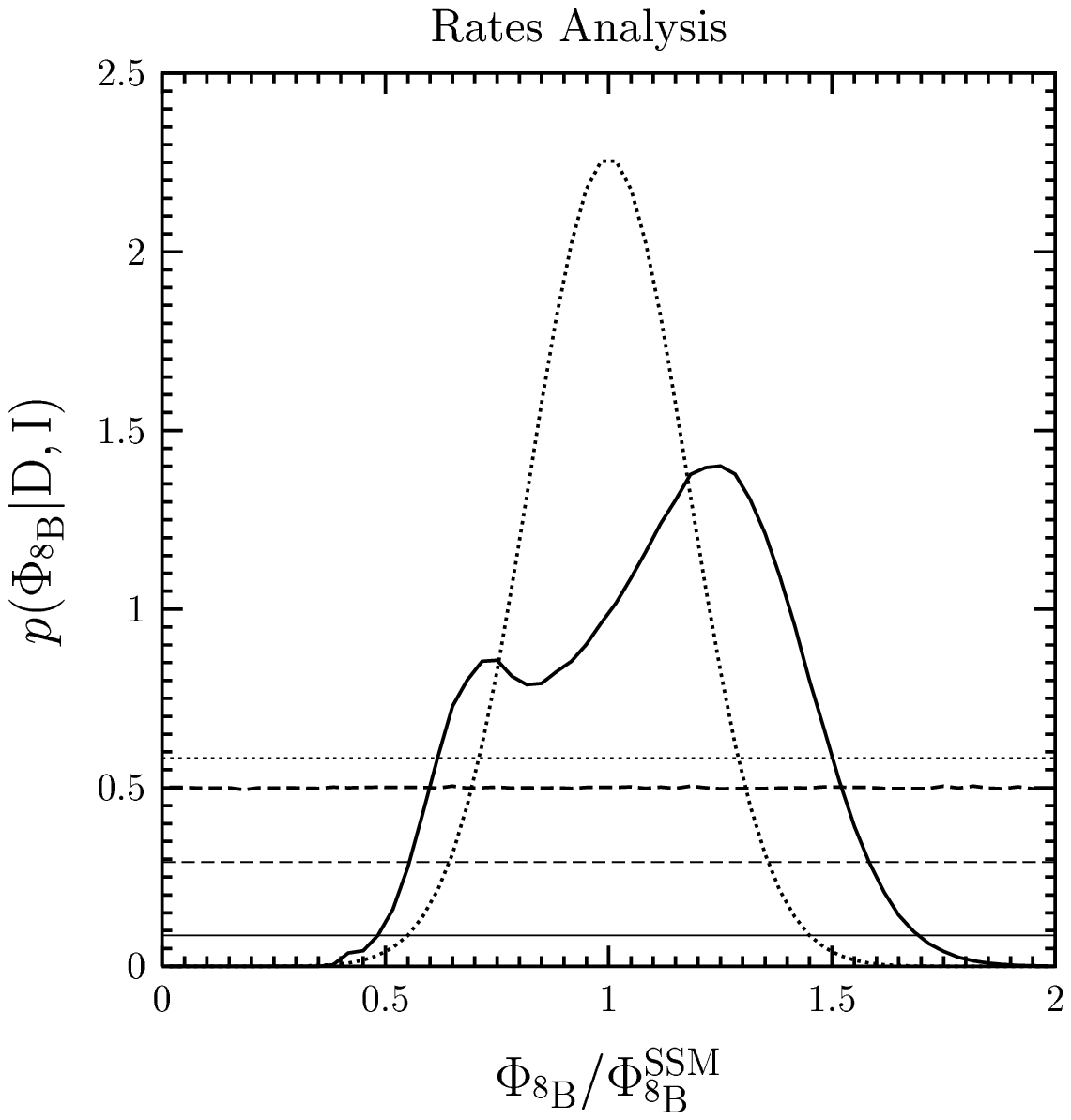}
\end{center}
\caption{ \label{misra_f5}
The thick solid line is the
posterior distribution
of the ${^8\mathrm{B}}$ flux obtained with the model independent Rates Analysis.
The intervals in which
the thick solid line lies above the horizontal thin
dotted, dashed, and solid lines
have, respectively,
90\%,
95\% and
99\%
posterior probability to contain the true value of
the $pp$ flux.
The gaussian thick dotted curve
is the BP2000 SSM distribution
and the thick dashed curve is the prior distribution
in the model independent analysis.
}
\end{figure}

\clearpage

\begin{figure}
\begin{center}
\includegraphics[bb=79 423 400 767, width=0.8\textwidth]{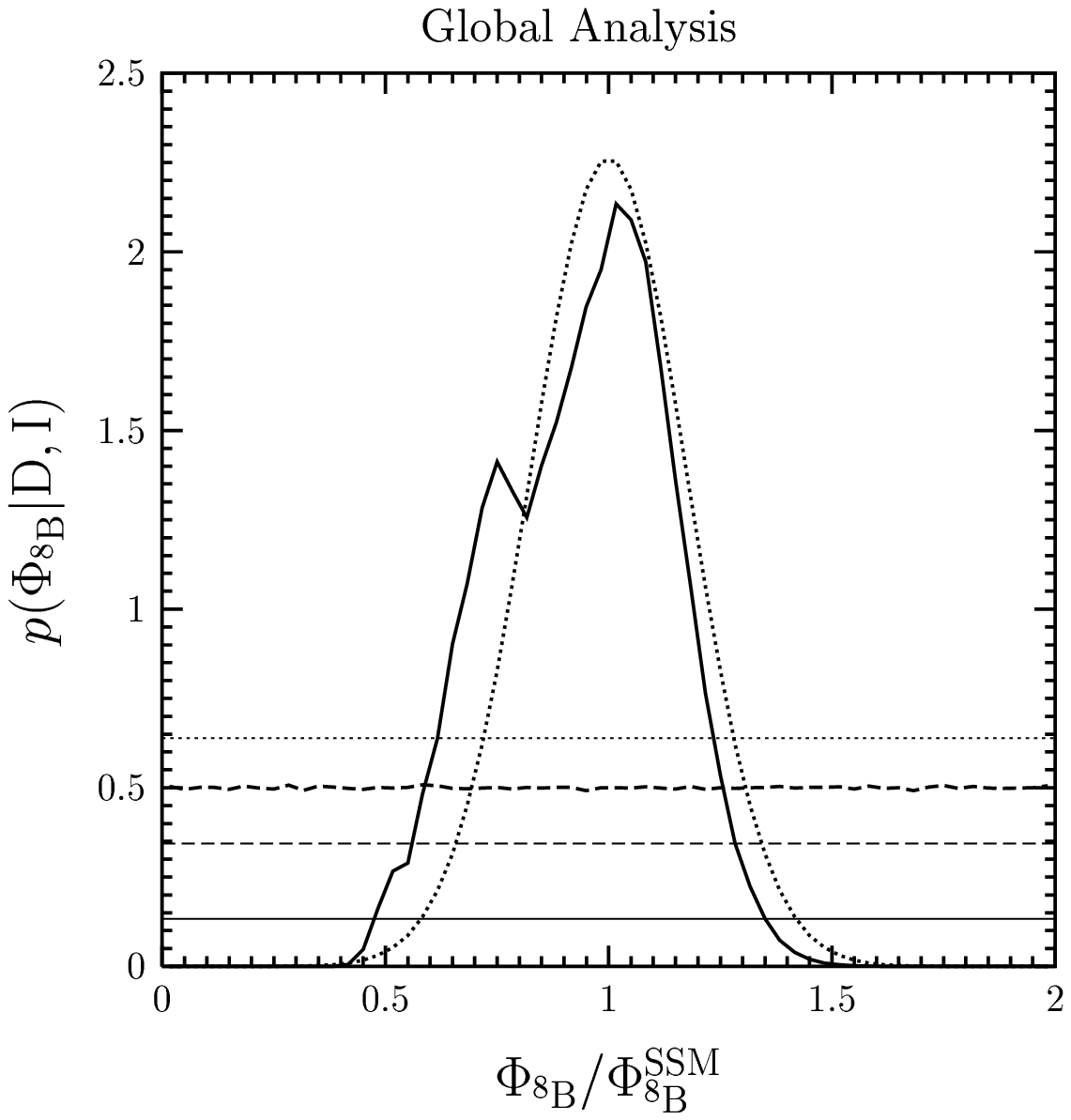}
\end{center}
\caption{ \label{misga_f5}
The thick solid line is the
posterior distribution
of the ${^8\mathrm{B}}$ flux
obtained with the model independent Global Analysis.
The intervals in which
the thick solid line lies above the horizontal thin
dotted, dashed, and solid lines
have, respectively,
90\%,
95\% and
99\%
posterior probability to contain the true value of
the $pp$ flux.
The gaussian thick dotted curve
is the BP2000 SSM distribution
and the thick dashed curve is the prior distribution
in the model independent analysis.
}
\end{figure}

\clearpage

%% file: fig/misra_fit_fig.tab
(A).
Solid line: $\tan^2\!\vartheta = 1.7 \times 10^{-3}$ and $\Delta{m}^2 = 8.0 \times 10^{-6} \, \mathrm{eV}^2$,
best fit point in the SMA region in the Rates Analysis;
long-dashed line: $\tan^2\!\vartheta = 0.20$ and $\Delta{m}^2 = 1.2 \times 10^{-5} \, \mathrm{eV}^2$,
best fit point in the LMA region in the Rates Analysis;
short-dashed line: $\tan^2\!\vartheta = 0.49$ and $\Delta{m}^2 = 1.4 \times 10^{-7} \, \mathrm{eV}^2$,
best fit point in the LOW region in the Rates Analysis;
dotted line: $\tan^2\!\vartheta = 3.59$ and $\Delta{m}^2 = 8.0 \times 10^{-11} \, \mathrm{eV}^2$,
best fit point in the VO region in the Rates Analysis.
(B).
Solid line: $\tan^2\!\vartheta = 1.0 \times 10^{-2}$ and $\Delta{m}^2 = 3.0 \times 10^{-5} \, \mathrm{eV}^2$;
long-dashed line: $\tan^2\!\vartheta = 5.0 \times 10^{-2}$ and $\Delta{m}^2 = 1.0 \times 10^{-6} \, \mathrm{eV}^2$;
short-dashed line: $\tan^2\!\vartheta = 1.00$ and $\Delta{m}^2 = 9.0 \times 10^{-11} \, \mathrm{eV}^2$;
dotted line: $\tan^2\!\vartheta = 1.00$ and $\Delta{m}^2 = 5.5 \times 10^{-12} \, \mathrm{eV}^2$;

%% file: fig/misga_fit_fig.tab
(A).
Solid line: $\tan^2\!\vartheta = 1.7 \times 10^{-3}$ and $\Delta{m}^2 = 8.0 \times 10^{-6} \, \mathrm{eV}^2$,
best fit point in the SMA region in the Rates Analysis;
long-dashed line: $\tan^2\!\vartheta = 0.20$ and $\Delta{m}^2 = 1.2 \times 10^{-5} \, \mathrm{eV}^2$,
best fit point in the LMA region in the Rates Analysis;
short-dashed line: $\tan^2\!\vartheta = 0.49$ and $\Delta{m}^2 = 1.4 \times 10^{-7} \, \mathrm{eV}^2$,
best fit point in the LOW region in the Rates Analysis;
dotted line: $\tan^2\!\vartheta = 3.59$ and $\Delta{m}^2 = 8.0 \times 10^{-11} \, \mathrm{eV}^2$,
best fit point in the VO region in the Rates Analysis.
(B).
Solid line: $\tan^2\!\vartheta = 3.8 \times 10^{-4}$ and $\Delta{m}^2 = 5.3 \times 10^{-6} \, \mathrm{eV}^2$,
best fit point in the SMA region in the Global Analysis;
long-dashed line: $\tan^2\!\vartheta = 0.34$ and $\Delta{m}^2 = 6.1 \times 10^{-5} \, \mathrm{eV}^2$,
best fit point in the LMA region in the Global Analysis;
short-dashed line: $\tan^2\!\vartheta = 0.60$ and $\Delta{m}^2 = 1.5 \times 10^{-7} \, \mathrm{eV}^2$,
best fit point in the LOW region in the Global Analysis;
dotted line: $\tan^2\!\vartheta = 3.16$ and $\Delta{m}^2 = 5.0 \times 10^{-10} \, \mathrm{eV}^2$,
best fit point in the VO region in the Global Analysis;

%% file: mis.bbl
\begin{thebibliography}{86}
\expandafter\ifx\csname natexlab\endcsname\relax\def\natexlab#1{#1}\fi
\expandafter\ifx\csname bibnamefont\endcsname\relax
  \def\bibnamefont#1{#1}\fi
\expandafter\ifx\csname bibfnamefont\endcsname\relax
  \def\bibfnamefont#1{#1}\fi
\expandafter\ifx\csname citenamefont\endcsname\relax
  \def\citenamefont#1{#1}\fi
\expandafter\ifx\csname url\endcsname\relax
  \def\url#1{\texttt{#1}}\fi
\expandafter\ifx\csname urlprefix\endcsname\relax\def\urlprefix{URL }\fi
\providecommand{\bibinfo}[2]{#2}
\providecommand{\eprint}[2][]{\url{#2}}

\bibitem[{\citenamefont{Fukuda et~al.}(2001{\natexlab{a}})}]{SK-sun-01}
\bibinfo{author}{\bibfnamefont{S.}~\bibnamefont{Fukuda}} \bibnamefont{et~al.}
  (\bibinfo{collaboration}{SuperKamiokande}), \bibinfo{journal}{Phys. Rev.
  Lett.} \textbf{\bibinfo{volume}{86}}, \bibinfo{pages}{5651}
  (\bibinfo{year}{2001}{\natexlab{a}}), \eprint{hep-ex/0103032}.

\bibitem[{\citenamefont{Ahmad et~al.}(2001)}]{SNO-01}
\bibinfo{author}{\bibfnamefont{Q.~R.} \bibnamefont{Ahmad}} \bibnamefont{et~al.}
  (\bibinfo{collaboration}{SNO}), \bibinfo{journal}{Phys. Rev. Lett.}
  \textbf{\bibinfo{volume}{87}}, \bibinfo{pages}{071301}
  (\bibinfo{year}{2001}), \eprint{nucl-ex/0106015}.

\bibitem[{\citenamefont{Fogli et~al.}(2001)\citenamefont{Fogli, Lisi,
  Montanino, and Palazzo}}]{Fogli:2001vr}
\bibinfo{author}{\bibfnamefont{G.~L.} \bibnamefont{Fogli}},
  \bibinfo{author}{\bibfnamefont{E.}~\bibnamefont{Lisi}},
  \bibinfo{author}{\bibfnamefont{D.}~\bibnamefont{Montanino}},
  \bibnamefont{and} \bibinfo{author}{\bibfnamefont{A.}~\bibnamefont{Palazzo}},
  \bibinfo{journal}{Phys. Rev.} \textbf{\bibinfo{volume}{D64}},
  \bibinfo{pages}{093007} (\bibinfo{year}{2001}), \eprint{hep-ph/0106247}.

\bibitem[{\citenamefont{Giunti}(2001)}]{Giunti-aoe-01}
\bibinfo{author}{\bibfnamefont{C.}~\bibnamefont{Giunti}}
  (\bibinfo{year}{2001}), \eprint{hep-ph/0107310}.

\bibitem[{\citenamefont{Fiorentini et~al.}(2001)\citenamefont{Fiorentini,
  Villante, and Ricci}}]{Fiorentini:2001jt}
\bibinfo{author}{\bibfnamefont{G.}~\bibnamefont{Fiorentini}},
  \bibinfo{author}{\bibfnamefont{F.~L.} \bibnamefont{Villante}},
  \bibnamefont{and} \bibinfo{author}{\bibfnamefont{B.}~\bibnamefont{Ricci}}
  (\bibinfo{year}{2001}), \eprint{hep-ph/0109275}.

\bibitem[{\citenamefont{Bahcall}(2001{\natexlab{a}})}]{Bahcall:2001pe}
\bibinfo{author}{\bibfnamefont{J.~N.} \bibnamefont{Bahcall}}
  (\bibinfo{year}{2001}{\natexlab{a}}),
  \eprint[http://arXiv.org/abs]{hep-ph/0108147}.

\bibitem[{\citenamefont{Cleveland et~al.}(1998)}]{Homestake-98}
\bibinfo{author}{\bibfnamefont{B.~T.} \bibnamefont{Cleveland}}
  \bibnamefont{et~al.}, \bibinfo{journal}{Astrophys. J.}
  \textbf{\bibinfo{volume}{496}}, \bibinfo{pages}{505} (\bibinfo{year}{1998}).

\bibitem[{\citenamefont{Hampel et~al.}(1999)}]{GALLEX-99}
\bibinfo{author}{\bibfnamefont{W.}~\bibnamefont{Hampel}} \bibnamefont{et~al.}
  (\bibinfo{collaboration}{GALLEX}), \bibinfo{journal}{Phys. Lett.}
  \textbf{\bibinfo{volume}{B447}}, \bibinfo{pages}{127} (\bibinfo{year}{1999}).

\bibitem[{\citenamefont{Gavrin}(2001)}]{SAGE-nu00}
\bibinfo{author}{\bibfnamefont{V.~N.} \bibnamefont{Gavrin}}
  (\bibinfo{collaboration}{SAGE}), \bibinfo{journal}{Nucl. Phys. Proc. Suppl.}
  \textbf{\bibinfo{volume}{91}}, \bibinfo{pages}{36} (\bibinfo{year}{2001}).

\bibitem[{\citenamefont{Altmann et~al.}(2000)}]{GNO-00}
\bibinfo{author}{\bibfnamefont{M.}~\bibnamefont{Altmann}} \bibnamefont{et~al.}
  (\bibinfo{collaboration}{GNO}), \bibinfo{journal}{Phys. Lett.}
  \textbf{\bibinfo{volume}{B490}}, \bibinfo{pages}{16} (\bibinfo{year}{2000}),
  \eprint{hep-ex/0006034}.

\bibitem[{\citenamefont{Bahcall
  et~al.}(2001{\natexlab{a}})\citenamefont{Bahcall, Pinsonneault, and
  Basu}}]{BP2000}
\bibinfo{author}{\bibfnamefont{J.~N.} \bibnamefont{Bahcall}},
  \bibinfo{author}{\bibfnamefont{M.~H.} \bibnamefont{Pinsonneault}},
  \bibnamefont{and} \bibinfo{author}{\bibfnamefont{S.}~\bibnamefont{Basu}},
  \bibinfo{journal}{Astrophys. J.} \textbf{\bibinfo{volume}{555}},
  \bibinfo{pages}{990} (\bibinfo{year}{2001}{\natexlab{a}}),
  \eprint{astro-ph/0010346}.

\bibitem[{\citenamefont{Bilenky and
  Pontecorvo}(1978)}]{Bilenky-Pontecorvo-PR-78}
\bibinfo{author}{\bibfnamefont{S.~M.} \bibnamefont{Bilenky}} \bibnamefont{and}
  \bibinfo{author}{\bibfnamefont{B.}~\bibnamefont{Pontecorvo}},
  \bibinfo{journal}{Phys. Rept.} \textbf{\bibinfo{volume}{41}},
  \bibinfo{pages}{225} (\bibinfo{year}{1978}).

\bibitem[{\citenamefont{Bilenky and Petcov}(1987)}]{Bilenky-Petcov-RMP-87}
\bibinfo{author}{\bibfnamefont{S.~M.} \bibnamefont{Bilenky}} \bibnamefont{and}
  \bibinfo{author}{\bibfnamefont{S.~T.} \bibnamefont{Petcov}},
  \bibinfo{journal}{Rev. Mod. Phys.} \textbf{\bibinfo{volume}{59}},
  \bibinfo{pages}{671} (\bibinfo{year}{1987}).

\bibitem[{\citenamefont{Kim and Pevsner}(1993)}]{CWKim-book-93}
\bibinfo{author}{\bibfnamefont{C.~W.} \bibnamefont{Kim}} \bibnamefont{and}
  \bibinfo{author}{\bibfnamefont{A.}~\bibnamefont{Pevsner}},
  \emph{\bibinfo{title}{Neutrinos in physics and astrophysics}}
  (\bibinfo{publisher}{Harwood Academic Press}, \bibinfo{address}{Chur,
  Switzerland}, \bibinfo{year}{1993}), \bibinfo{note}{{Contemporary Concepts in
  Physics, Vol. 8}}.

\bibitem[{\citenamefont{Bilenky et~al.}(1999)\citenamefont{Bilenky, Giunti, and
  Grimus}}]{BGG-review-98}
\bibinfo{author}{\bibfnamefont{S.~M.} \bibnamefont{Bilenky}},
  \bibinfo{author}{\bibfnamefont{C.}~\bibnamefont{Giunti}}, \bibnamefont{and}
  \bibinfo{author}{\bibfnamefont{W.}~\bibnamefont{Grimus}},
  \bibinfo{journal}{Prog. Part. Nucl. Phys.} \textbf{\bibinfo{volume}{43}},
  \bibinfo{pages}{1} (\bibinfo{year}{1999}), \eprint{hep-ph/9812360}.

\bibitem[{\citenamefont{Barger et~al.}(2001)\citenamefont{Barger, Marfatia, and
  Whisnant}}]{Barger:2001zs}
\bibinfo{author}{\bibfnamefont{V.}~\bibnamefont{Barger}},
  \bibinfo{author}{\bibfnamefont{D.}~\bibnamefont{Marfatia}}, \bibnamefont{and}
  \bibinfo{author}{\bibfnamefont{K.}~\bibnamefont{Whisnant}}
  (\bibinfo{year}{2001}), \eprint{hep-ph/0106207}.

\bibitem[{\citenamefont{Bahcall
  et~al.}(2001{\natexlab{b}})\citenamefont{Bahcall, Gonzalez-Garcia, and
  Pena-Garay}}]{Bahcall:2001zu}
\bibinfo{author}{\bibfnamefont{J.~N.} \bibnamefont{Bahcall}},
  \bibinfo{author}{\bibfnamefont{M.~C.} \bibnamefont{Gonzalez-Garcia}},
  \bibnamefont{and}
  \bibinfo{author}{\bibfnamefont{C.}~\bibnamefont{Pena-Garay}},
  \bibinfo{journal}{JHEP} \textbf{\bibinfo{volume}{08}}, \bibinfo{pages}{014}
  (\bibinfo{year}{2001}{\natexlab{b}}),
  \eprint[http://arXiv.org/abs]{hep-ph/0106258}.

\bibitem[{\citenamefont{Bandyopadhyay et~al.}(2001)\citenamefont{Bandyopadhyay,
  Choubey, Goswami, and Kar}}]{Bandyopadhyay:2001aa}
\bibinfo{author}{\bibfnamefont{A.}~\bibnamefont{Bandyopadhyay}},
  \bibinfo{author}{\bibfnamefont{S.}~\bibnamefont{Choubey}},
  \bibinfo{author}{\bibfnamefont{S.}~\bibnamefont{Goswami}}, \bibnamefont{and}
  \bibinfo{author}{\bibfnamefont{K.}~\bibnamefont{Kar}} (\bibinfo{year}{2001}),
  \eprint{hep-ph/0106264}.

\bibitem[{\citenamefont{Creminelli
  et~al.}(2001{\natexlab{a}})\citenamefont{Creminelli, Signorelli, and
  Strumia}}]{Creminelli:2001ij-hepph}
\bibinfo{author}{\bibfnamefont{P.}~\bibnamefont{Creminelli}},
  \bibinfo{author}{\bibfnamefont{G.}~\bibnamefont{Signorelli}},
  \bibnamefont{and} \bibinfo{author}{\bibfnamefont{A.}~\bibnamefont{Strumia}}
  (\bibinfo{year}{2001}{\natexlab{a}}),
  \eprint[http://arXiv.org/abs]{hep-ph/0102234}.

\bibitem[{\citenamefont{Barbieri and Strumia}(2000)}]{Barbieri:2000sv-hepph}
\bibinfo{author}{\bibfnamefont{R.}~\bibnamefont{Barbieri}} \bibnamefont{and}
  \bibinfo{author}{\bibfnamefont{A.}~\bibnamefont{Strumia}}
  (\bibinfo{year}{2000}), \eprint[http://arXiv.org/abs]{hep-ph/0011307}.

\bibitem[{\citenamefont{Kachelriess et~al.}(2001)\citenamefont{Kachelriess,
  Strumia, Tomas, and Valle}}]{Kachelriess:2001sg}
\bibinfo{author}{\bibfnamefont{M.}~\bibnamefont{Kachelriess}},
  \bibinfo{author}{\bibfnamefont{A.}~\bibnamefont{Strumia}},
  \bibinfo{author}{\bibfnamefont{R.}~\bibnamefont{Tomas}}, \bibnamefont{and}
  \bibinfo{author}{\bibfnamefont{J.~W.~F.} \bibnamefont{Valle}}
  (\bibinfo{year}{2001}), \eprint[http://arXiv.org/abs]{hep-ph/0108100}.

\bibitem[{\citenamefont{Berezinsky and Lissia}(2001)}]{Berezinsky-Lissia-01}
\bibinfo{author}{\bibfnamefont{V.}~\bibnamefont{Berezinsky}} \bibnamefont{and}
  \bibinfo{author}{\bibfnamefont{M.}~\bibnamefont{Lissia}}
  (\bibinfo{year}{2001}), \eprint{hep-ph/0108108}.

\bibitem[{\citenamefont{Berezinsky}(2001)}]{Berezinsky:2001se}
\bibinfo{author}{\bibfnamefont{V.}~\bibnamefont{Berezinsky}}
  (\bibinfo{year}{2001}), \eprint{hep-ph/0108166}.

\bibitem[{\citenamefont{Krastev and Smirnov}(2001)}]{Krastev:2001tv}
\bibinfo{author}{\bibfnamefont{P.~I.} \bibnamefont{Krastev}} \bibnamefont{and}
  \bibinfo{author}{\bibfnamefont{A.~Y.} \bibnamefont{Smirnov}}
  (\bibinfo{year}{2001}), \eprint{hep-ph/0108177}.

\bibitem[{\citenamefont{Garzelli and Giunti}(2001)}]{Garzelli:2001zu}
\bibinfo{author}{\bibfnamefont{M.~V.} \bibnamefont{Garzelli}} \bibnamefont{and}
  \bibinfo{author}{\bibfnamefont{C.}~\bibnamefont{Giunti}}
  (\bibinfo{year}{2001}), \eprint{hep-ph/0108191}.

\bibitem[{\citenamefont{Smy}(2001)}]{Smy:2001yn}
\bibinfo{author}{\bibfnamefont{M.~B.} \bibnamefont{Smy}}
  (\bibinfo{year}{2001}), \eprint{arXiv:hep-ex/0108053}.

\bibitem[{\citenamefont{Bilenky et~al.}(2001)\citenamefont{Bilenky,
  Lachenmaier, Potzel, and von Feilitzsch}}]{Bilenky:2001tf}
\bibinfo{author}{\bibfnamefont{S.~M.} \bibnamefont{Bilenky}},
  \bibinfo{author}{\bibfnamefont{T.}~\bibnamefont{Lachenmaier}},
  \bibinfo{author}{\bibfnamefont{W.}~\bibnamefont{Potzel}}, \bibnamefont{and}
  \bibinfo{author}{\bibfnamefont{F.}~\bibnamefont{von Feilitzsch}}
  (\bibinfo{year}{2001}), \eprint[http://arXiv.org/abs]{hep-ph/0109200}.

\bibitem[{\citenamefont{Strumia and Vissani}(2001)}]{Strumia:2001gi}
\bibinfo{author}{\bibfnamefont{A.}~\bibnamefont{Strumia}} \bibnamefont{and}
  \bibinfo{author}{\bibfnamefont{F.}~\bibnamefont{Vissani}}
  (\bibinfo{year}{2001}), \eprint[http://arXiv.org/abs]{hep-ph/0109172}.

\bibitem[{\citenamefont{Bahcall
  et~al.}(2001{\natexlab{c}})\citenamefont{Bahcall, Gonzalez-Garcia, and
  Pena-Garay}}]{Bahcall:2001cb}
\bibinfo{author}{\bibfnamefont{J.~N.} \bibnamefont{Bahcall}},
  \bibinfo{author}{\bibfnamefont{M.~C.} \bibnamefont{Gonzalez-Garcia}},
  \bibnamefont{and}
  \bibinfo{author}{\bibfnamefont{C.}~\bibnamefont{Pena-Garay}}
  (\bibinfo{year}{2001}{\natexlab{c}}),
  \eprint[http://arXiv.org/abs]{hep-ph/0111150}.

\bibitem[{\citenamefont{Bahcall}(1989)}]{Bahcall-book-89}
\bibinfo{author}{\bibfnamefont{J.~N.} \bibnamefont{Bahcall}},
  \emph{\bibinfo{title}{Neutrino Astrophysics}} (\bibinfo{publisher}{Cambridge
  University Press}, \bibinfo{address}{Cambridge, UK}, \bibinfo{year}{1989}).

\bibitem[{\citenamefont{Bilenky and Giunti}(1998)}]{Bilenky-Giunti-CHOOZ-98}
\bibinfo{author}{\bibfnamefont{S.~M.} \bibnamefont{Bilenky}} \bibnamefont{and}
  \bibinfo{author}{\bibfnamefont{C.}~\bibnamefont{Giunti}},
  \bibinfo{journal}{Phys. Lett.} \textbf{\bibinfo{volume}{B444}},
  \bibinfo{pages}{379} (\bibinfo{year}{1998}), \eprint{hep-ph/9802201}.

\bibitem[{\citenamefont{Fogli et~al.}(2000{\natexlab{a}})\citenamefont{Fogli,
  Lisi, Montanino, and Palazzo}}]{Fogli:1999zg}
\bibinfo{author}{\bibfnamefont{G.~L.} \bibnamefont{Fogli}},
  \bibinfo{author}{\bibfnamefont{E.}~\bibnamefont{Lisi}},
  \bibinfo{author}{\bibfnamefont{D.}~\bibnamefont{Montanino}},
  \bibnamefont{and} \bibinfo{author}{\bibfnamefont{A.}~\bibnamefont{Palazzo}},
  \bibinfo{journal}{Phys. Rev.} \textbf{\bibinfo{volume}{D62}},
  \bibinfo{pages}{013002} (\bibinfo{year}{2000}{\natexlab{a}}),
  \eprint{hep-ph/9912231}.

\bibitem[{\citenamefont{Apollonio et~al.}(1999)}]{Apollonio:1999ae}
\bibinfo{author}{\bibfnamefont{M.}~\bibnamefont{Apollonio}}
  \bibnamefont{et~al.} (\bibinfo{collaboration}{CHOOZ}),
  \bibinfo{journal}{Phys. Lett.} \textbf{\bibinfo{volume}{B466}},
  \bibinfo{pages}{415} (\bibinfo{year}{1999}),
  \eprint[http://arXiv.org/abs]{hep-ex/9907037}.

\bibitem[{\citenamefont{Barger et~al.}(2000)\citenamefont{Barger, Kayser,
  Learned, Weiler, and Whisnant}}]{Barger-Fate-2000}
\bibinfo{author}{\bibfnamefont{V.}~\bibnamefont{Barger}},
  \bibinfo{author}{\bibfnamefont{B.}~\bibnamefont{Kayser}},
  \bibinfo{author}{\bibfnamefont{J.}~\bibnamefont{Learned}},
  \bibinfo{author}{\bibfnamefont{T.}~\bibnamefont{Weiler}}, \bibnamefont{and}
  \bibinfo{author}{\bibfnamefont{K.}~\bibnamefont{Whisnant}},
  \bibinfo{journal}{Phys. Lett.} \textbf{\bibinfo{volume}{B489}},
  \bibinfo{pages}{345} (\bibinfo{year}{2000}), \eprint{hep-ph/0008019}.

\bibitem[{\citenamefont{Giunti and Laveder}(2001)}]{Giunti-Laveder-3+1-00}
\bibinfo{author}{\bibfnamefont{C.}~\bibnamefont{Giunti}} \bibnamefont{and}
  \bibinfo{author}{\bibfnamefont{M.}~\bibnamefont{Laveder}},
  \bibinfo{journal}{JHEP} \textbf{\bibinfo{volume}{02}}, \bibinfo{pages}{001}
  (\bibinfo{year}{2001}), \eprint{hep-ph/0010009}.

\bibitem[{\citenamefont{Bhat et~al.}(1998)\citenamefont{Bhat, Bhat, Paterno,
  and Prosper}}]{Bhat:1998qq}
\bibinfo{author}{\bibfnamefont{C.~M.} \bibnamefont{Bhat}},
  \bibinfo{author}{\bibfnamefont{P.~C.} \bibnamefont{Bhat}},
  \bibinfo{author}{\bibfnamefont{M.}~\bibnamefont{Paterno}}, \bibnamefont{and}
  \bibinfo{author}{\bibfnamefont{H.~B.} \bibnamefont{Prosper}},
  \bibinfo{journal}{Phys. Rev. Lett.} \textbf{\bibinfo{volume}{81}},
  \bibinfo{pages}{5056} (\bibinfo{year}{1998}),
  \eprint[http://arXiv.org/abs]{astro-ph/9804252}.

\bibitem[{\citenamefont{Jeffreys}(1961)}]{Jeffreys-book-39}
\bibinfo{author}{\bibfnamefont{H.}~\bibnamefont{Jeffreys}},
  \emph{\bibinfo{title}{Theory of Probability}} (\bibinfo{publisher}{Oxford
  University Press}, \bibinfo{address}{New York, USA}, \bibinfo{year}{1961}),
  \bibinfo{note}{first published in 1939}.

\bibitem[{\citenamefont{Loredo}(1990)}]{Loredo-90}
\bibinfo{author}{\bibfnamefont{T.~J.} \bibnamefont{Loredo}}
  (\bibinfo{year}{1990}), \bibinfo{note}{in Maximum-Entropy and Bayesian
  Methods, Dartmouth, 1989, ed. P. Fougere, Kluwer Academic Publishers,
  Dordrecht, The Netherlands, 1990, pp. 81--142,
  http://{\-}astrosun.{\-}tn.{\-}cornell.{\-}edu/{\-}staff/{\-}loredo/{\-}baye%
s/{\-}tjl.{\-}html}.

\bibitem[{\citenamefont{Loredo}(1992)}]{Loredo-92}
\bibinfo{author}{\bibfnamefont{T.~J.} \bibnamefont{Loredo}}
  (\bibinfo{year}{1992}), \bibinfo{note}{in Statistical Challenges in Modern
  Astronomy, ed. E.D. Feigelson and G.J. Babu, Springer-Verlag, New York, 1992,
  pp. 275--297,
  http://{\-}astrosun.{\-}tn.{\-}cornell.{\-}edu/{\-}staff/{\-}loredo/{\-}baye%
s/{\-}tjl.{\-}html}.

\bibitem[{\citenamefont{Jaynes}(June 1994)}]{Jaynes-book-95}
\bibinfo{author}{\bibfnamefont{E.~T.} \bibnamefont{Jaynes}},
  \emph{\bibinfo{title}{Probability Theory: The Logic of Science}}
  (\bibinfo{publisher}{Fragmentary Edition}, \bibinfo{year}{June 1994}),
  \bibinfo{note}{http://{\-}bayes.{\-}wustl.{\-}edu/{\-}etj/{\-}prob.{\-}html}.

\bibitem[{\citenamefont{D'Agostini}(1999)}]{D'Agostini-99}
\bibinfo{author}{\bibfnamefont{G.}~\bibnamefont{D'Agostini}},
  \bibinfo{journal}{CERN Yellow Report} \textbf{\bibinfo{volume}{99-03}}
  (\bibinfo{year}{1999}).

\bibitem[{\citenamefont{Bilenky and Giunti}(1994)}]{Bilenky:1994ti}
\bibinfo{author}{\bibfnamefont{S.~M.} \bibnamefont{Bilenky}} \bibnamefont{and}
  \bibinfo{author}{\bibfnamefont{C.}~\bibnamefont{Giunti}}
  (\bibinfo{year}{1994}), \eprint{hep-ph/9407379}.

\bibitem[{\citenamefont{Gribov and Pontecorvo}(1969)}]{Gribov:1969kq}
\bibinfo{author}{\bibfnamefont{V.~N.} \bibnamefont{Gribov}} \bibnamefont{and}
  \bibinfo{author}{\bibfnamefont{B.}~\bibnamefont{Pontecorvo}},
  \bibinfo{journal}{Phys. Lett.} \textbf{\bibinfo{volume}{B28}},
  \bibinfo{pages}{493} (\bibinfo{year}{1969}).

\bibitem[{\citenamefont{Wolfenstein}(1978)}]{Wolfenstein:1978ue}
\bibinfo{author}{\bibfnamefont{L.}~\bibnamefont{Wolfenstein}},
  \bibinfo{journal}{Phys. Rev.} \textbf{\bibinfo{volume}{D17}},
  \bibinfo{pages}{2369} (\bibinfo{year}{1978}).

\bibitem[{\citenamefont{Mikheev and Smirnov}(1985)}]{Mikheev:1985gs}
\bibinfo{author}{\bibfnamefont{S.~P.} \bibnamefont{Mikheev}} \bibnamefont{and}
  \bibinfo{author}{\bibfnamefont{A.~Y.} \bibnamefont{Smirnov}},
  \bibinfo{journal}{Sov. J. Nucl. Phys.} \textbf{\bibinfo{volume}{42}},
  \bibinfo{pages}{913} (\bibinfo{year}{1985}).

\bibitem[{\citenamefont{Creminelli
  et~al.}(2001{\natexlab{b}})\citenamefont{Creminelli, Signorelli, and
  Strumia}}]{Creminelli:2001ij}
\bibinfo{author}{\bibfnamefont{P.}~\bibnamefont{Creminelli}},
  \bibinfo{author}{\bibfnamefont{G.}~\bibnamefont{Signorelli}},
  \bibnamefont{and} \bibinfo{author}{\bibfnamefont{A.}~\bibnamefont{Strumia}},
  \bibinfo{journal}{JHEP} \textbf{\bibinfo{volume}{05}}, \bibinfo{pages}{052}
  (\bibinfo{year}{2001}{\natexlab{b}}),
  \eprint[http://arXiv.org/abs]{hep-ph/0102234}.

\bibitem[{\citenamefont{Apollonio et~al.}(1998)}]{Apollonio:1998xe}
\bibinfo{author}{\bibfnamefont{M.}~\bibnamefont{Apollonio}}
  \bibnamefont{et~al.} (\bibinfo{collaboration}{CHOOZ}),
  \bibinfo{journal}{Phys. Lett.} \textbf{\bibinfo{volume}{B420}},
  \bibinfo{pages}{397} (\bibinfo{year}{1998}),
  \eprint[http://arXiv.org/abs]{hep-ex/9711002}.

\bibitem[{\citenamefont{Bahcall}(2001{\natexlab{b}})}]{Bahcall-WWW}
\bibinfo{author}{\bibfnamefont{J.}~\bibnamefont{Bahcall}}
  (\bibinfo{year}{2001}{\natexlab{b}}), \bibinfo{note}{{WWW} page:
  http://{\-}www.{\-}sns.{\-}ias.{\-}edu/{\-}\~{}jnb}.

\bibitem[{\citenamefont{Nakamura et~al.}(2001)\citenamefont{Nakamura, Sato,
  Gudkov, and Kubodera}}]{Kubodera-deuteron-01}
\bibinfo{author}{\bibfnamefont{S.}~\bibnamefont{Nakamura}},
  \bibinfo{author}{\bibfnamefont{T.}~\bibnamefont{Sato}},
  \bibinfo{author}{\bibfnamefont{V.}~\bibnamefont{Gudkov}}, \bibnamefont{and}
  \bibinfo{author}{\bibfnamefont{K.}~\bibnamefont{Kubodera}},
  \bibinfo{journal}{Phys. Rev.} \textbf{\bibinfo{volume}{C63}},
  \bibinfo{pages}{034617} (\bibinfo{year}{2001}), \eprint{nucl-th/0009012}.

\bibitem[{\citenamefont{Kubodera}(2001)}]{Kubodera-www}
\bibinfo{author}{\bibfnamefont{K.}~\bibnamefont{Kubodera}}
  (\bibinfo{year}{2001}), \bibinfo{note}{{WWW} page:
  http://{\-}nuc003.{\-}psc.{\-}sc.{\-}edu/{\-}\~{}kubodera/{\-}NU-D-NSGK}.

\bibitem[{\citenamefont{Butler et~al.}(2001)\citenamefont{Butler, Chen, and
  Kong}}]{Butler:2000zp}
\bibinfo{author}{\bibfnamefont{M.}~\bibnamefont{Butler}},
  \bibinfo{author}{\bibfnamefont{J.-W.} \bibnamefont{Chen}}, \bibnamefont{and}
  \bibinfo{author}{\bibfnamefont{X.}~\bibnamefont{Kong}},
  \bibinfo{journal}{Phys. Rev.} \textbf{\bibinfo{volume}{C63}},
  \bibinfo{pages}{035501} (\bibinfo{year}{2001}),
  \eprint[http://arXiv.org/abs]{nucl-th/0008032}.

\bibitem[{\citenamefont{Beacom and Parke}(2001)}]{Beacom:2001hr}
\bibinfo{author}{\bibfnamefont{J.~F.} \bibnamefont{Beacom}} \bibnamefont{and}
  \bibinfo{author}{\bibfnamefont{S.~J.} \bibnamefont{Parke}},
  \bibinfo{journal}{Phys. Rev.} \textbf{\bibinfo{volume}{D64}},
  \bibinfo{pages}{091302} (\bibinfo{year}{2001}),
  \eprint[http://arXiv.org/abs]{hep-ph/0106128}.

\bibitem[{\citenamefont{Kurylov et~al.}(2001)\citenamefont{Kurylov,
  Ramsey-Musolf, and Vogel}}]{Kurylov:2001av}
\bibinfo{author}{\bibfnamefont{A.}~\bibnamefont{Kurylov}},
  \bibinfo{author}{\bibfnamefont{M.~J.} \bibnamefont{Ramsey-Musolf}},
  \bibnamefont{and} \bibinfo{author}{\bibfnamefont{P.}~\bibnamefont{Vogel}}
  (\bibinfo{year}{2001}), \eprint[http://arXiv.org/abs]{nucl-th/0110051}.

\bibitem[{\citenamefont{Poon}(2001)}]{Poon:2001ee}
\bibinfo{author}{\bibfnamefont{A.~W.~P.} \bibnamefont{Poon}}
  (\bibinfo{collaboration}{SNO}) (\bibinfo{year}{2001}),
  \eprint[http://arXiv.org/abs]{nucl-ex/0110005}.

\bibitem[{\citenamefont{Friedland}(2000)}]{Friedland-vo-00}
\bibinfo{author}{\bibfnamefont{A.}~\bibnamefont{Friedland}},
  \bibinfo{journal}{Phys. Rev. Lett.} \textbf{\bibinfo{volume}{85}},
  \bibinfo{pages}{936} (\bibinfo{year}{2000}), \eprint{hep-ph/0002063}.

\bibitem[{\citenamefont{Fogli et~al.}(2000{\natexlab{b}})\citenamefont{Fogli,
  Lisi, Montanino, and Palazzo}}]{Fogli-Lisi-Montanino-Palazzo-Quasi-vacuum-00}
\bibinfo{author}{\bibfnamefont{G.~L.} \bibnamefont{Fogli}},
  \bibinfo{author}{\bibfnamefont{E.}~\bibnamefont{Lisi}},
  \bibinfo{author}{\bibfnamefont{D.}~\bibnamefont{Montanino}},
  \bibnamefont{and} \bibinfo{author}{\bibfnamefont{A.}~\bibnamefont{Palazzo}},
  \bibinfo{journal}{Phys. Rev.} \textbf{\bibinfo{volume}{D62}},
  \bibinfo{pages}{113004} (\bibinfo{year}{2000}{\natexlab{b}}),
  \eprint{hep-ph/0005261}.

\bibitem[{\citenamefont{Lisi et~al.}(2001)\citenamefont{Lisi, Marrone,
  Montanino, Palazzo, and Petcov}}]{Lisi:2000su}
\bibinfo{author}{\bibfnamefont{E.}~\bibnamefont{Lisi}},
  \bibinfo{author}{\bibfnamefont{A.}~\bibnamefont{Marrone}},
  \bibinfo{author}{\bibfnamefont{D.}~\bibnamefont{Montanino}},
  \bibinfo{author}{\bibfnamefont{A.}~\bibnamefont{Palazzo}}, \bibnamefont{and}
  \bibinfo{author}{\bibfnamefont{S.~T.} \bibnamefont{Petcov}},
  \bibinfo{journal}{Phys. Rev.} \textbf{\bibinfo{volume}{D63}},
  \bibinfo{pages}{093002} (\bibinfo{year}{2001}), \eprint{hep-ph/0011306}.

\bibitem[{\citenamefont{Petcov}(1988{\natexlab{a}})}]{Petcov:1988wv}
\bibinfo{author}{\bibfnamefont{S.~T.} \bibnamefont{Petcov}},
  \bibinfo{journal}{Phys. Lett.} \textbf{\bibinfo{volume}{B214}},
  \bibinfo{pages}{139} (\bibinfo{year}{1988}{\natexlab{a}}).

\bibitem[{\citenamefont{Petcov and Rich}(1989)}]{Petcov:1989du}
\bibinfo{author}{\bibfnamefont{S.~T.} \bibnamefont{Petcov}} \bibnamefont{and}
  \bibinfo{author}{\bibfnamefont{J.}~\bibnamefont{Rich}},
  \bibinfo{journal}{Phys. Lett.} \textbf{\bibinfo{volume}{B224}},
  \bibinfo{pages}{426} (\bibinfo{year}{1989}).

\bibitem[{\citenamefont{Gonzalez-Garcia and
  Pena-Garay}(2000)}]{Gonzalez-Garcia:2000sk}
\bibinfo{author}{\bibfnamefont{M.~C.} \bibnamefont{Gonzalez-Garcia}}
  \bibnamefont{and}
  \bibinfo{author}{\bibfnamefont{C.}~\bibnamefont{Pena-Garay}},
  \bibinfo{journal}{Nucl. Phys. Proc. Suppl.} \textbf{\bibinfo{volume}{91}},
  \bibinfo{pages}{80} (\bibinfo{year}{2000}), \eprint{hep-ph/0009041}.

\bibitem[{\citenamefont{Petcov}(1988{\natexlab{b}})}]{Petcov-analytic-87}
\bibinfo{author}{\bibfnamefont{S.~T.} \bibnamefont{Petcov}},
  \bibinfo{journal}{Phys. Lett.} \textbf{\bibinfo{volume}{B200}},
  \bibinfo{pages}{373} (\bibinfo{year}{1988}{\natexlab{b}}).

\bibitem[{\citenamefont{Kuo and Pantaleone}(1989)}]{Kuo-Pantaleone-RMP-89}
\bibinfo{author}{\bibfnamefont{T.~K.} \bibnamefont{Kuo}} \bibnamefont{and}
  \bibinfo{author}{\bibfnamefont{J.}~\bibnamefont{Pantaleone}},
  \bibinfo{journal}{Rev. Mod. Phys.} \textbf{\bibinfo{volume}{61}},
  \bibinfo{pages}{937} (\bibinfo{year}{1989}).

\bibitem[{\citenamefont{Liu et~al.}(1997)\citenamefont{Liu, Maris, and
  Petcov}}]{Liu-Maris-Petcov-earth1-97}
\bibinfo{author}{\bibfnamefont{Q.~Y.} \bibnamefont{Liu}},
  \bibinfo{author}{\bibfnamefont{M.}~\bibnamefont{Maris}}, \bibnamefont{and}
  \bibinfo{author}{\bibfnamefont{S.~T.} \bibnamefont{Petcov}},
  \bibinfo{journal}{Phys. Rev.} \textbf{\bibinfo{volume}{D56}},
  \bibinfo{pages}{5991} (\bibinfo{year}{1997}), \eprint{hep-ph/9702361}.

\bibitem[{\citenamefont{Petcov}(1998)}]{Petcov-diffractive-98}
\bibinfo{author}{\bibfnamefont{S.~T.} \bibnamefont{Petcov}},
  \bibinfo{journal}{Phys. Lett.} \textbf{\bibinfo{volume}{B434}},
  \bibinfo{pages}{321} (\bibinfo{year}{1998}), \eprint{hep-ph/9805262}.

\bibitem[{\citenamefont{Akhmedov}(1999)}]{Akhmedov-parametric-99}
\bibinfo{author}{\bibfnamefont{E.~K.} \bibnamefont{Akhmedov}},
  \bibinfo{journal}{Nucl. Phys.} \textbf{\bibinfo{volume}{B538}},
  \bibinfo{pages}{25} (\bibinfo{year}{1999}), \eprint{hep-ph/9805272}.

\bibitem[{\citenamefont{Chizhov and Petcov}(1999)}]{Chizhov-Petcov-earth-1-99}
\bibinfo{author}{\bibfnamefont{M.~V.} \bibnamefont{Chizhov}} \bibnamefont{and}
  \bibinfo{author}{\bibfnamefont{S.~T.} \bibnamefont{Petcov}},
  \bibinfo{journal}{Phys. Rev. Lett.} \textbf{\bibinfo{volume}{83}},
  \bibinfo{pages}{1096} (\bibinfo{year}{1999}), \eprint{hep-ph/9903399}.

\bibitem[{\citenamefont{Chizhov and Petcov}(2001)}]{Chizhov-Petcov-earth-2-99}
\bibinfo{author}{\bibfnamefont{M.~V.} \bibnamefont{Chizhov}} \bibnamefont{and}
  \bibinfo{author}{\bibfnamefont{S.~T.} \bibnamefont{Petcov}},
  \bibinfo{journal}{Phys. Rev.} \textbf{\bibinfo{volume}{D63}},
  \bibinfo{pages}{073003} (\bibinfo{year}{2001}), \eprint{hep-ph/9903424}.

\bibitem[{\citenamefont{Fukuda
  et~al.}(2001{\natexlab{b}})}]{SK-sun-hep-ex-0103032}
\bibinfo{author}{\bibfnamefont{S.}~\bibnamefont{Fukuda}} \bibnamefont{et~al.}
  (\bibinfo{collaboration}{SuperKamiokande})
  (\bibinfo{year}{2001}{\natexlab{b}}), \eprint{hep-ex/0103032}.

\bibitem[{\citenamefont{Ortiz et~al.}(2000)\citenamefont{Ortiz, Garcia, Waltz,
  Bhattacharya, and Komives}}]{Ortiz:2000nf}
\bibinfo{author}{\bibfnamefont{C.~E.} \bibnamefont{Ortiz}},
  \bibinfo{author}{\bibfnamefont{A.}~\bibnamefont{Garcia}},
  \bibinfo{author}{\bibfnamefont{R.~A.} \bibnamefont{Waltz}},
  \bibinfo{author}{\bibfnamefont{M.}~\bibnamefont{Bhattacharya}},
  \bibnamefont{and} \bibinfo{author}{\bibfnamefont{A.~K.}
  \bibnamefont{Komives}}, \bibinfo{journal}{Phys. Rev. Lett.}
  \textbf{\bibinfo{volume}{85}}, \bibinfo{pages}{2909} (\bibinfo{year}{2000}),
  \eprint{nucl-ex/0003006}.

\bibitem[{\citenamefont{Bahcall et~al.}(1996)}]{Bahcall:1996qv}
\bibinfo{author}{\bibfnamefont{J.~N.} \bibnamefont{Bahcall}}
  \bibnamefont{et~al.}, \bibinfo{journal}{Phys. Rev.}
  \textbf{\bibinfo{volume}{C54}}, \bibinfo{pages}{411} (\bibinfo{year}{1996}),
  \eprint{nucl-th/9601044}.

\bibitem[{\citenamefont{Garzelli and Giunti}(2000)}]{Garzelli-Giunti-cs-00}
\bibinfo{author}{\bibfnamefont{M.~V.} \bibnamefont{Garzelli}} \bibnamefont{and}
  \bibinfo{author}{\bibfnamefont{C.}~\bibnamefont{Giunti}},
  \bibinfo{journal}{Phys. Lett.} \textbf{\bibinfo{volume}{B488}},
  \bibinfo{pages}{339} (\bibinfo{year}{2000}), \eprint{hep-ph/0006026}.

\bibitem[{\citenamefont{Fogli and Lisi}(1995)}]{Fogli-Lisi-correlations-95}
\bibinfo{author}{\bibfnamefont{G.~L.} \bibnamefont{Fogli}} \bibnamefont{and}
  \bibinfo{author}{\bibfnamefont{E.}~\bibnamefont{Lisi}},
  \bibinfo{journal}{Astropart. Phys.} \textbf{\bibinfo{volume}{3}},
  \bibinfo{pages}{185} (\bibinfo{year}{1995}).

\bibitem[{\citenamefont{Bahcall}(2000)}]{Bahcall:2000ue}
\bibinfo{author}{\bibfnamefont{J.~N.} \bibnamefont{Bahcall}},
  \bibinfo{journal}{Nucl. Phys. Proc. Suppl.} \textbf{\bibinfo{volume}{91}},
  \bibinfo{pages}{9} (\bibinfo{year}{2000}),
  \eprint[http://arXiv.org/abs]{hep-ph/0009044}.

\bibitem[{\citenamefont{Bahcall and Krastev}(1998)}]{Bahcall:1998se}
\bibinfo{author}{\bibfnamefont{J.~N.} \bibnamefont{Bahcall}} \bibnamefont{and}
  \bibinfo{author}{\bibfnamefont{P.~I.} \bibnamefont{Krastev}},
  \bibinfo{journal}{Phys. Lett.} \textbf{\bibinfo{volume}{B436}},
  \bibinfo{pages}{243} (\bibinfo{year}{1998}), \eprint{hep-ph/9807525}.

\bibitem[{\citenamefont{Bahcall}(2001{\natexlab{c}})}]{Bahcall:2001pf}
\bibinfo{author}{\bibfnamefont{J.~N.} \bibnamefont{Bahcall}}
  (\bibinfo{year}{2001}{\natexlab{c}}),
  \eprint[http://arXiv.org/abs]{hep-ph/0108148}.

\bibitem[{\citenamefont{Bahcall
  et~al.}(2001{\natexlab{d}})\citenamefont{Bahcall, Krastev, and
  Smirnov}}]{Bahcall:2001hv}
\bibinfo{author}{\bibfnamefont{J.~N.} \bibnamefont{Bahcall}},
  \bibinfo{author}{\bibfnamefont{P.~I.} \bibnamefont{Krastev}},
  \bibnamefont{and} \bibinfo{author}{\bibfnamefont{A.~Y.}
  \bibnamefont{Smirnov}}, \bibinfo{journal}{JHEP}
  \textbf{\bibinfo{volume}{05}}, \bibinfo{pages}{015}
  (\bibinfo{year}{2001}{\natexlab{d}}),
  \eprint[http://arXiv.org/abs]{hep-ph/0103179}.

\bibitem[{\citenamefont{de~Gouvea et~al.}(2000)\citenamefont{de~Gouvea,
  Friedland, and Murayama}}]{deGouvea:2000cq}
\bibinfo{author}{\bibfnamefont{A.}~\bibnamefont{de~Gouvea}},
  \bibinfo{author}{\bibfnamefont{A.}~\bibnamefont{Friedland}},
  \bibnamefont{and} \bibinfo{author}{\bibfnamefont{H.}~\bibnamefont{Murayama}},
  \bibinfo{journal}{Phys. Lett.} \textbf{\bibinfo{volume}{B490}},
  \bibinfo{pages}{125} (\bibinfo{year}{2000}), \eprint{hep-ph/0002064}.

\bibitem[{\citenamefont{Piepke}(2001)}]{KAMLAND}
\bibinfo{author}{\bibfnamefont{A.}~\bibnamefont{Piepke}}
  (\bibinfo{collaboration}{KamLAND}), \bibinfo{journal}{Nucl. Phys. Proc.
  Suppl.} \textbf{\bibinfo{volume}{91}}, \bibinfo{pages}{99}
  (\bibinfo{year}{2001}).

\bibitem[{\citenamefont{Ranucci et~al.}(2001)}]{BOREXINO}
\bibinfo{author}{\bibfnamefont{G.}~\bibnamefont{Ranucci}} \bibnamefont{et~al.}
  (\bibinfo{collaboration}{BOREXINO}), \bibinfo{journal}{Nucl. Phys. Proc.
  Suppl.} \textbf{\bibinfo{volume}{91}}, \bibinfo{pages}{58}
  (\bibinfo{year}{2001}).

\bibitem[{\citenamefont{Ricci and Villante}(2000)}]{Ricci:2000jd}
\bibinfo{author}{\bibfnamefont{B.}~\bibnamefont{Ricci}} \bibnamefont{and}
  \bibinfo{author}{\bibfnamefont{F.~L.} \bibnamefont{Villante}},
  \bibinfo{journal}{Phys. Lett.} \textbf{\bibinfo{volume}{B488}},
  \bibinfo{pages}{123} (\bibinfo{year}{2000}), \eprint{astro-ph/0005538}.

\bibitem[{\citenamefont{Bahcall et~al.}(1998)\citenamefont{Bahcall, Basu, and
  Pinsonneault}}]{BP98}
\bibinfo{author}{\bibfnamefont{J.~N.} \bibnamefont{Bahcall}},
  \bibinfo{author}{\bibfnamefont{S.}~\bibnamefont{Basu}}, \bibnamefont{and}
  \bibinfo{author}{\bibfnamefont{M.~H.} \bibnamefont{Pinsonneault}},
  \bibinfo{journal}{Phys. Lett.} \textbf{\bibinfo{volume}{B433}},
  \bibinfo{pages}{1} (\bibinfo{year}{1998}), \eprint{astro-ph/9805135}.

\bibitem[{\citenamefont{Bahcall and Pinsonneault}(1995)}]{BP95}
\bibinfo{author}{\bibfnamefont{J.~N.} \bibnamefont{Bahcall}} \bibnamefont{and}
  \bibinfo{author}{\bibfnamefont{M.~H.} \bibnamefont{Pinsonneault}},
  \bibinfo{journal}{Rev. Mod. Phys.} \textbf{\bibinfo{volume}{67}},
  \bibinfo{pages}{781} (\bibinfo{year}{1995}), \eprint{hep-ph/9505425}.

\bibitem[{\citenamefont{Chieze and Lopes}(1993)}]{TurckChieze-Lopes-93}
\bibinfo{author}{\bibfnamefont{S.~T.} \bibnamefont{Chieze}} \bibnamefont{and}
  \bibinfo{author}{\bibfnamefont{I.}~\bibnamefont{Lopes}},
  \bibinfo{journal}{Astrophys. J.} \textbf{\bibinfo{volume}{408}},
  \bibinfo{pages}{347} (\bibinfo{year}{1993}).

\bibitem[{\citenamefont{Chieze et~al.}(1993)\citenamefont{Chieze, Dappen,
  Fossat, Provostand, Schatzman, and Vignaud}}]{TurckChieze-PR230-93}
\bibinfo{author}{\bibfnamefont{S.~T.} \bibnamefont{Chieze}},
  \bibinfo{author}{\bibfnamefont{W.}~\bibnamefont{Dappen}},
  \bibinfo{author}{\bibfnamefont{E.}~\bibnamefont{Fossat}},
  \bibinfo{author}{\bibfnamefont{J.}~\bibnamefont{Provostand}},
  \bibinfo{author}{\bibfnamefont{E.}~\bibnamefont{Schatzman}},
  \bibnamefont{and} \bibinfo{author}{\bibfnamefont{D.}~\bibnamefont{Vignaud}},
  \bibinfo{journal}{Phys. Rept.} \textbf{\bibinfo{volume}{230}},
  \bibinfo{pages}{57} (\bibinfo{year}{1993}).

\bibitem[{\citenamefont{Bahcall and Pinsonneault}(1992)}]{BP92}
\bibinfo{author}{\bibfnamefont{J.~N.} \bibnamefont{Bahcall}} \bibnamefont{and}
  \bibinfo{author}{\bibfnamefont{M.~H.} \bibnamefont{Pinsonneault}},
  \bibinfo{journal}{Rev. Mod. Phys.} \textbf{\bibinfo{volume}{64}},
  \bibinfo{pages}{885} (\bibinfo{year}{1992}).

\bibitem[{\citenamefont{Bahcall and Ulrich}(1988)}]{BU88}
\bibinfo{author}{\bibfnamefont{J.~N.} \bibnamefont{Bahcall}} \bibnamefont{and}
  \bibinfo{author}{\bibfnamefont{R.~K.} \bibnamefont{Ulrich}},
  \bibinfo{journal}{Rev. Mod. Phys.} \textbf{\bibinfo{volume}{60}},
  \bibinfo{pages}{297} (\bibinfo{year}{1988}).

\end{thebibliography}
